\documentclass{aa}     

\usepackage{txfonts}
\usepackage{graphicx}
\usepackage{amssymb}
\usepackage{lscape}
\usepackage{natbib}

\newcommand{\eqn} [1] {
\begin{equation}
#1
\end{equation}}

\def\vsini {{v\!\sin\!i}}
\def\kms {\rm{km}\,\rm{s}^{-1}}

\def\teff {T_{\mathrm{eff}}}
\def\msol {{\mathrm{M}_{\odot}}}

\def\logg {\log g}
\def\Strom {Str\"omgren}
\def\muHz {\mu\mbox{Hz}}
\def\cd   {\mbox{cd}^{-1}}
\def\nuf  {\nu_{\mathrm{f}}}

\def\xc   {X_{\mathrm{c}}}
\def\amlt {\alpha_{\mathrm{MLT}}}

\def\numin {\nu_{\mathrm{min}}}
\def\numax {\nu_{\mathrm{max}}}
\def\nufund {\nu_{\mathrm{f}}}

\def\ds {$\delta$ Scuti}
\def\dss {$\delta$ Scuti stars}
\def\rvari {RV Ari}

\def\ratio {{\Pi_{1/0}}}

\def\po {{\Pi_0}}

\begin{document}

   \title{A comprehensive asteroseismic modelling of the high-amplitude \ds\ star RV Arietis}
   \authorrunning{Casas et al.}

   \author{R. Casas\inst{1}
   \and J.C. Su\'arez\inst{1,2}
   \thanks{Associate researcher at institute (2), with financial support from Spanish "Consejer\'{\i}a de Innovaci\'on, Ciencia y Empresa" from the "Junta de Andaluc\'{\i}a" local government.}
   \and A. Moya\inst{2}
   \and R. Garrido\inst{1}}

   \offprints{R. Casas\,\email{ricardo@iaa.es}}

   \institute{Instituto de Astrof\'{\i}sica de Andaluc\'{\i}a (CSIC), CP3004, Granada, Spain \and Observatoire de Paris, LESIA, UMR 8109, Meudon, France}

   \date{Received ... / Accepted ...}

   \abstract{We present a comprehensive asteroseismic study of the double-mode high-amplitude \ds\ star \object{HD\,187642} (\object{\rvari}). The modelling includes some of the most recent techniques: 1) effects of rotation on both equilibrium models and adiabatic oscillation spectrum, 2) non-adiabatic study of radial and non-radial modes, 3) relationship between the fundamental radial mode and the first overtone in the framework of Petersen diagrams. The analysis reveals that two of the observed frequencies are very probably identified as the fundamental and first overtone radial modes. Analysis of the colour index variations, together with theoretical non-adiabatic calculations, points to models in the range of [7065,7245] K in effective temperature and of [1190, 1270] Myr in stellar age. These values were found to be compatible with those obtained using the three other asteroseismic techniques.
            }

\maketitle


\section{Introduction\label{sec:intro}}


The \dss\ are intermediate-mass pulsating variables. With spectral types ranging from A2 to F0, they are located on and just off the main sequence in the lower part of the Cepheid instability strip (luminosity classes V \& IV). Nowadays, the \dss\ are considered as particularly suitable to the asteroseismic study of poorly known hydrodynamical process occurring in stellar interiors, such as convective overshoot, mixing of chemical elements, and redistribution of angular momentum \citep{Zahn92}, to mention only the most important. Due to the complexity of the oscillation spectra, their pulsating behaviour is not understood very well: see \citet{Cox02} for a complete review of unsolved problems in stellar pulsation physics.


With two observed frequencies \citep{Eloy92rvari}, the \ds\ star \object{RV~Arietis} (\object{GSC\,01217-01057}) belongs to the subclass known as high-amplitude \dss\ (HADS). Such characteristics were first attributed to \rvari\ by \citet{Eloy92rvari}, after they analysed $uvby\beta$ \Strom\ photometric observations, and then identified the observed frequencies (Table \ref{tab:freqobs_mean}) as the fundamental radial mode and its first overtone. More recently, \citet{Pocs02} examined the amplitude variation of both modes and found an additional frequency at $f_3=13.6116\,\cd$. However, the same authors claim that more observations are needed in order to confirm this third frequency and we have not used it in the present work.

Such stars have long been considered similar to double-mode Cepheids, i.e., evolved stars pulsating only in the first two radial modes. However, in the last years, low-amplitude, non-radial modes have been found together with the radial ones, making these stars more similar to the unevolved \dss, which are quite commonly distributed between the zero-age and the terminal-age main sequence stars \citep{Poretti03}. Nevertheless, \citet{Poretti05hads} have provided very recent evidence for the existence of pure radial double-mode pulsators within the HADS class and have stressed that they consitute a very homogeneous class of variable stars.


\citet{Poretti05hads} and \citet{Sua06pdrot} studied RV Ari from a theoretical point of view and in the framework of the Petersen Diagrams. While the former considered it within the subset of classic Pop. I pure radial double-mode pulsators, the latter used it (along with other similar objects of the class) to study the impact of rotation on the determination of masses and metallicities as derived from the Petersen Diagrams. The \emph{independent} information that such diagrams can provide turns out to be very helpful for modelling this kind of star. However, neither of these two works made a specific study of RV Ari.

\begin{table}
   \begin{center}
    \caption{Observed mean periods of the three detected oscillation frequencies with their corresponding mean amplitudes. The first two periods are reported by \citet{Eloy92rvari}, and the third one is proposed by \citet{Pocs02}. The last column gives the frequency ratio $f_1/f_{i=2,3}$.}
    \vspace{1em}
    \renewcommand{\arraystretch}{1.2}
    \begin{tabular}[ht!]{cccccc}
    \hline
    \hline
  Frequency     &   P(d)   &   $\nu$(c/d) & $\nu$($\muHz$)   & $A_v$ (mag)  &  $f_1/f_{i=2,3}$  \\
    \hline
  $f_1$ &  0.0931   &    10.738   &      124.233      &    0.308     &                    \\
  $f_2$ &  0.0720   &    13.898   &      160.856      &    0.091     &  0.772             \\
  $f_3$ &  0.0735   &    13.611   &      157.542      &    0.016     &  0.788             \\
    \hline
  \end{tabular}
    \label{tab:freqobs_mean}
   \end{center}
\end{table}


The present work aims at a comprehensive modelling of the HADS star \rvari. To do so, some of most updated tools adapted for this purpose are used: 1) the evolutionary code CESAM \citep{Morel97} and 2) the pulsation codes GRACO \citep{Moya04} and FILOU \citep{filou,SuaThesis}. The GRACO provides the non-adiabatic quantities related to pulsation and includes the atmosphere-pulsation interaction described in \citet{Dupret02}. The FILOU code includes the effects of rotation on adiabatic oscillations up
to the second order in a perturbative theory. Using these tools we performed a massive numerical study of \rvari\ to attempt to constrain physical and theoretical parameters with a precision never obtained for any other \ds\ star. Particularly, the following aspects are studied:
\begin{itemize}
   \item[$\bullet$] the effects of rotation on both the oscillation frequencies and the equilibrium models;

   \item[$\bullet$] the use of the mixing-length theory for describing convection in equilibrium models;

   \item[$\bullet$] the calculation of non-adiabatic observables, needed for instability analysis, are also used for mode identification within the framework of multicolour photometry;

   \item[$\bullet$] the use of updated Kurucz models for a detailed study of the atmosphere-pulsation interaction,

   \item[$\bullet$] analysis of \rvari\ in the framework of the Petersen diagrams (PD), including the rotation effects.
\end{itemize}
Such an exhaustive modelling intends not only to better constrain the set of representative models of the star but also to introduce a comprehensive and self-consistent methodology for interpreting the oscillation spectra of \dss.

\begin{table*}
   \begin{center}
    \caption{Observed amplitudes and phases (referred to $y$) and their corresponding uncertainties in the four filters of the \Strom\ system.}
    \vspace{1em}
    \renewcommand{\arraystretch}{1.2}
    \begin{tabular}{ccccccc}
      \hline
      \hline
        & $\phi_u-\phi_y$ & $\phi_v-\phi_y$ & $\phi_b-\phi_y$ & $A_u/A_y$ & $A_v/A_y$ & $A_b/A_y$ \\
      \hline
      $f_1$ & $6.2\pm2.0$ & $1.8\pm1.7$ & $1.2\pm1.7$ & $1.086\pm0.033$ & $1.394\pm0.037$ & $1.204\pm0.034$ \\
      $f_2$ & $3.8\pm6.7$ & $-1.7\pm5.8$ & $-2.1\pm5.7$ & $1.197\pm0.124$ & $1.492\pm0.139$ & $1.328\pm0.131$ \\
      \hline
    \end{tabular}
    \label{tab:freqobs_uvbV}
   \end{center}
\end{table*}
The paper is structured as follows: equilibrium models and oscillation computations are described in Sect. \ref{sec:models}. The fundamental parameters of \rvari\ and the location of the star in the HR diagram are given in Sect. \ref{sec:HRdiag}. Then, a non-adiabatic analysis is performed in Sect. \ref{sec:nonad}, including studies of mode-instability ranges and convective efficiency. Section \ref{sec:identif} focuses on the problem of mode identification, which is undertaken by considering amplitude/phase diagrams in the framework of multicolour photometry, and examining the variation of colour
indices as a function of the star position in the HR diagram. In Sect. \ref{sec:effrot} the effects of rotation are considered, and in Sect. \ref{sec:pd}, Petersen diagrams for this star are analysed. Finally, conclusions are reported in Sect. \ref{sec:conclusions}.

\section{The modelling \label{sec:models}}
In this section we describe the equilibrium models, the theoretical oscillation computations and the
linear stability analysis performed for the complete modelling of \rvari.

\subsection{Equilibrium models\label{ssec:eqmodels}}
The stellar models were computed with the evolutionary code CESAM \citep{Morel97}. A grid of around 2000 B-spline basis points was used in order to optimise the numerical precision of oscillation
frequencies and eigenfunctions.

The equation of state CEFF \citep{ceff} was used, for which the Coulombian correction to the classical EFF \citep{Eggleton73} was included. The $pp$ chain, as well as the CNO cycle nuclear reactions, were considered and standard species from $^1$H to $^{17}$O included. Opacity tables were taken from the OPAL package \citep{Igle96}, complemented at low temperatures ($T\leq10^3\,K$) by the tables provided by \citet{AlexFergu94}. In the atmosphere, the Eddington $T(\tau)$ law (grey approximation) was considered for models used to compute rotating adiabatic oscillations. For non-rotating, non-adiabatic calculations, a \citet{Kurucz98} model atmosphere was adopted. A metallicity of $Z=0.018$ was used, but allowed to vary when PD were analysed.
\begin{figure}
 \begin{center}
   \includegraphics[width=8.5cm]{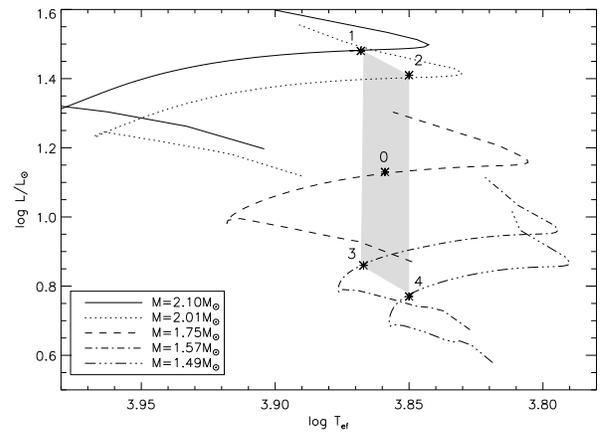}
   \caption{HR diagram containing the observational error box (shaded area) for \rvari. Asterisks correspond to models representative of the star, labelled from 0 (which corresponds to the observed mean values) to 4, covering the extreme regions of the error box (more details in the text).}
   \label{fig:HR-rvari}
 \end{center}
\end{figure}

Rotation effects on equilibrium models (\emph{pseudo}-rotating models) were considered by modifying the equations \citep{KipWeig90} to include the spherically symmetric contribution of the centrifugal acceleration by means of an effective gravity $g_{\mathrm{eff}}=g-{\cal A}_c(r)$, where $g$ corresponds to the local gravity and ${\cal A}_c(r)$ represents the radial component of the centrifugal acceleration.
During evolution, models were assumed to rotate as a rigid body, and their total angular momentum was conserved.
\begin{figure*}
 \begin{center}
  \scalebox{.35}{\includegraphics{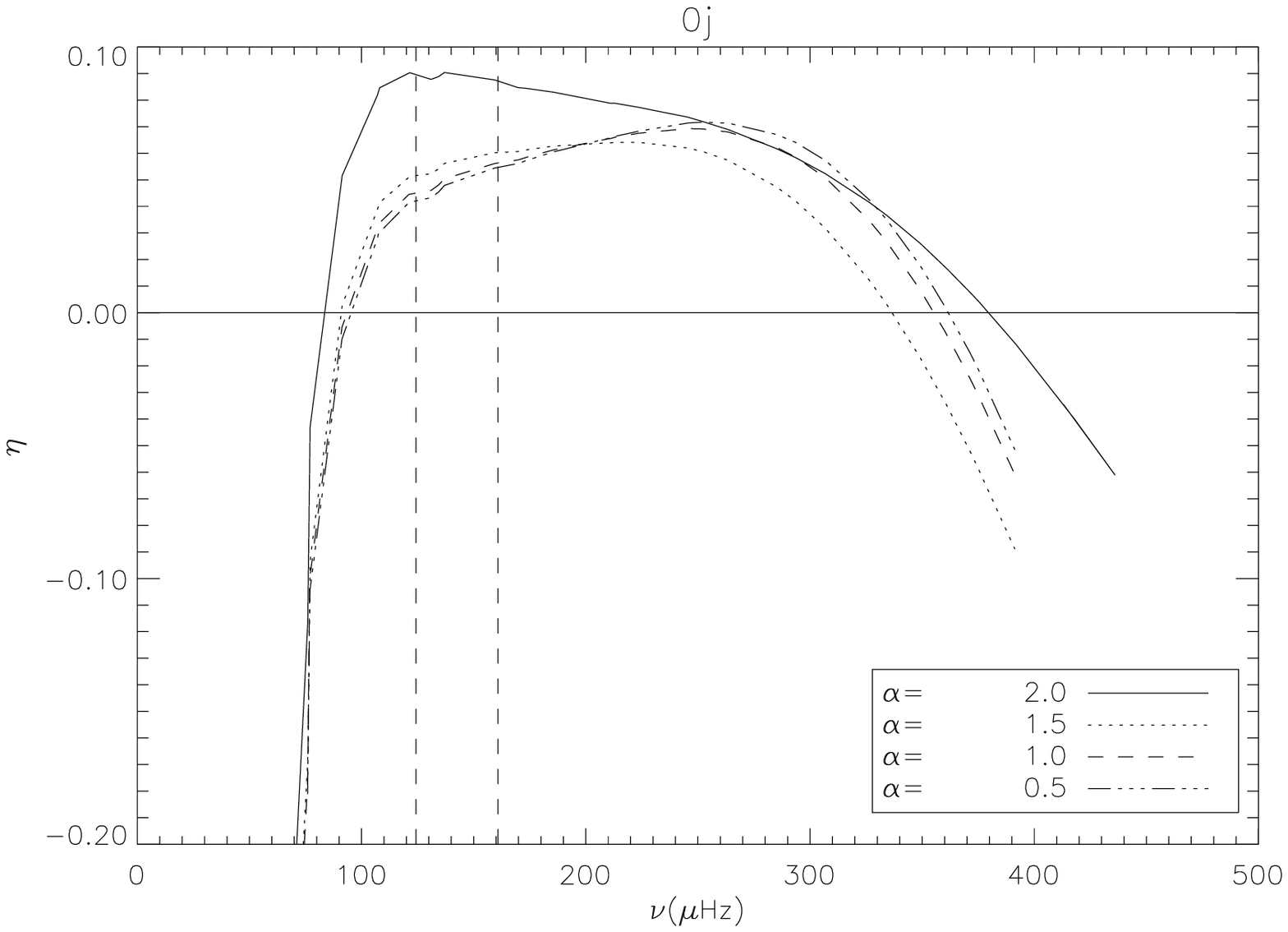}}
  \scalebox{.35}{\includegraphics{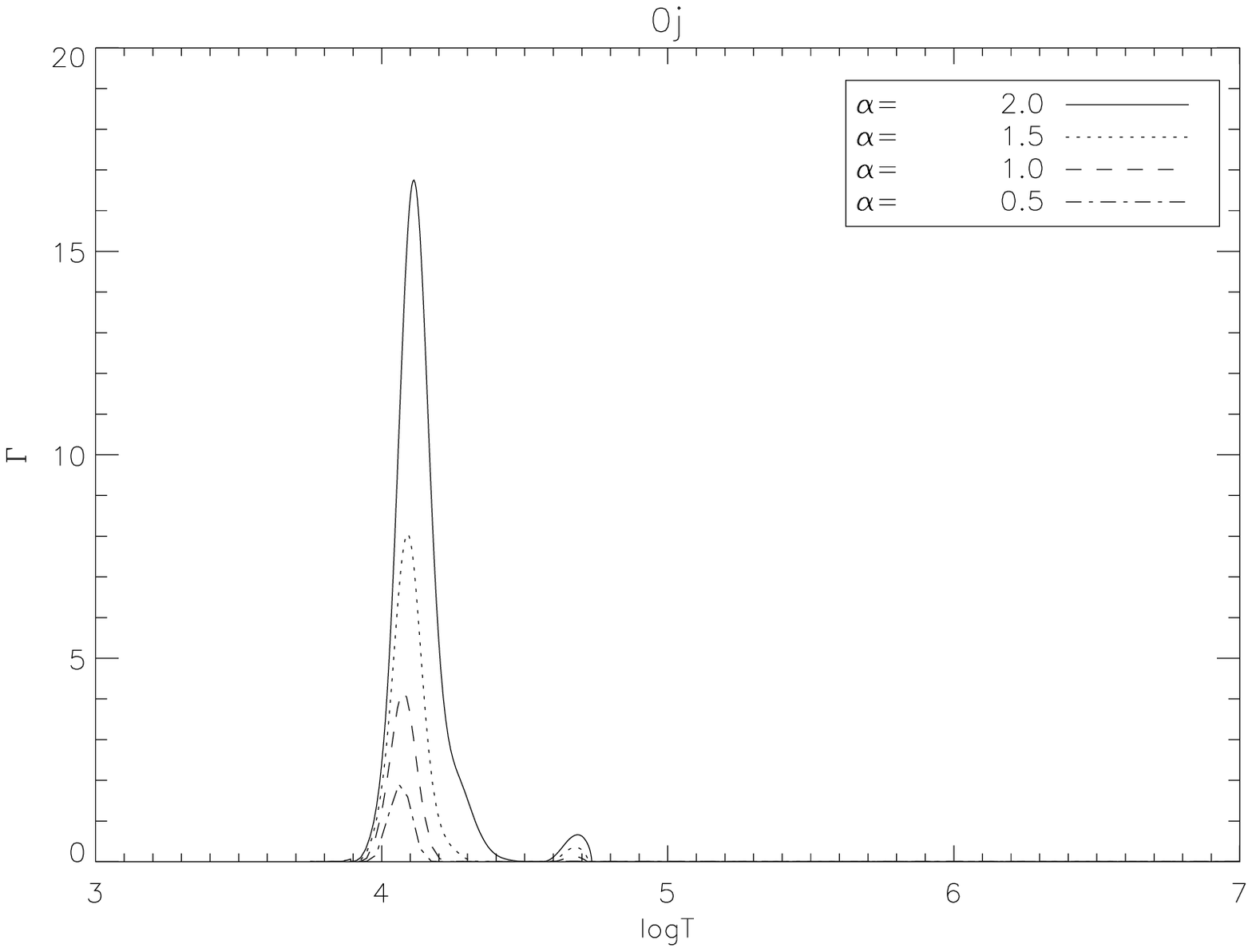}}
  \scalebox{.35}{\includegraphics{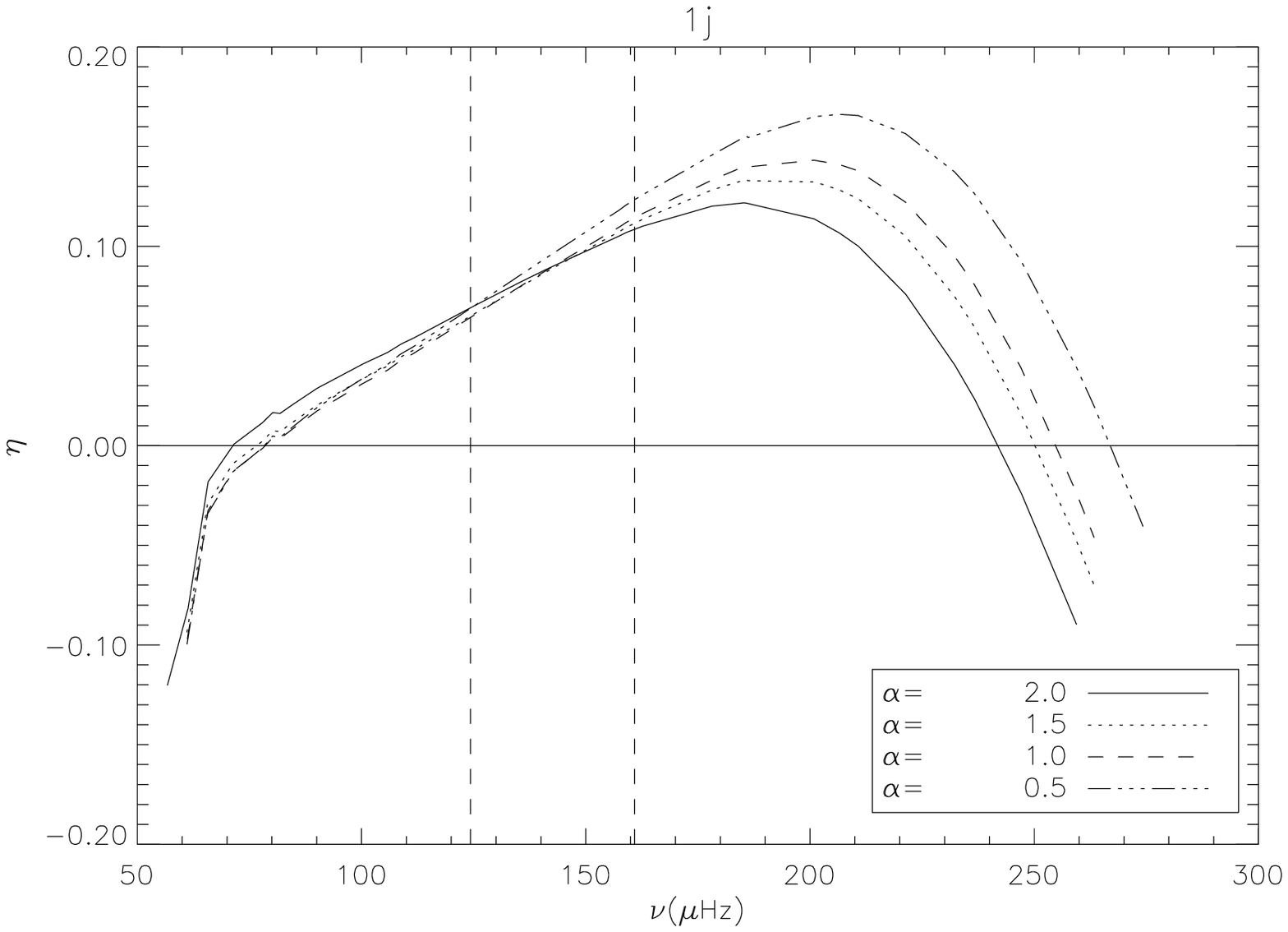}}
  \scalebox{.35}{\includegraphics{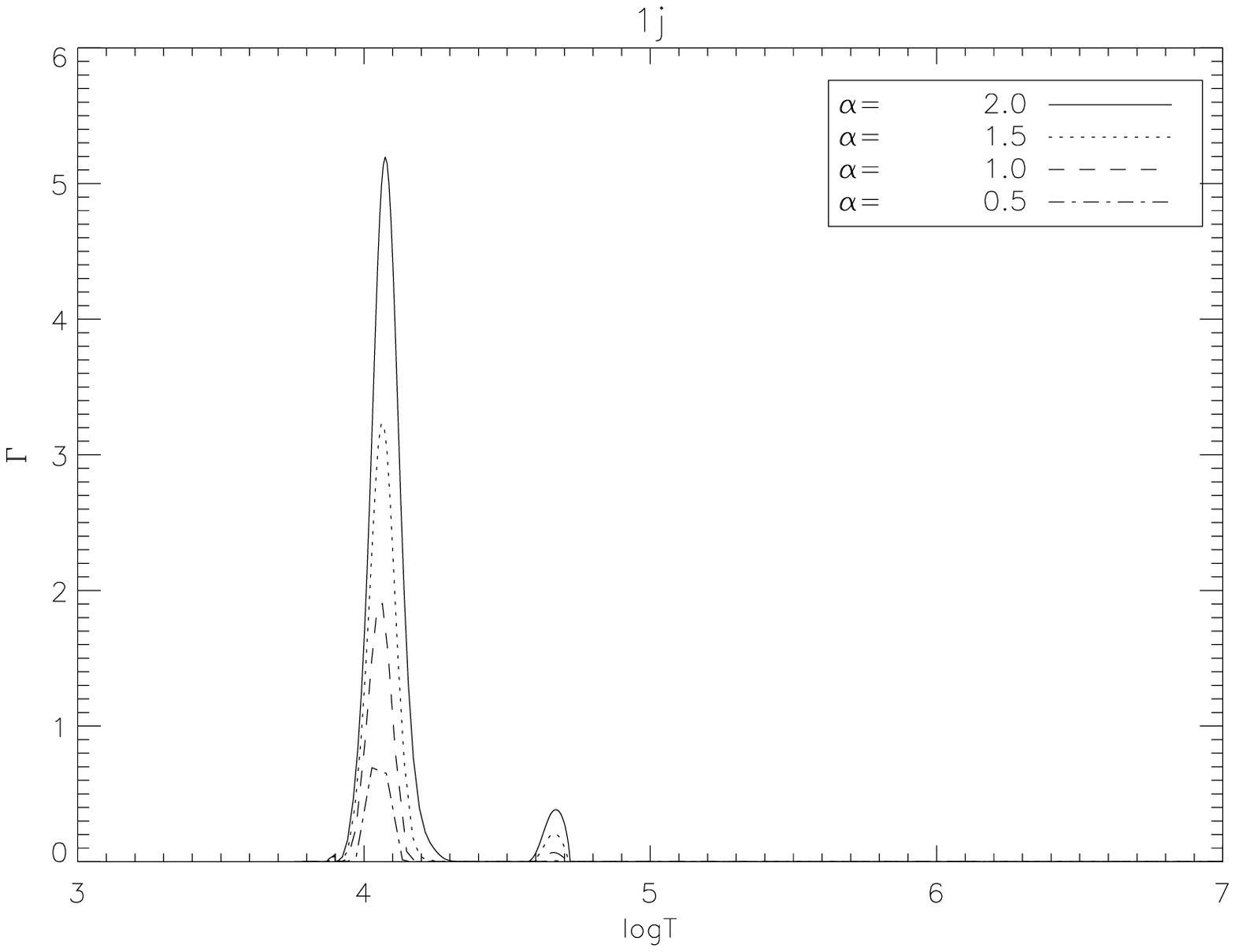}}
  \scalebox{.35}{\includegraphics{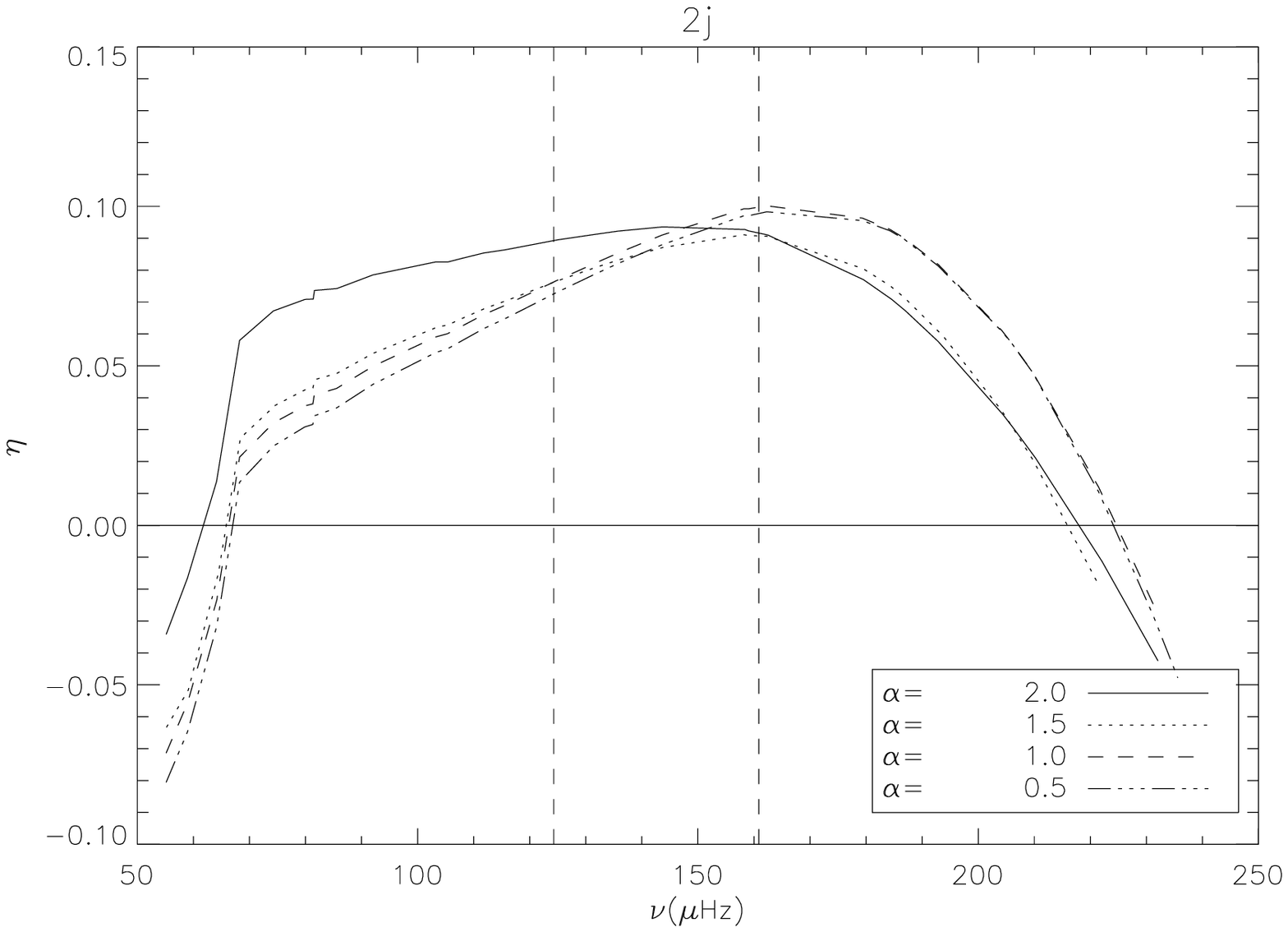}}
  \scalebox{.35}{\includegraphics{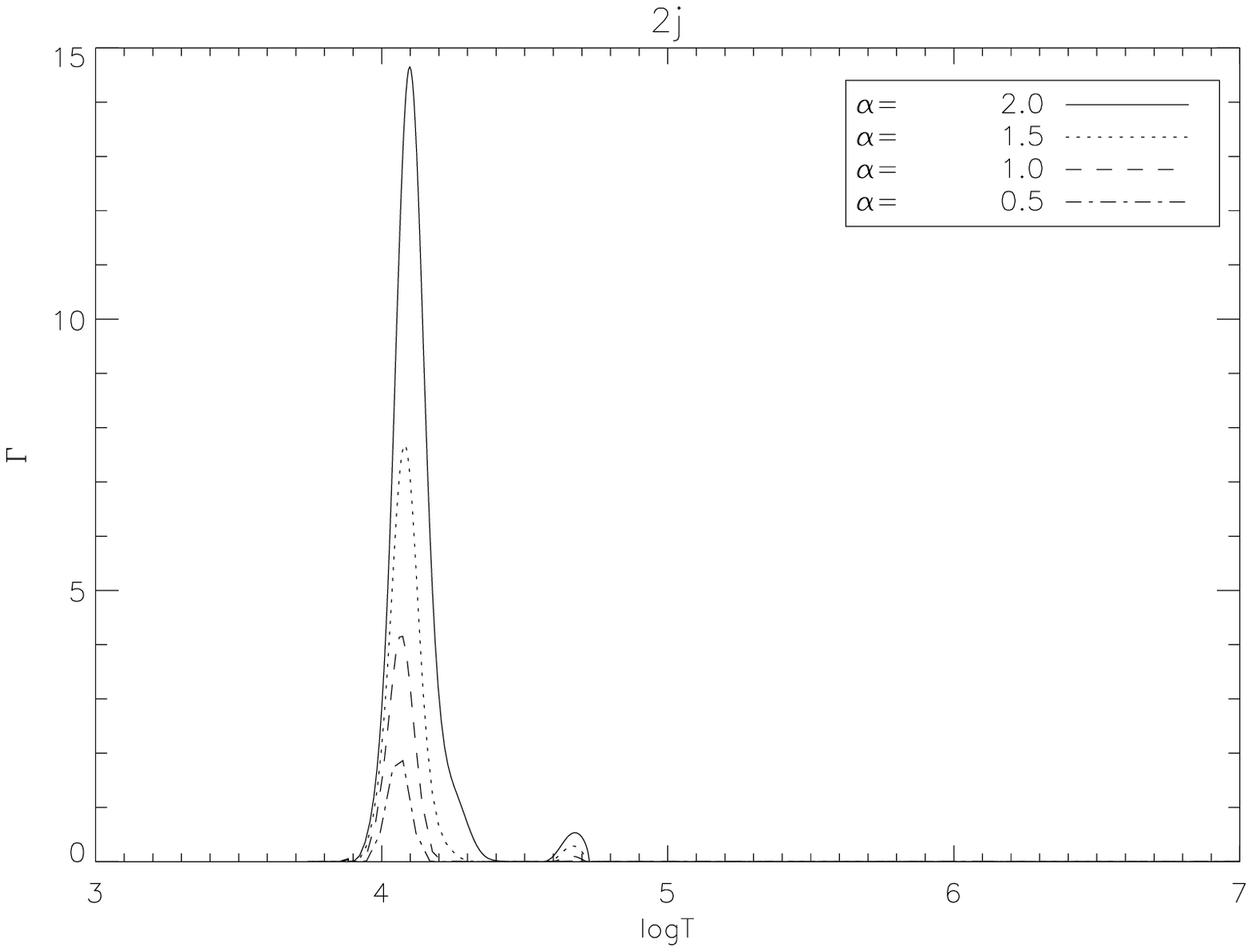}}
  \scalebox{.35}{\includegraphics{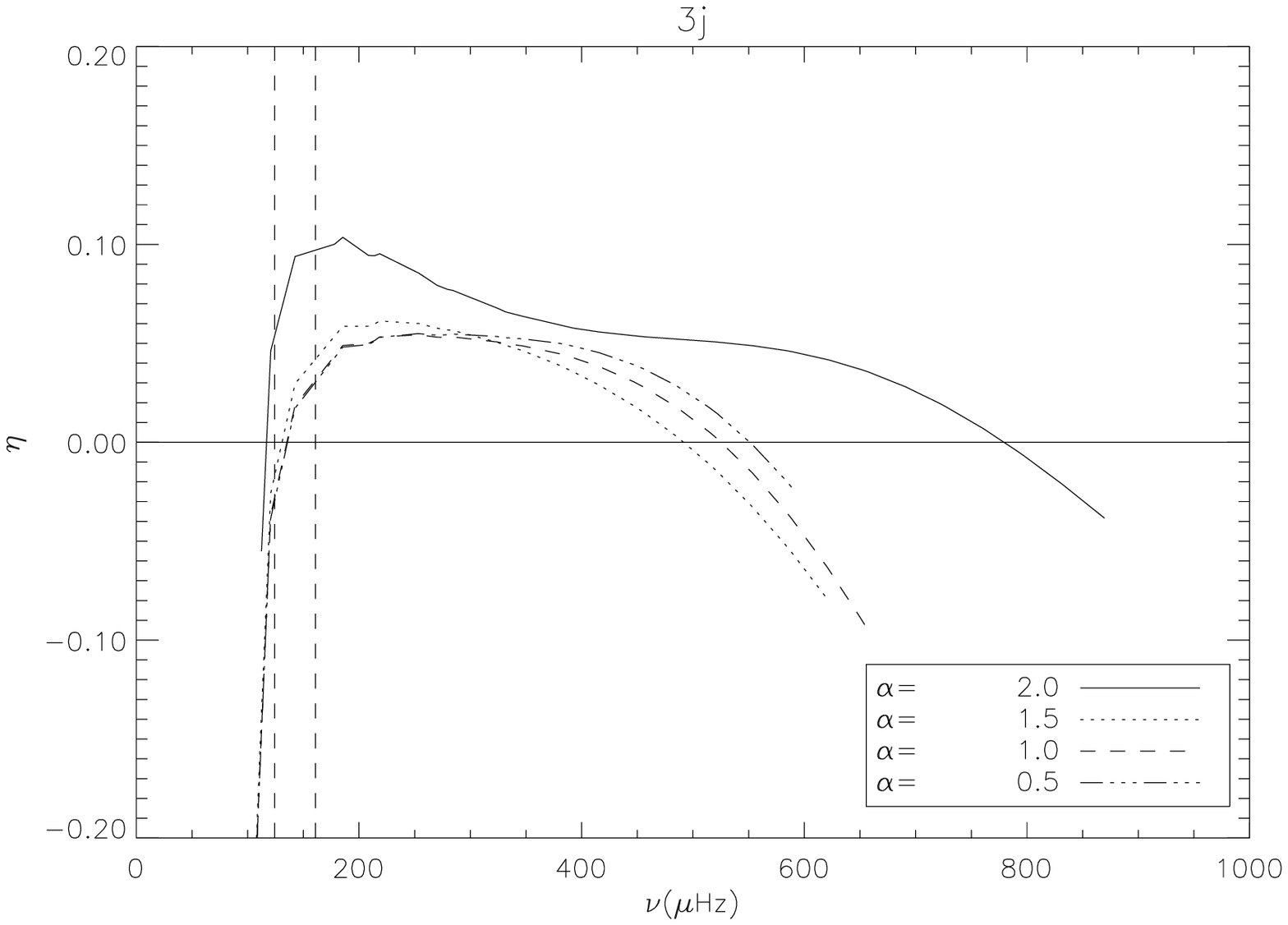}}
  \scalebox{.35}{\includegraphics{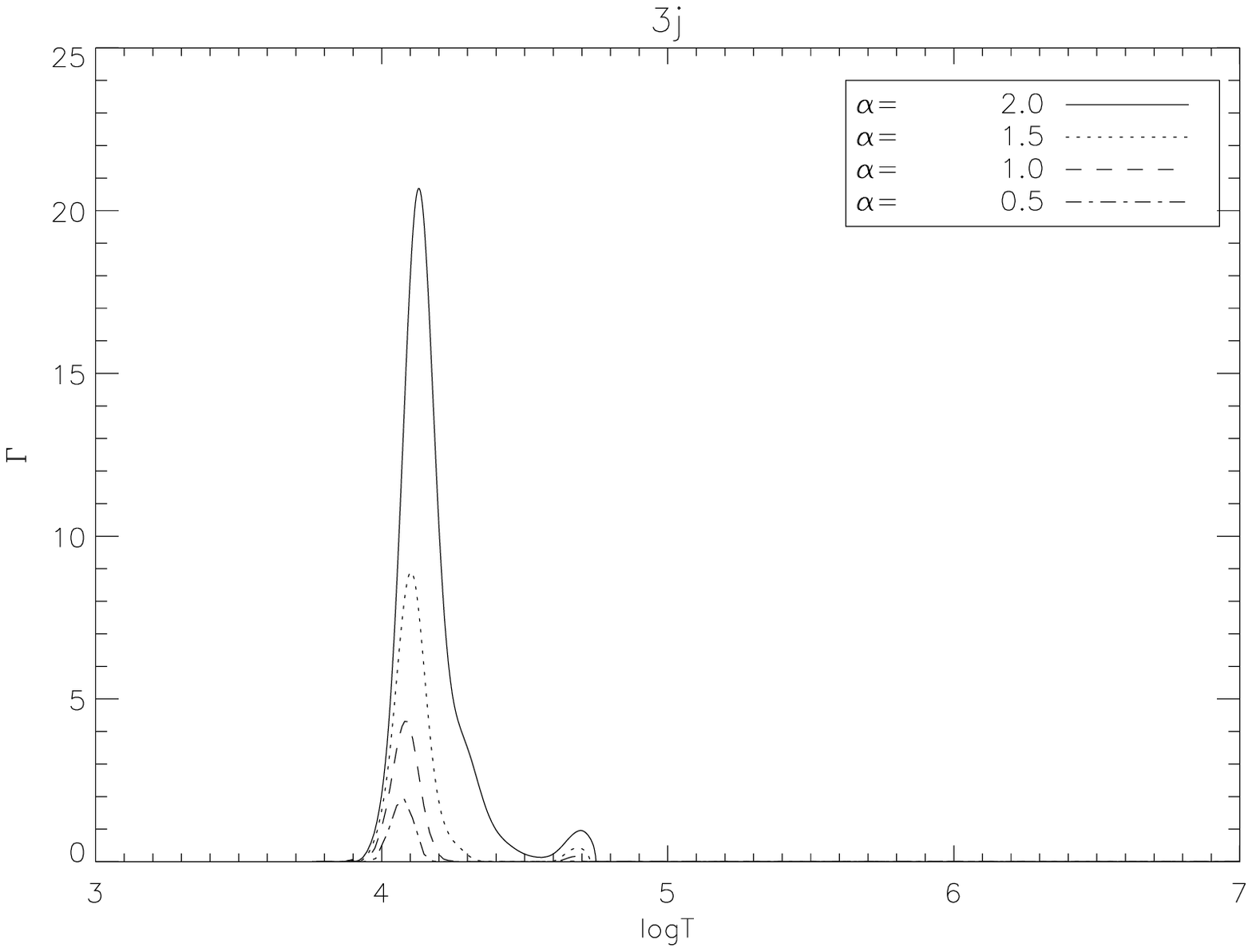}}
  \scalebox{.35}{\includegraphics{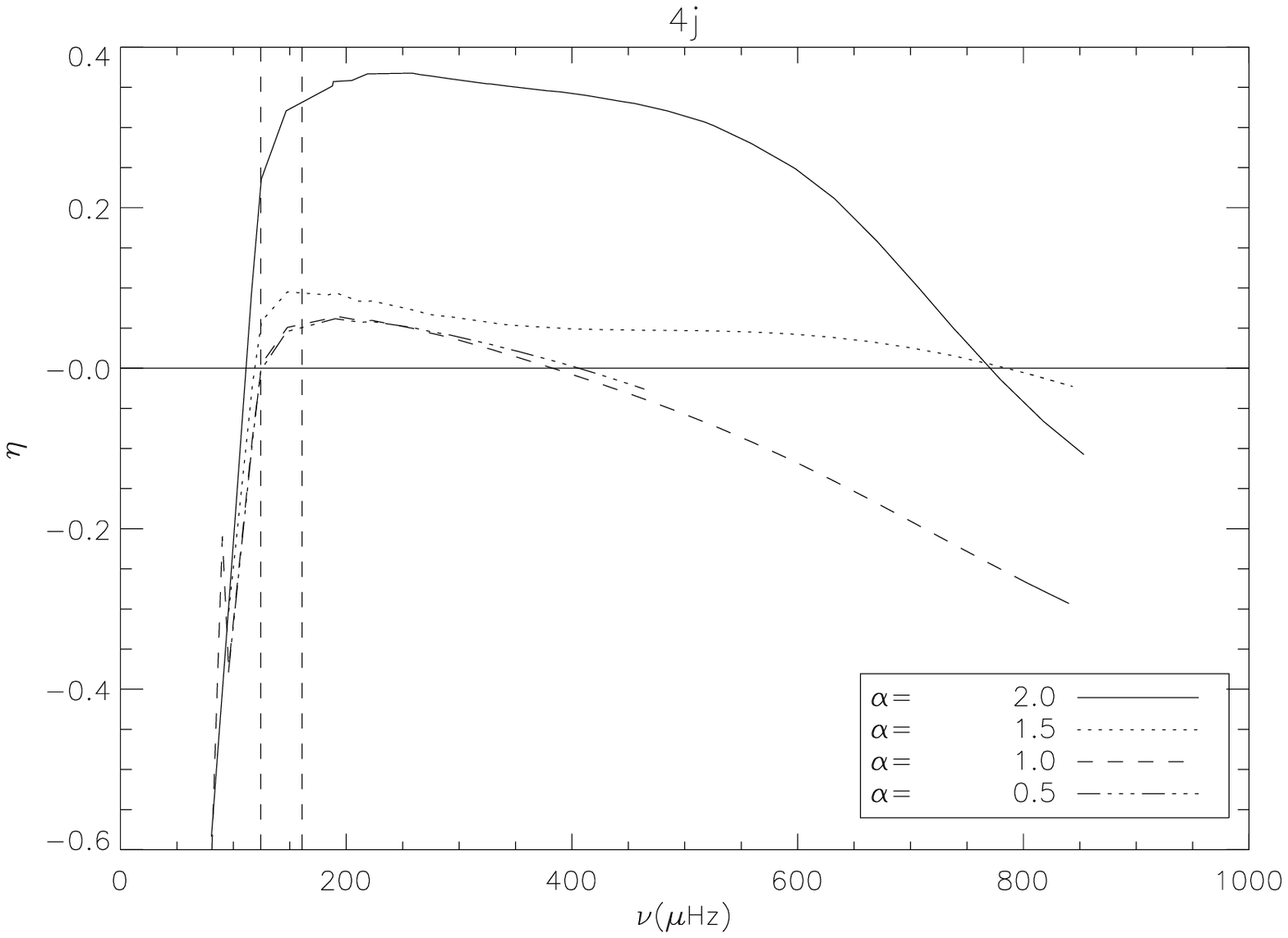}}
  \scalebox{.35}{\includegraphics{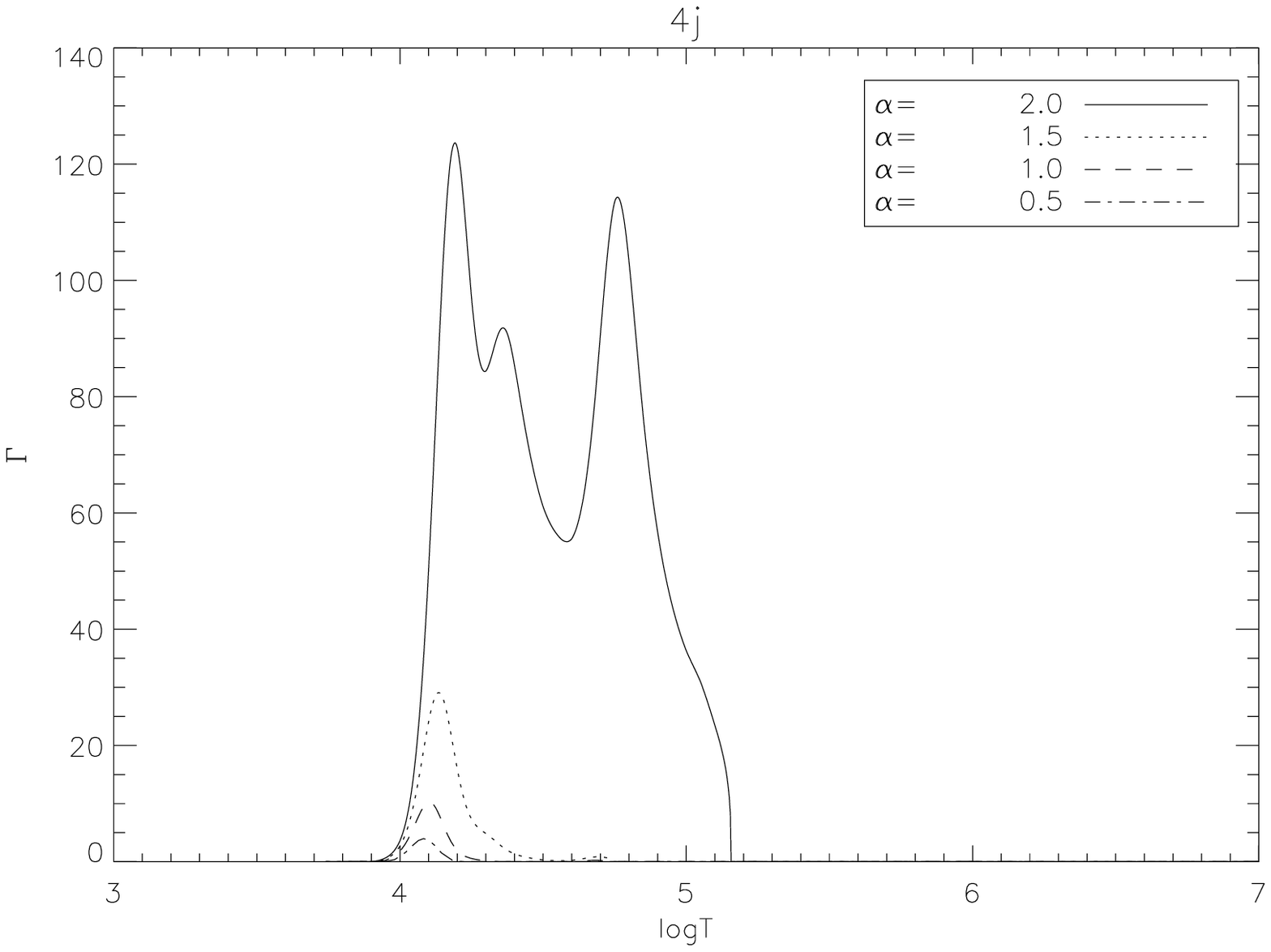}}
  \caption{Predicted growth rates for modes in the range of the observed frequencies (left panels), and convective efficiency (defined in Sect. \ref{ssec:instab}) as a function of the temperature (in logarithmic scale) along the star radius (right panels). Different line types correspond to different values of the mixing-length parameter $\alpha$. The degree of modes are considered up to $\ell=3$ and radial orders $n$ are those required to cover the instability range. Vertical lines represent the observed frequencies $f_1$ and $f_2$.}
   \label{fig:eta_conveff}
 \end{center}
\end{figure*}

\subsection{Oscillation computations \& linear stability analysis \label{ssec:oscil-insta}}

Theoretical oscillation spectra were computed from the equilibrium models described in the previous
section. For this purpose two oscillation codes were used:
\begin{enumerate}
  \item The Granada Oscillation COde \citep[GRACO,][]{Moya04} provided diagnostics on the instability and non-adiabatic observables required for fitting the multicolour photometric observations. In this code the non-adiabatic pulsation equations are solved mainly following \citet{Unno89}, including the non-adiabatic pulsation treatment of the atmosphere proposed by \citet{Dupret02}. Although including this atmosphere-pulsation interaction does not have any significant influence on the modal stability, it does allow theoretical predictions to be compared with photometric colour observations \citep[see][ for more details]{Moya04}. The atmospheres were reconstructed at specific Rosseland optical depths ($\tau_{\mathrm{Ross}}= 1$) until the last photospheric edge of the star was reached.

  \item For rotating adiabatic oscillations computations, the oscillation code FILOU \citep{SuaThesis} provides adiabatic oscillation modes corrected for the effects of rotation up to second order (centrifugal and Coriolis forces), including second-order effects of near degeneracy that are expected to be significant for moderately high rotational velocities \citep{Soufi98}. Indeed, two or more modes, close in frequency, are rendered \emph{degenerate} by rotation under certain conditions, corresponding to selection rules. In particular, these rules select modes with the same azimuthal order $m$ and degrees $\ell$ differing by 0 or 2 \citep{Soufi98}. If we consider two generic modes $\alpha_1\equiv(n,\ell,m)$ and $\alpha_2\equiv(n^\prime,\ell^\prime,m^\prime)$ under these conditions, near degeneracy occurs for $|\sigma_{\alpha_1}-\sigma_{\alpha_2}|\leq\sigma_{\Omega}$, where $\sigma_{\alpha_1}$ and $\sigma_{\alpha_2}$ represent the eigenfrequency associated to modes $\alpha_1$ and $\alpha_2$, respectively, and $\sigma_\Omega$ represents the stellar rotational frequency \citep[see][ for more details]{Goupil00}.
\end{enumerate}

\section{Locating \rvari\ in the HR diagram\label{sec:HRdiag}}

The fundamental parameters of \rvari\ ($\log\teff=3.859\pm0.09$, $\logg=3.94\pm0.25$, $\log L/L_{\sun}=1.11\pm0.25$, [M/H]= 0.01$\pm0.20$ and $\vsini\lesssim 18\,\kms$) were kindly provided by E. Rodriguez, who derived them from data reported in Rodriguez et al. (1992, 2000; and references therein)\nocite{Eloy92rvari,eloy00catalog}. This allows us to depict an error box in which our models can in principle be considered as representative of the star (see Fig. \ref{fig:HR-rvari}). On the other hand, rotational velocity, as well as the inclination angle, modifies the location of stars in the HR diagram. \citet{MiHer99} propose a method of determining such effects on photometric parameters.
\begin{table*}
   \begin{center}
    \caption{Main characteristics of computed models representative of \rvari. Labels ranging from 0 to 4 indicate the location of the models in the HR diagram displayed in Fig. \ref{fig:HR-rvari}. From left to right, $M$ represents the stellar mass in solar masses $\msol$; $\teff$ the effective temperature in K (on a logarithmic scale); $g$ the surface gravity in cgs (on a logarithmic scale); $\xc$ the central hydrogen mass fraction; the age in Myr; $\alpha$ the mixing-length parameter; $\numin$--$\numax$ represents the frequency range ($\muHz$) of predicted unstable modes; $\Delta z/H_p$ the size of the external convective zone normalised by the local pressure scaleheight measured at the bottom of the convection zone; $\Delta z/R_*$ the fraction of stellar radius occupied by the external convective zone; $\nufund$ the fundamental radial mode (in $\muHz$); and finally, $\bar \rho$ represents the stellar mean density (in cgs).}
    \vspace{1em}
    \renewcommand{\arraystretch}{1.2}
    \begin{tabular}[ht!]{ccccccccccccc}
    \hline
    \hline
 ID (ij) & $M$ & $\teff$ & $g$ & $\xc$ & Age & $\alpha$ & $\numin$ & $\numax$ & $\Delta z/H_p$ & 
     $\Delta z/R_*$ & $\nufund$ & $\bar \rho$ \\
      \hline
      00 & 1.75 & 3.859 & 3.94 & 0.302 & 1186 & 2.0 & 78 & 380 & 1.16 & 0.33 & 130.98 & 0.192\\
      01 & 1.75 & 3.859 & 3.94 &  0.302 & 1186 & 1.5 & 79 & 336 & 0.82 & 0.18 & 131.03 & 0.192\\
      02 & 1.75 & 3.859 & 3.94 &  0.303 & 1186 & 1.0 & 95 & 355 & 0.71 & 0.14 & 131.10 & 0.192\\
      03 & 1.75 & 3.859 & 3.94 &  0.303 & 1186 & 0.5 & 98 & 361 & 0.66 & 0.13 & 131.13 & 0.192\\
      \hline
      10 & 2.10 & 3.867 & 3.70 &  0.132 & 860 & 2.0 & 71 & 242 & 0.81 & 0.20 & 81.82 & 0.076\\
      11 & 2.10 & 3.868 & 3.70 &  0.132 & 860 & 1.5 & 77 & 250 & 0.75 & 0.18 & 81.82 & 0.076\\
      12 & 2.10 & 3.868 & 3.70 &  0.132 & 860 & 1.0 & 78 & 255 & 0.74 & 0.17 & 81.82 & 0.076\\
      13 & 2.10 & 3.868 & 3.70 &  0.132 & 860 & 0.5 & 78 & 267 & 0.71 & 0.16 & 81.83 & 0.076\\
      \hline
      20 & 2.01 & 3.850 & 3.69 &  0.121 & 985 & 2.0 & 63 & 218 & 1.03 & 0.33 & 81.32 & 0.075 \\
      21 & 2.01 & 3.850 & 3.69 &  0.120 & 985 & 1.5 & 66 & 216 & 0.70 & 0.18 & 81.32 & 0.075\\
      22 & 2.01 & 3.850 & 3.69 &  0.120 & 985 & 1.0 & 66 & 225 & 0.70 & 0.18 & 81.32 & 0.075\\
      23 & 2.01 & 3.850 & 3.69 & 0.120 & 985 & 0.5 & 67 & 224  & 0.63 & 0.15 & 81.32 & 0.075\\
      \hline
      30 & 1.57 & 3.868 & 4.19 &  0.538 & 850 & 2.0 & 117 & 780 & 1.95 & 0.80 & 208.59 & 0.483\\
      31 & 1.57 & 3.868 & 4.19 &  0.538 & 850 & 1.5 & 124 & 491 & 0.87 & 0.16 & 208.50 & 0.483\\
      32 & 1.57 & 3.867 & 4.19 &  0.538 & 850 & 1.0 & 127 & 525 & 0.74 & 0.12 & 208.59 & 0.483\\
      33 & 1.57 & 3.867 & 4.19 &  0.538 & 850 & 0.5 & 127 & 551 & 0.69 & 0.11 & 208.49 & 0.483\\
      \hline
      40 & 1.49 & 3.850 & 4.19 &  0.521 & 1080 & 2.0 & 111 & 770 & 2.58 & 2.71 & 205.05 & 0.485\\
      41 & 1.49 & 3.850 & 4.19 &  0.540 & 985 & 1.5 & 119 & 785 & 1.83 & 0.77 & 211.31 & 0.494\\
      42 & 1.49 & 3.849 & 4.19 &  0.542 & 980 & 1.0 & 125 & 383 & 0.80 & 0.15 & 211.61 & 0.494\\
      43 & 1.49 & 3.850 & 4.19 &  0.543 & 975 & 0.5 & 125 & 406 & 0.67 & 0.12 & 212.00 & 0.494\\
      \hline
    \hline
  \end{tabular}
    \label{tab:models-instab}
   \end{center}
\end{table*}
This method was further developed by \citet{Pe99} for \dss, showing that uncertainties around $100-150\,K$ in effective temperature and $\sim0.10\,\mbox{dex}$ in $\logg$ can be found for moderately rotating stars. That result was later confirmed by \citet{Sua02aa}. In the present case, considering the absence of additional information on the inclination angle of the star, we considered an error box of $\sim150\,\mbox{K}$ in $\teff$ and $\sim0.25\,\mbox{dex}$ in $\logg$. Moreover, since \rvari\ shows a relatively small projected rotational velocity, these uncertainties remain within the typical values
for \dss.

\section{Non-adiabatic analysis\label{sec:nonad}}

Equilibrium non-rotating models were computed in the manner described in Sect. \ref{sec:models}.
In order to avoid redundant (and thereby unnecessary) model computations, the centre and the four corners of the photometric error box, whose characteristics are summarised in Table \ref{tab:models-instab}, were considered. Models are identified (ID) with labels $ij$ with $i=0,...,4$ and $j=0,...,3$. The index $i$ indicates the location in the HR diagram ($0j$ corresponds to central models represented in Fig. \ref{fig:HR-rvari}). For each $i$, i.e., for each selected location, $j$ indicates a different value of the convective efficiency (MLT) parameter $\alpha\equiv\amlt$ used (see Sect. \ref{sec:models}).
This procedure allowed us to discard non-satisfactory models, thereby bounding the region in which models can be considered as representative of \rvari.

\subsection{Instability ranges \& convective efficiency\label{ssec:instab}}

We basically attempted to significantly reduce the set of representative models by examining the region of the photometric box where modes are predicted to be overstable.
\begin{figure}
 \begin{center}
  \scalebox{.50}{\includegraphics{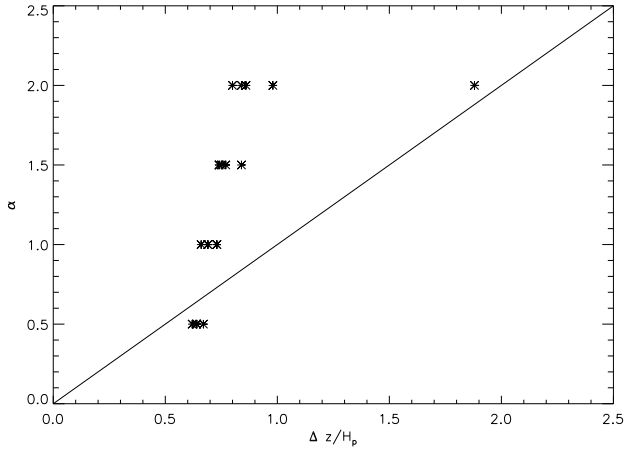}}
  \caption{Values of $\alpha$ used in models of Table \ref{tab:models-instab} as a function of their corresponding $\Delta z/H_p$ value. The straight line represents the limit of validity of MLT (see discussion in the text).}
  \label{fig:a-zhp}
 \end{center}
\end{figure}
Predicted growth rates, $\eta$ (positive for unstable modes), obtained for modes in the range of the observed frequencies, are depicted in Fig. \ref{fig:eta_conveff} (left panels) for the models considered. Since $\eta$ is a discrete function of $\nu$, the instability ranges given in Table \ref{tab:models-instab} were obtained by interpolating the value for which $\eta$ changes its sign. According to this, almost all the selected models predict an unstable observed frequency range. The main difference concerns the size of the region (in frequency) of unstable modes, while the modes at the upper boundary of the instability region may have very different frequencies. The total number of unstable modes decreases when the mass of the model increases. For the less massive ones, the predictions for $\alpha=2$ are quite different to those obtained for $\alpha\leq 1$. In particular, for the last value, large radial order $n$ modes become stable. This can be explained by examining the convective efficiency defined \citep{Cox80} as
\eqn{\Gamma=\Big[\frac{A^2}{a_0}(\nabla_{\mathrm{rad}}-\nabla)\Big]^{1/3}, \label{eq:effconv}}
where $\nabla_{\mathrm{rad}}$ and $\nabla$ represent the radiative and temperature gradients, respectively. Here, $A$ basically represents the ratio between the convective and radiative conductivities, and $a_0$ is a constant.

As shown in Fig. \ref{fig:eta_conveff} (right panels), the convective efficiency obtained from models $0j$ and $2j$ varies significantly with respect to the $4j$ models. On the other hand, models $4j$ present a very different convective efficiency for $\alpha=2$ compared to the other values. In general, the larger $\alpha$, the larger the size of the convective zone, the more efficient the convection, and the more important the influence of convection in the pulsations. Intermediate stars in the temperature range of \rvari, i.e., approximately from 7380 K to 7080 K, show two thin external convective zones. High $\alpha$ values may give rise to extend the size of the convective zones artificially, which are then overestimated. Such a scenario should be carefully considered since the limit of validity for MLT may be exceeded.
\begin{figure*}
 \begin{center}
    \scalebox{.25}{\includegraphics{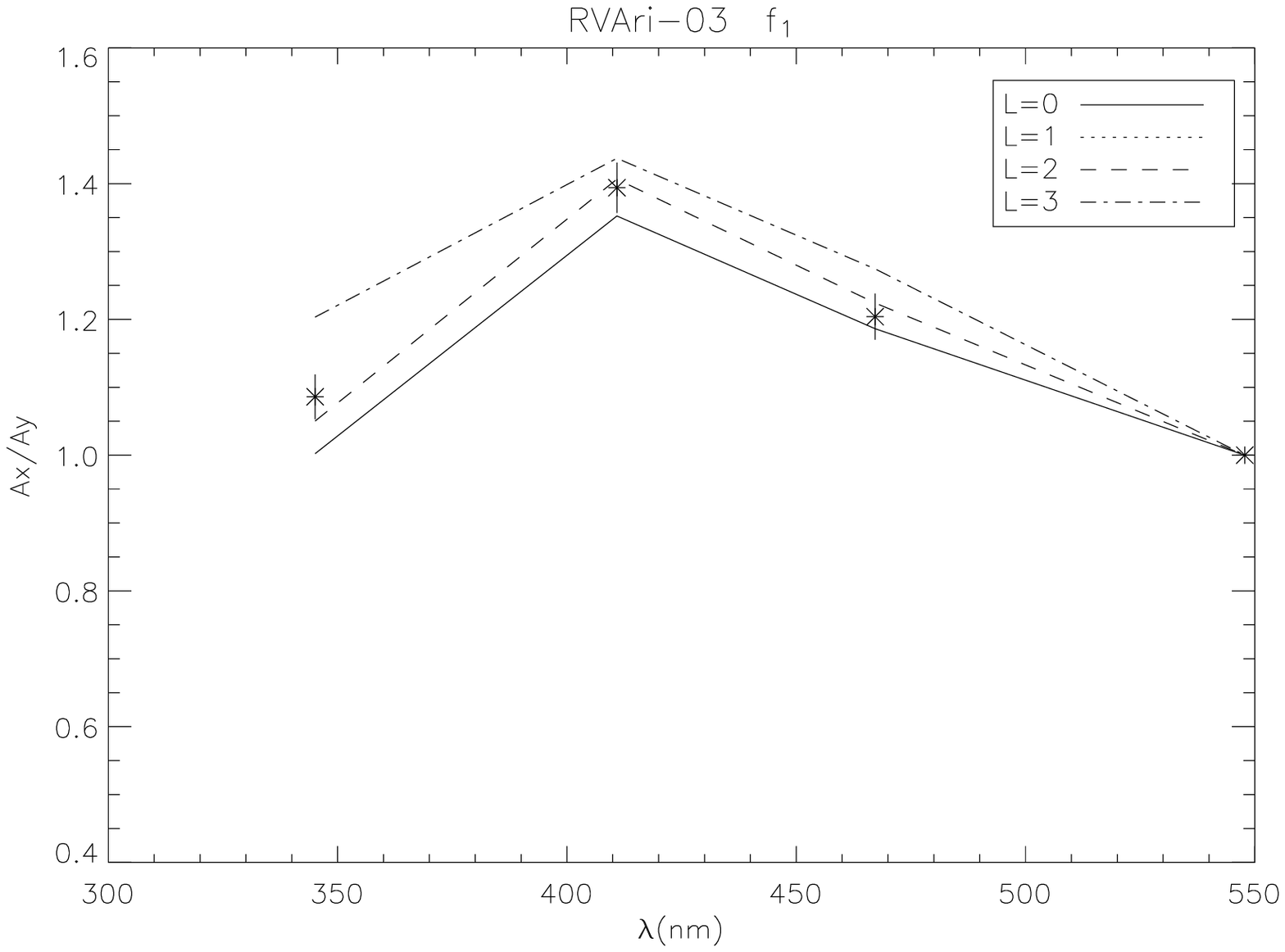}}
    \scalebox{.25}{\includegraphics{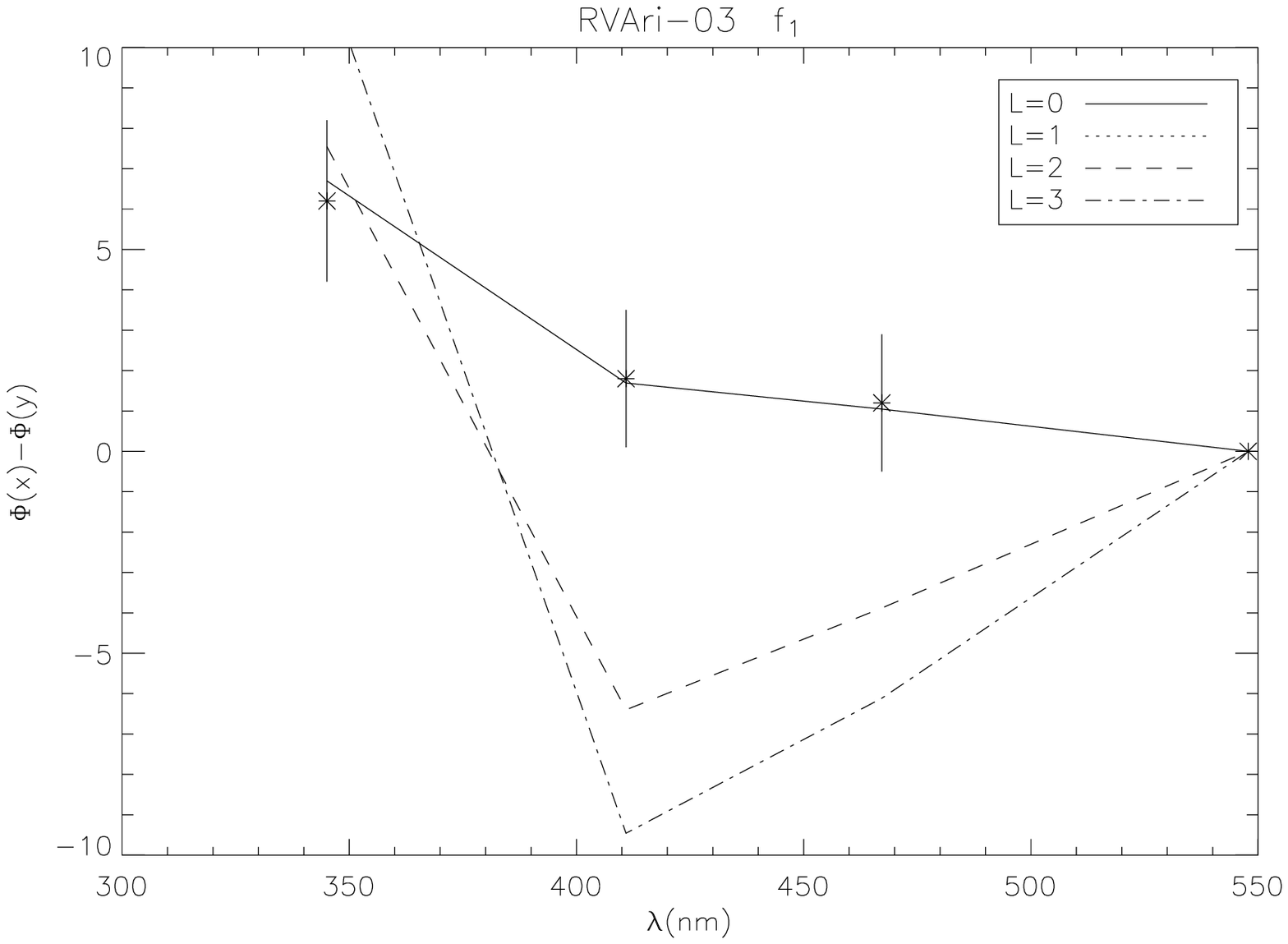}}
    \scalebox{.25}{\includegraphics{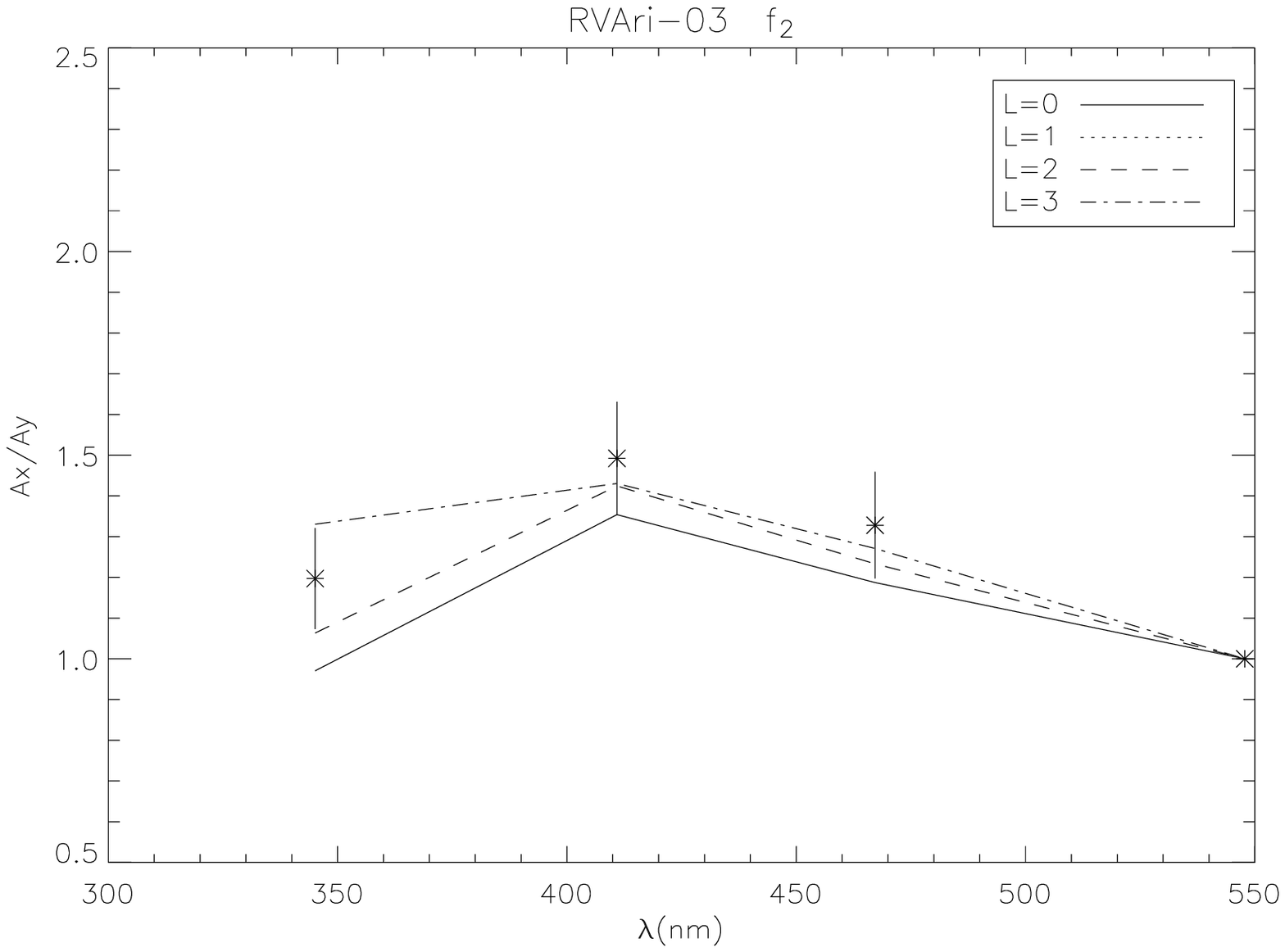}}
    \scalebox{.25}{\includegraphics{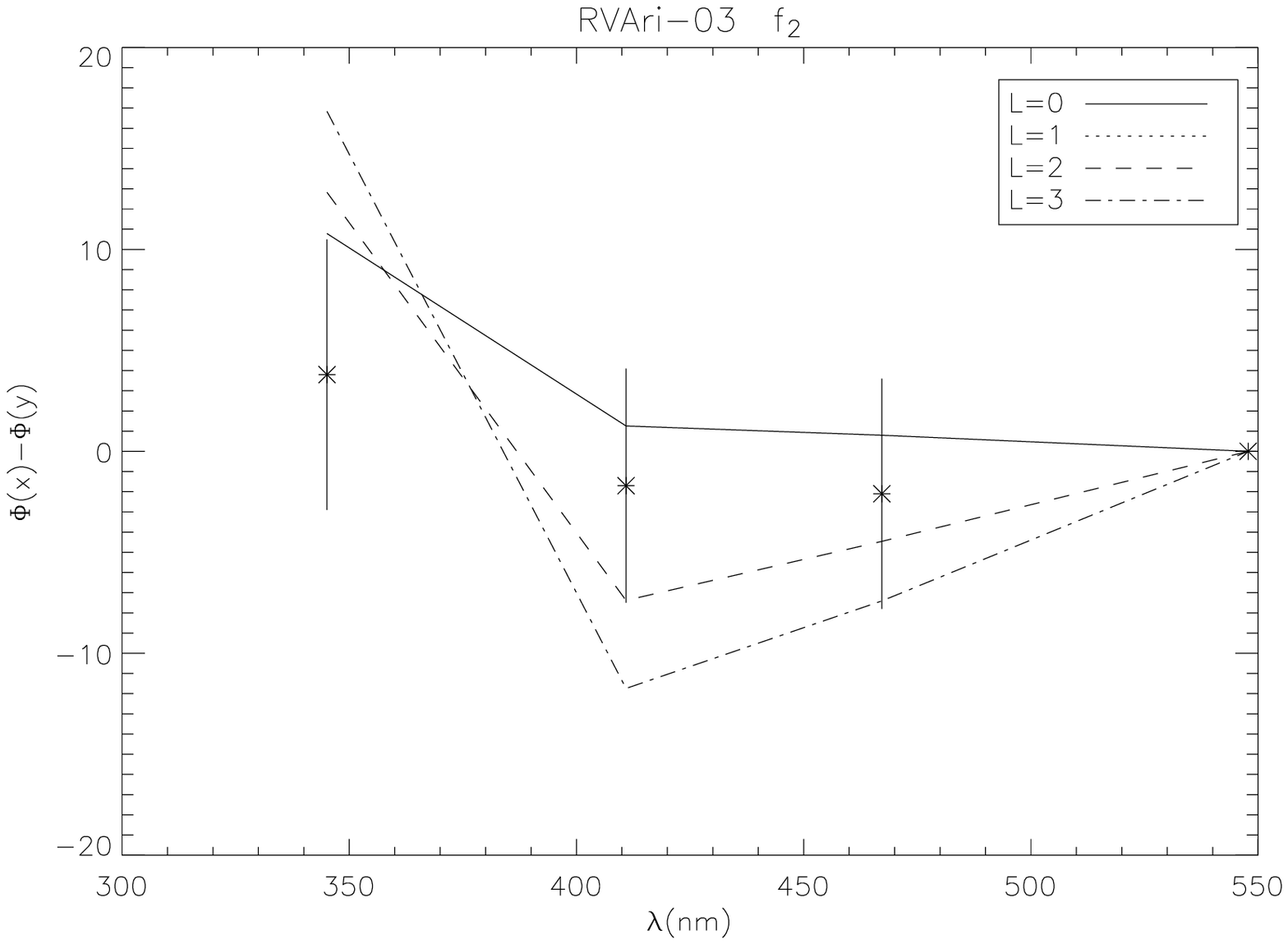}}

    \scalebox{.25}{\includegraphics{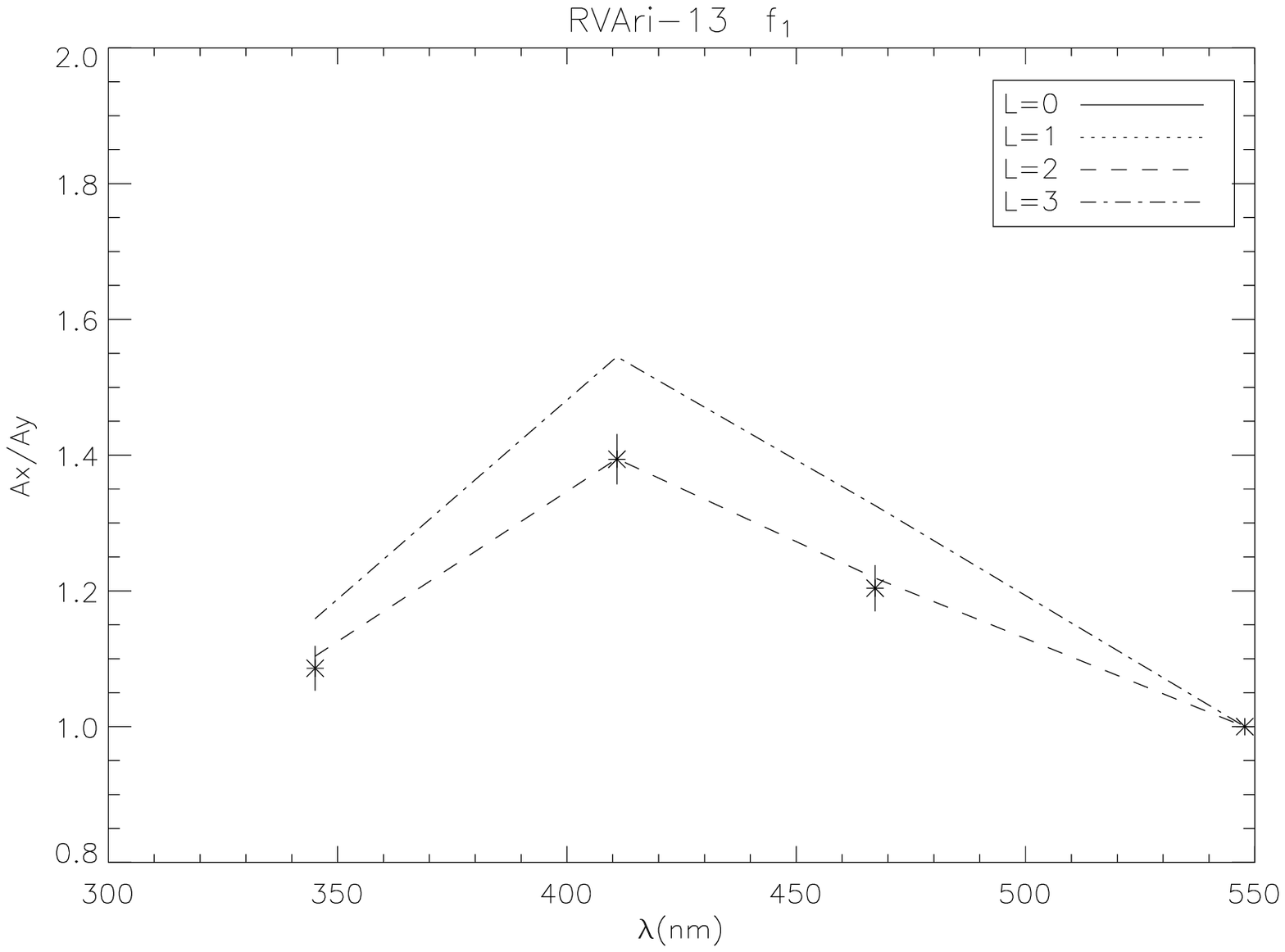}}
    \scalebox{.25}{\includegraphics{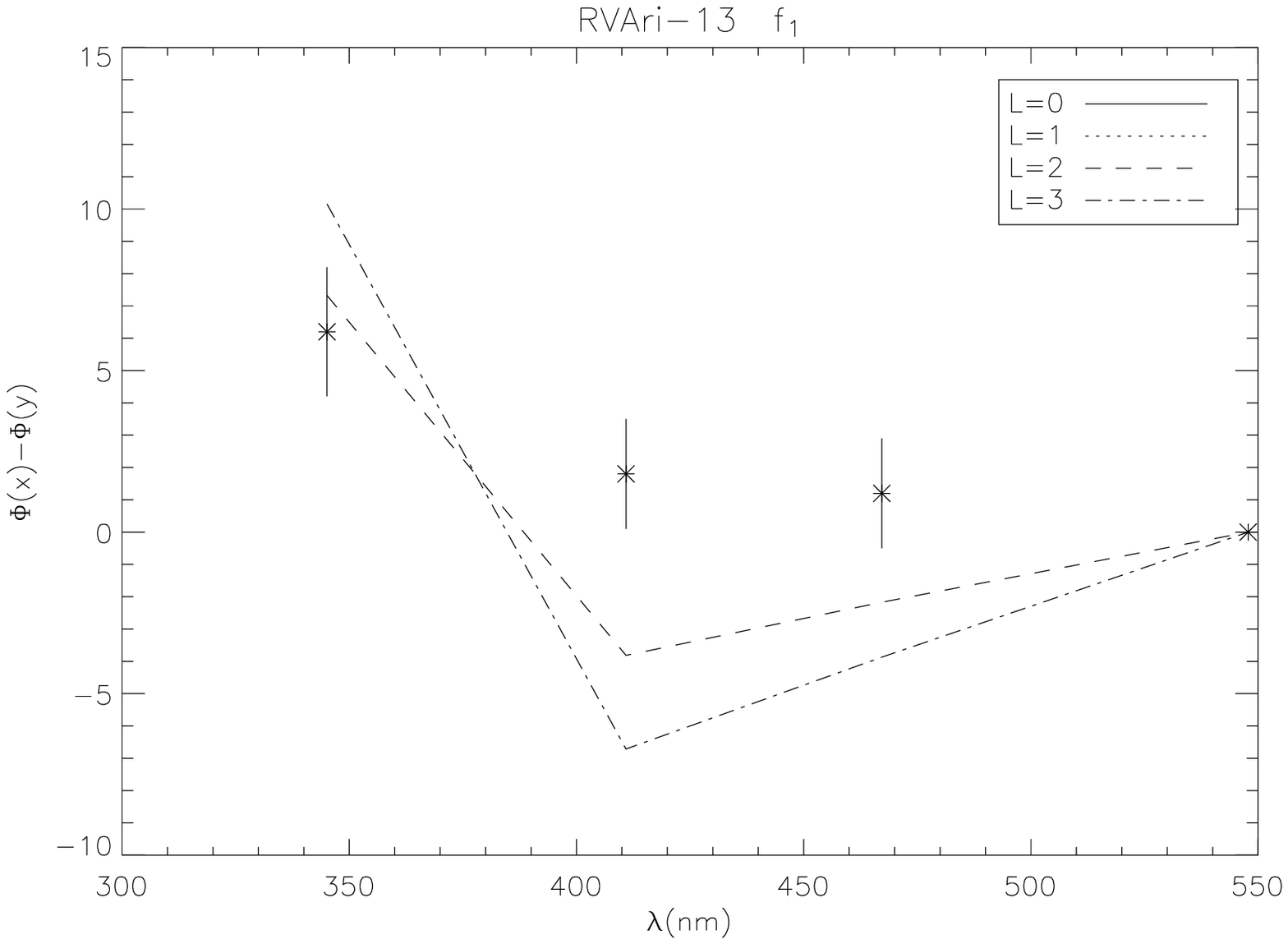}}
    \scalebox{.25}{\includegraphics{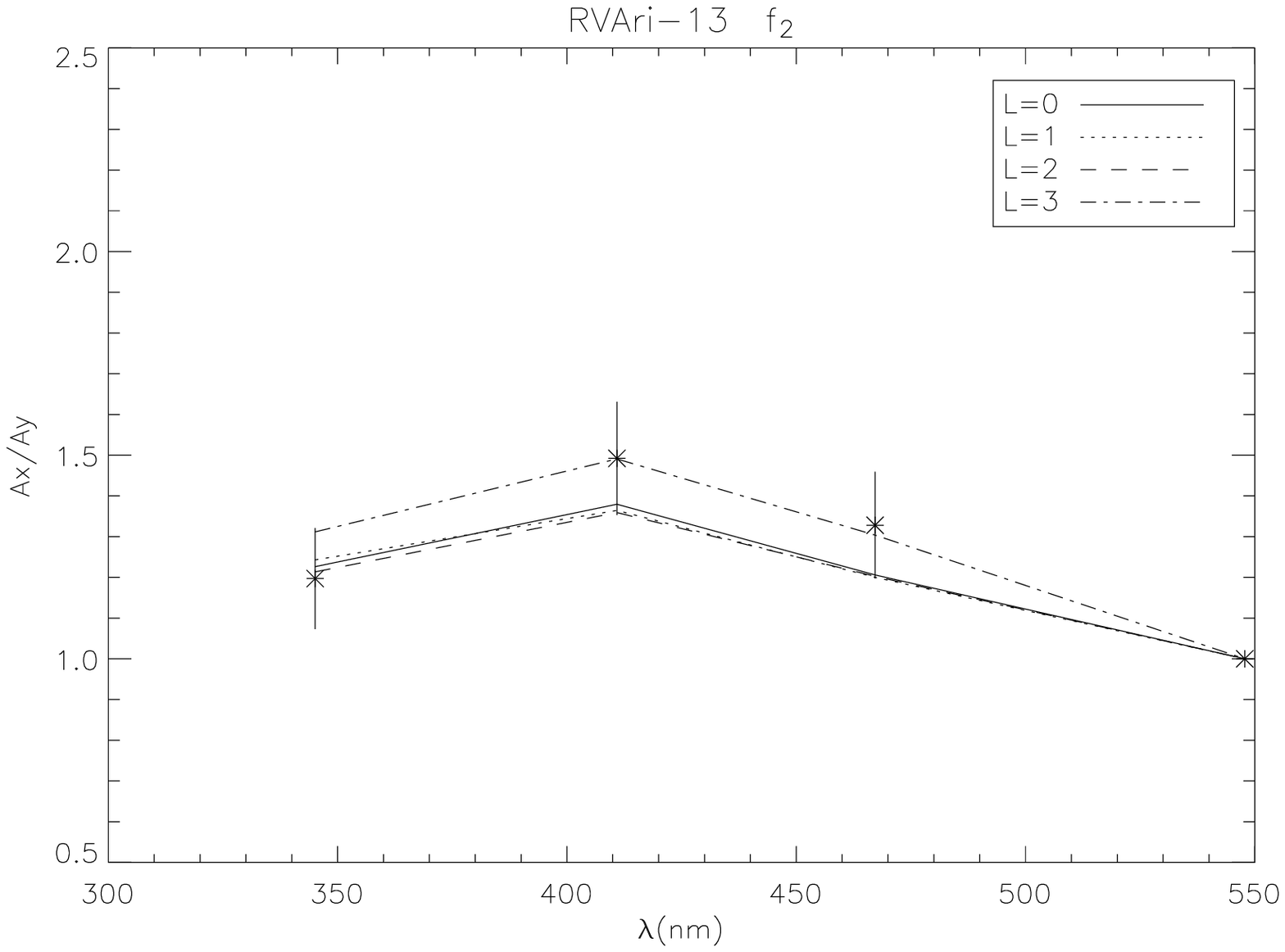}}
    \scalebox{.25}{\includegraphics{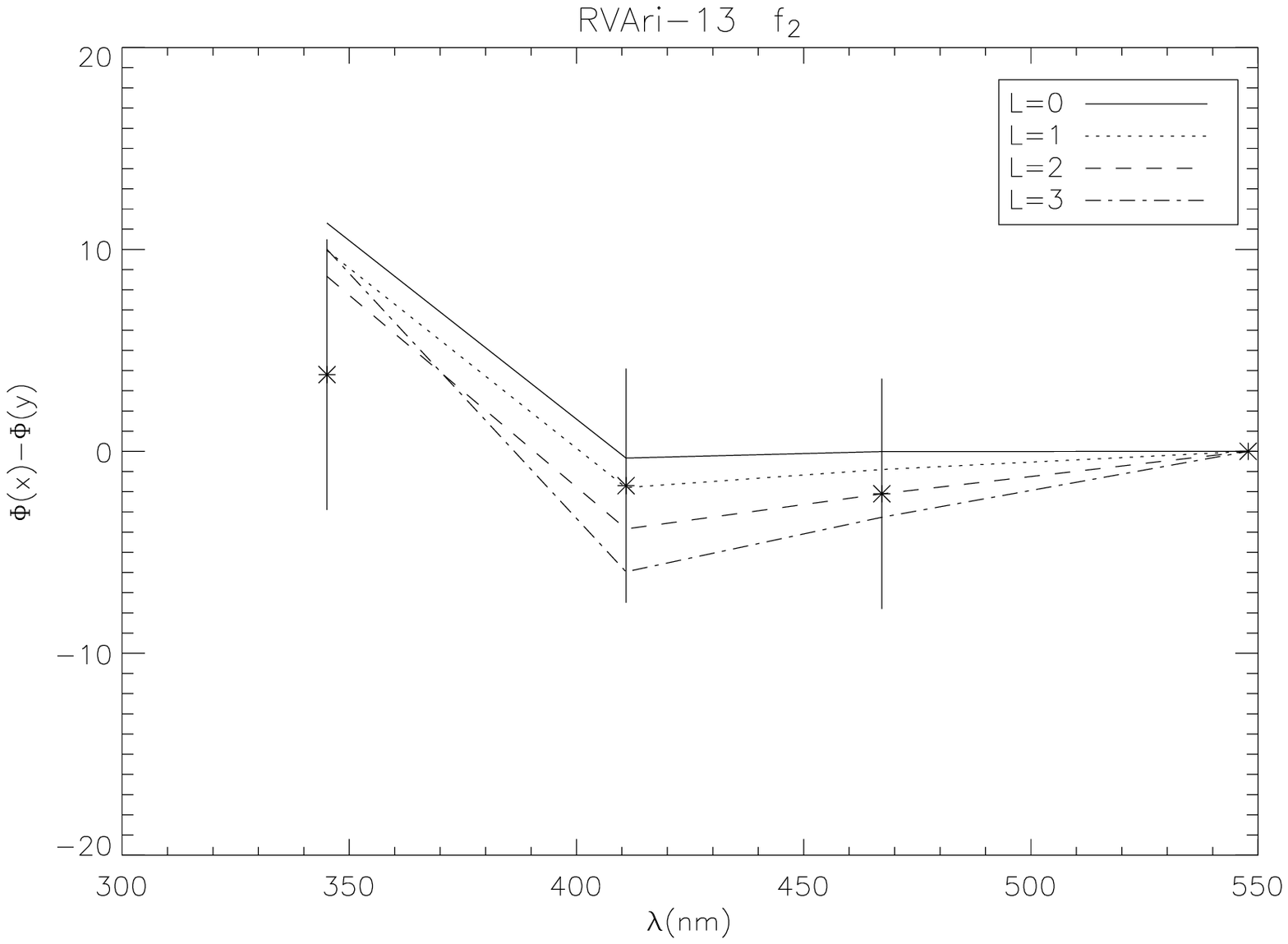}}

    \scalebox{.25}{\includegraphics{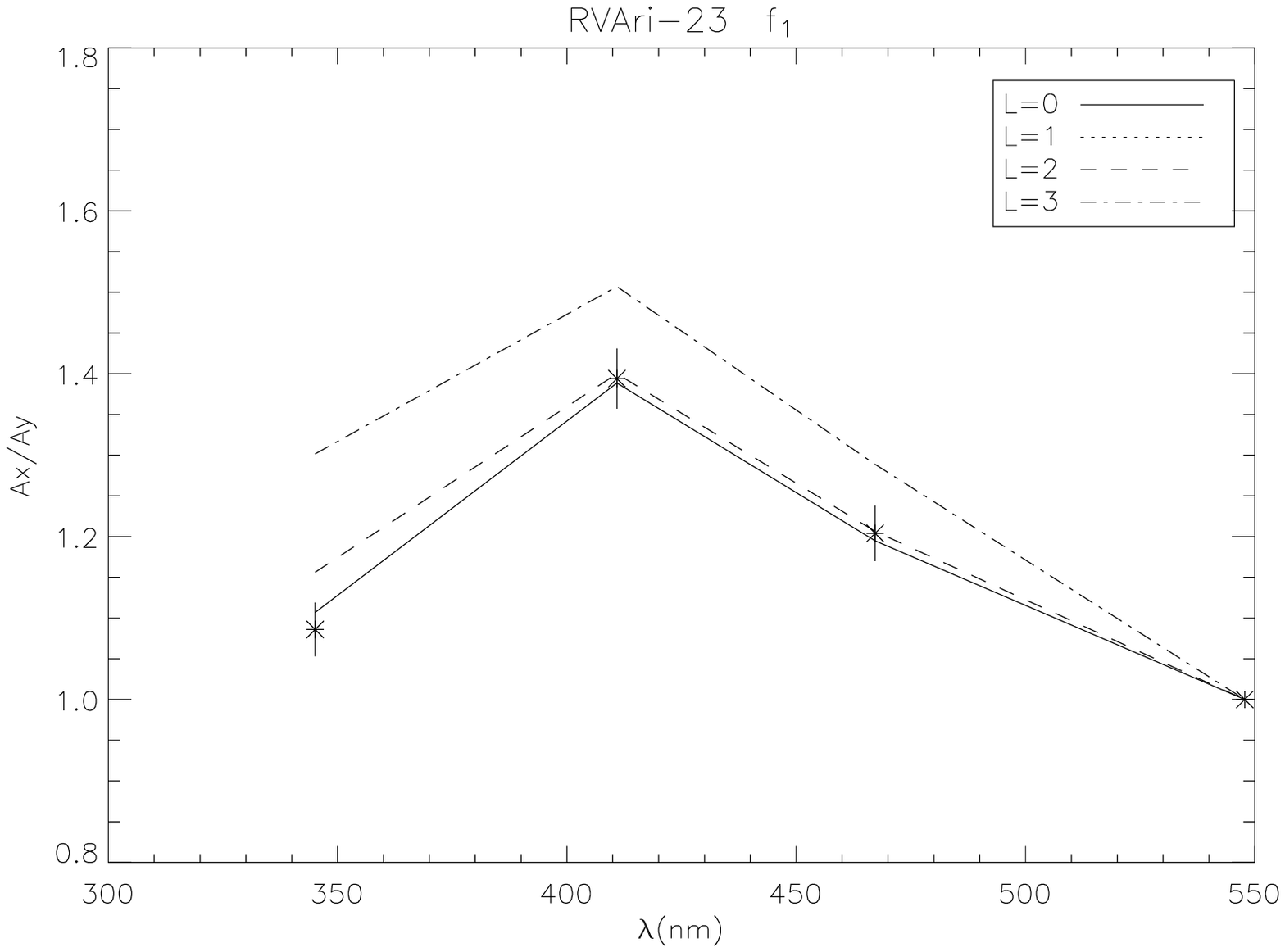}}
    \scalebox{.25}{\includegraphics{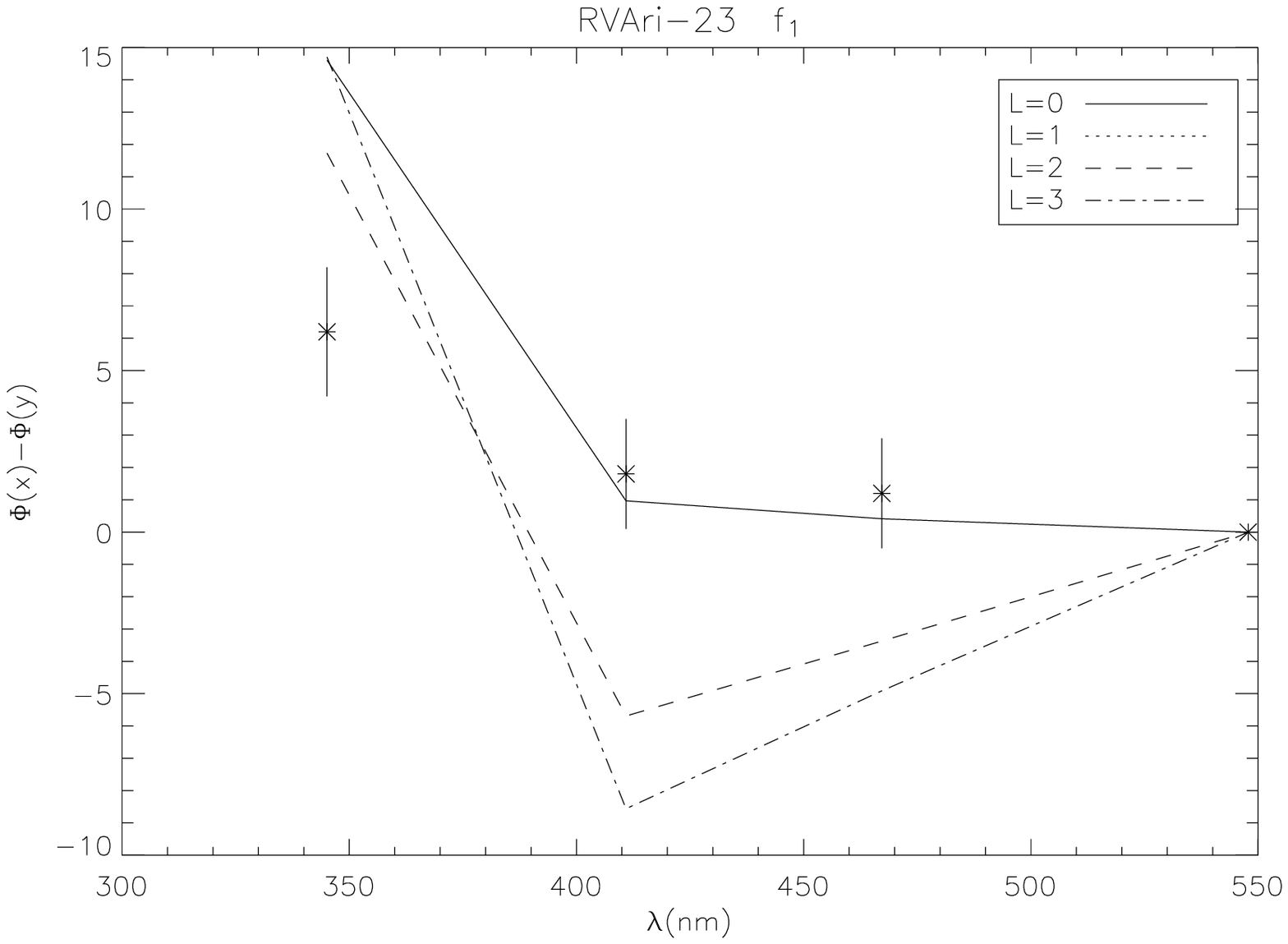}}
    \scalebox{.25}{\includegraphics{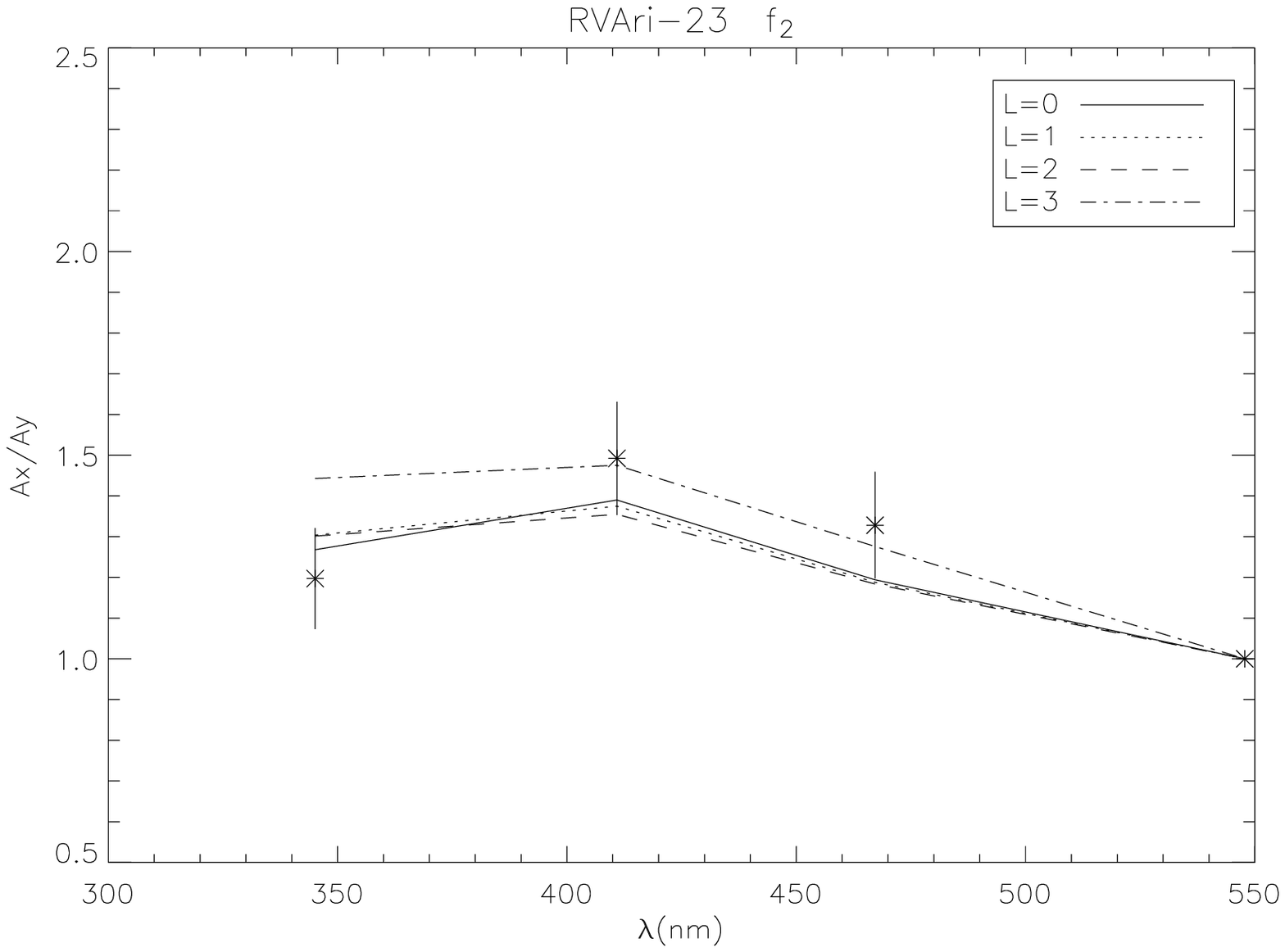}}
    \scalebox{.25}{\includegraphics{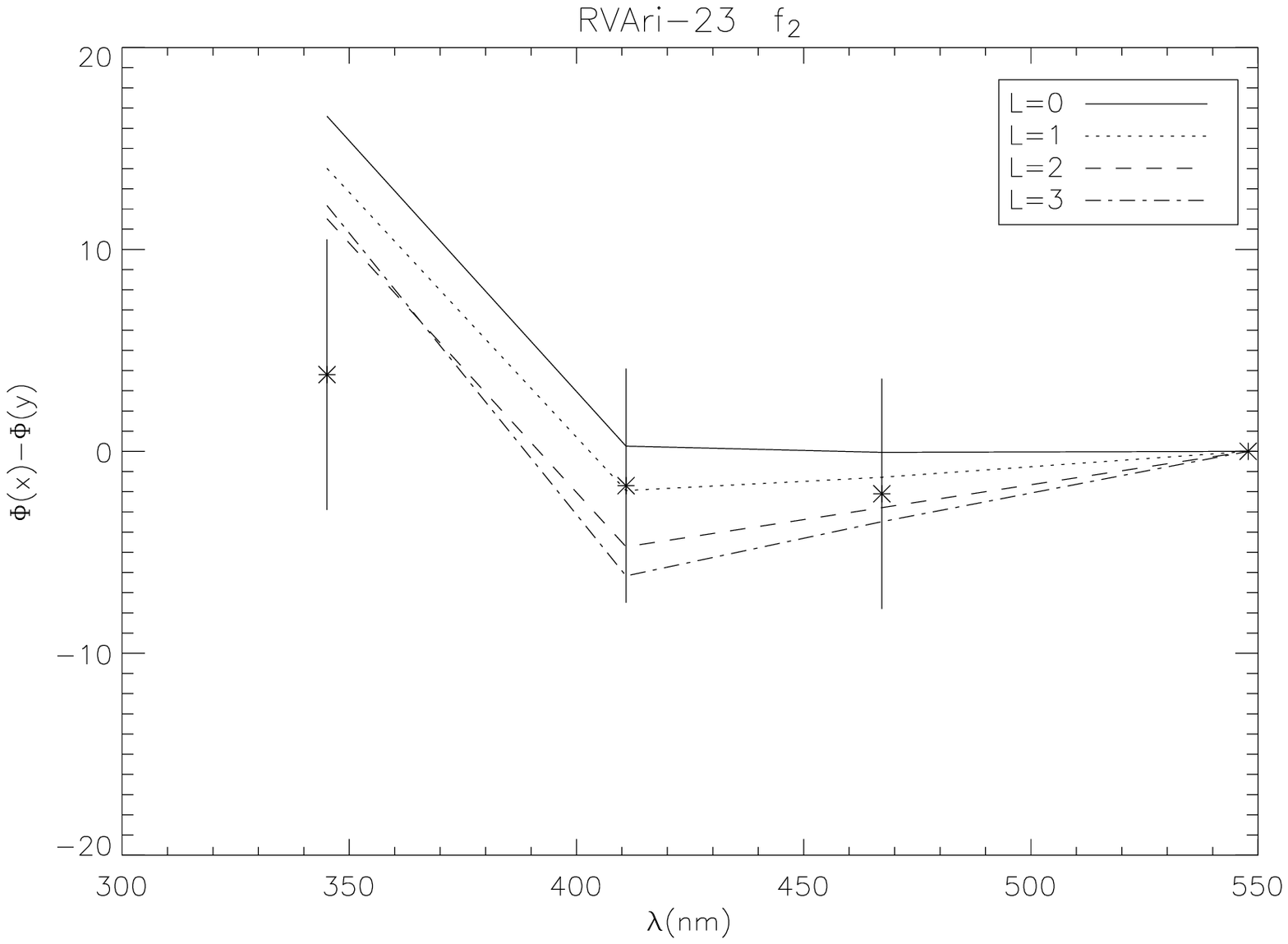}}

    \scalebox{.25}{\includegraphics{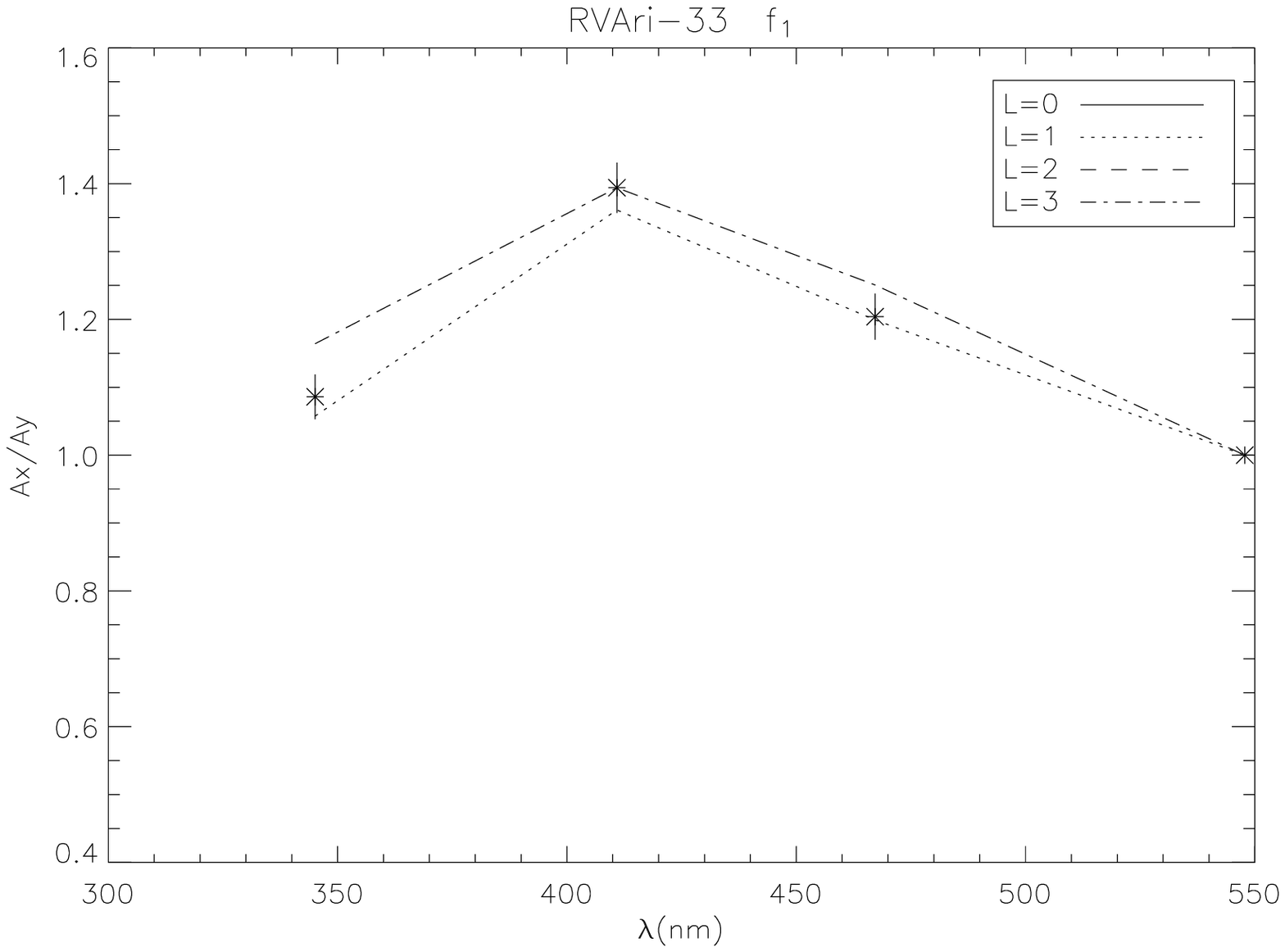}}
    \scalebox{.25}{\includegraphics{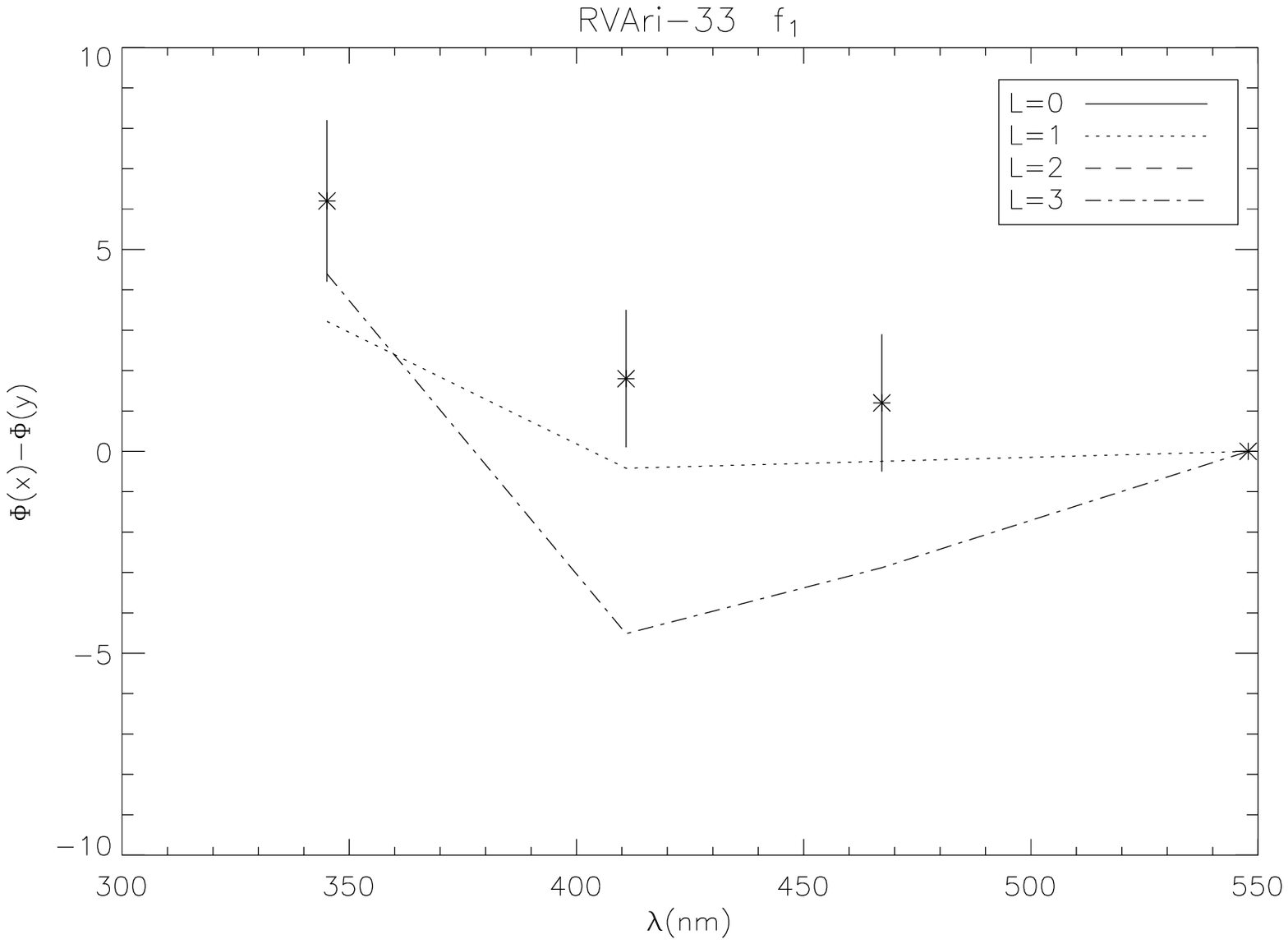}}
    \scalebox{.25}{\includegraphics{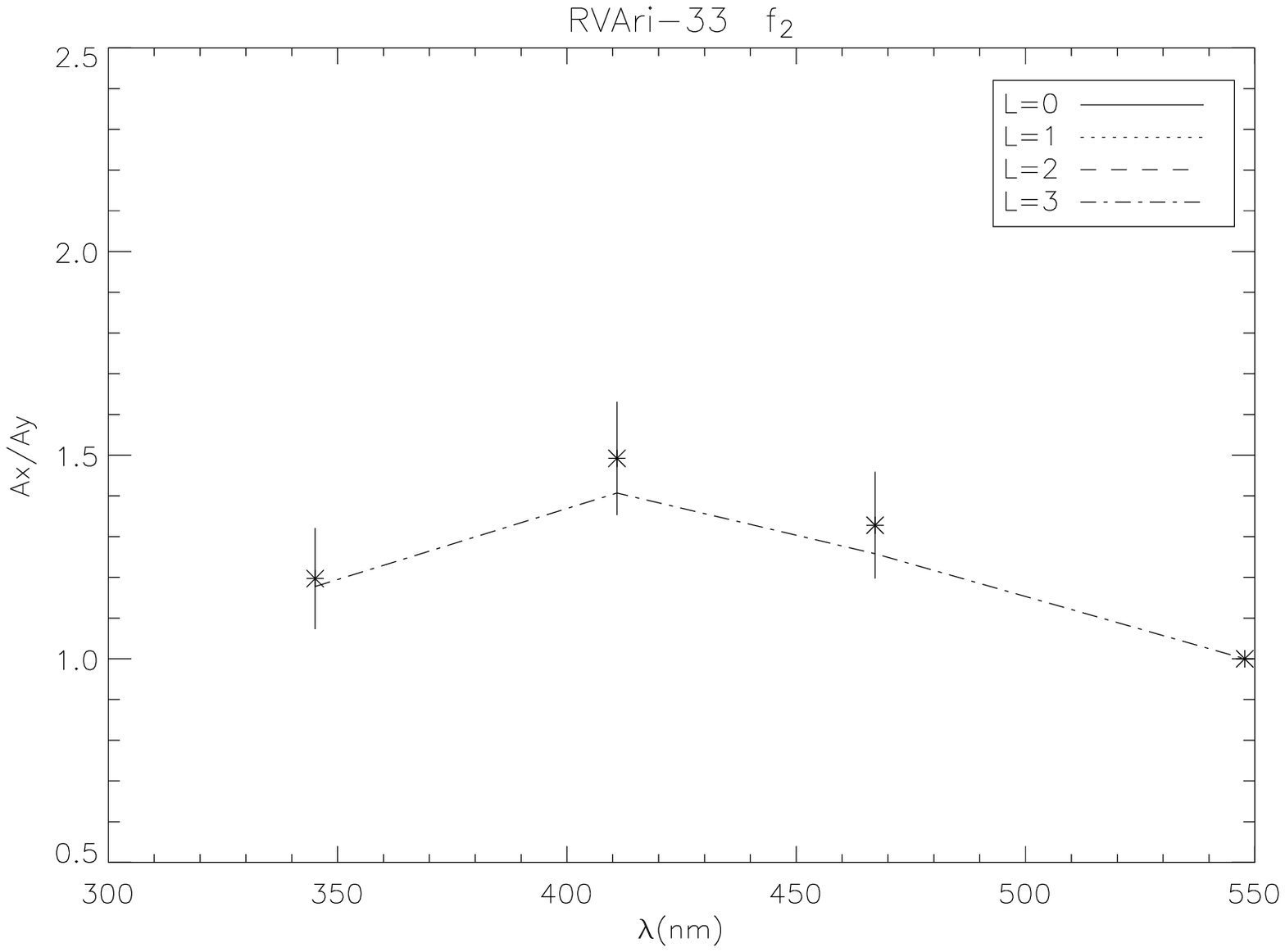}}
    \scalebox{.25}{\includegraphics{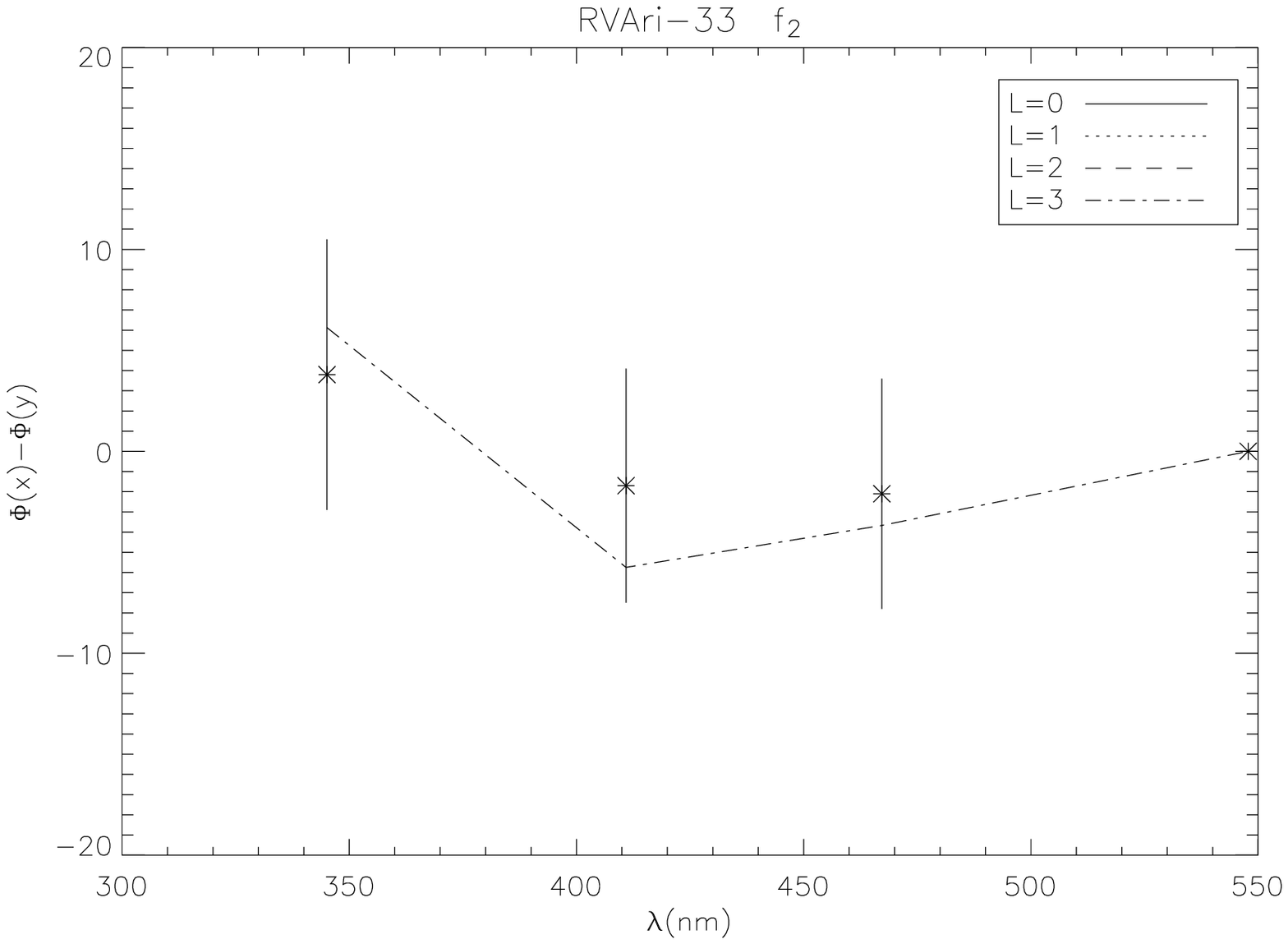}}

    \scalebox{.25}{\includegraphics{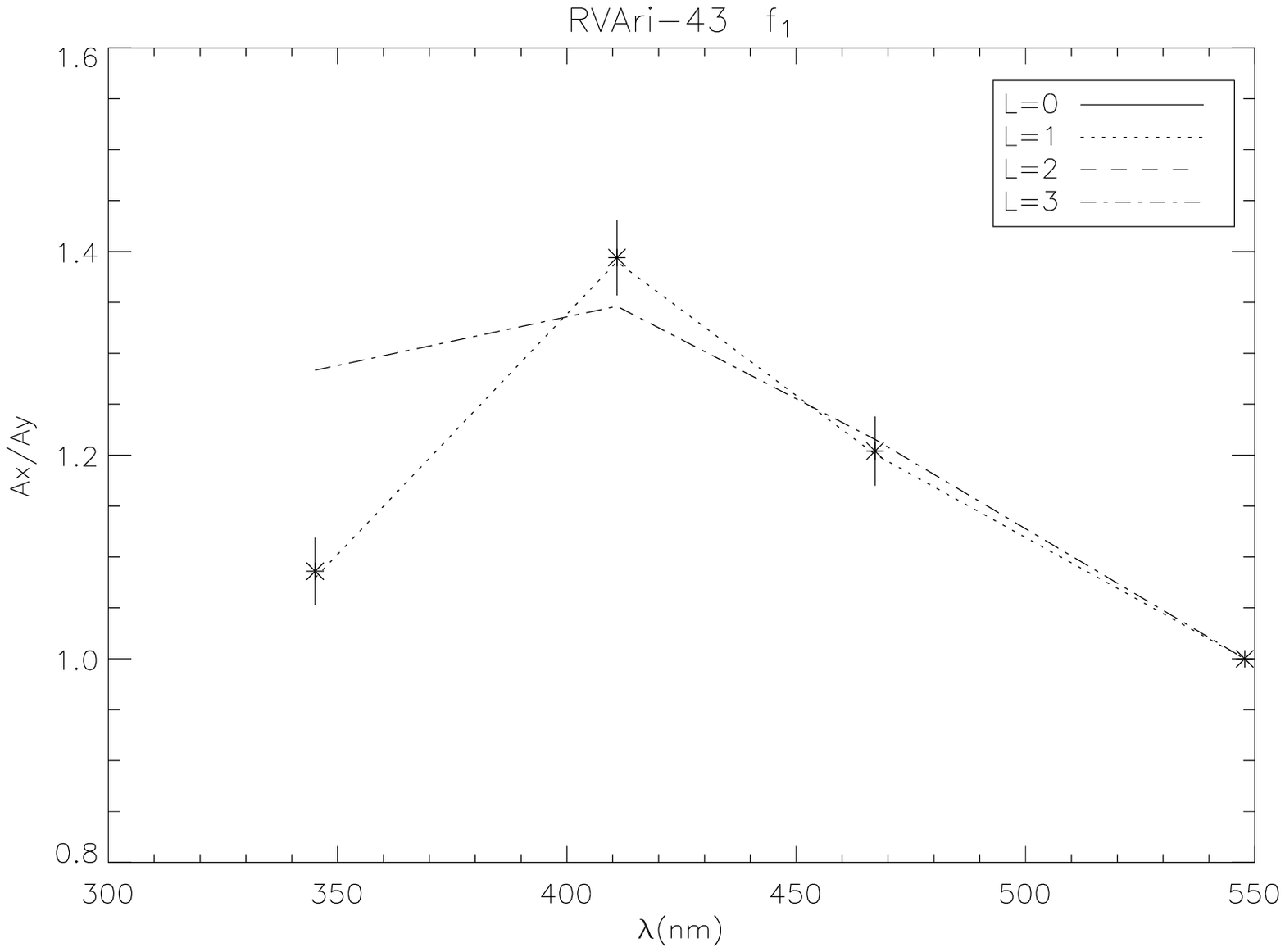}}
    \scalebox{.25}{\includegraphics{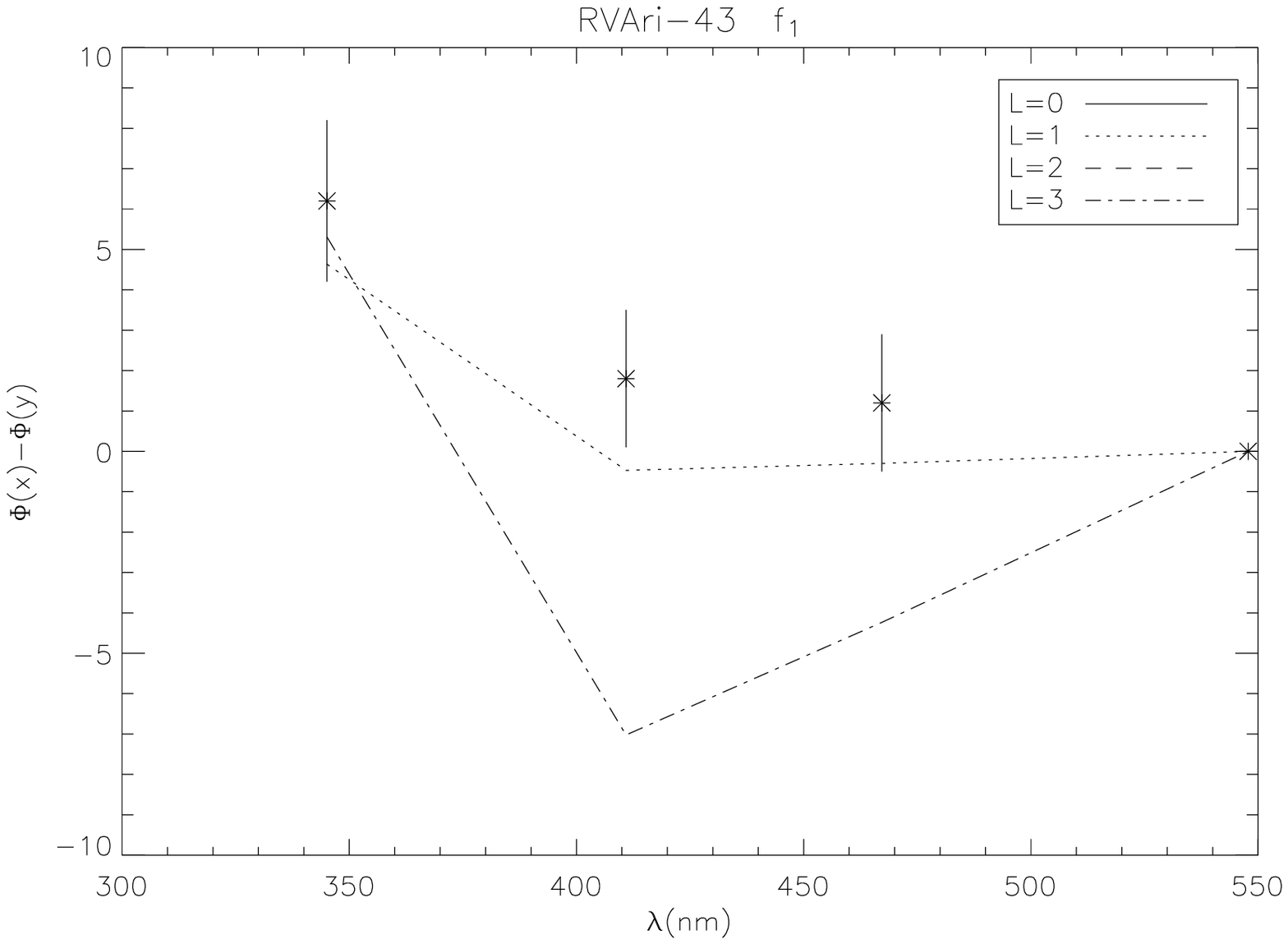}}
    \scalebox{.25}{\includegraphics{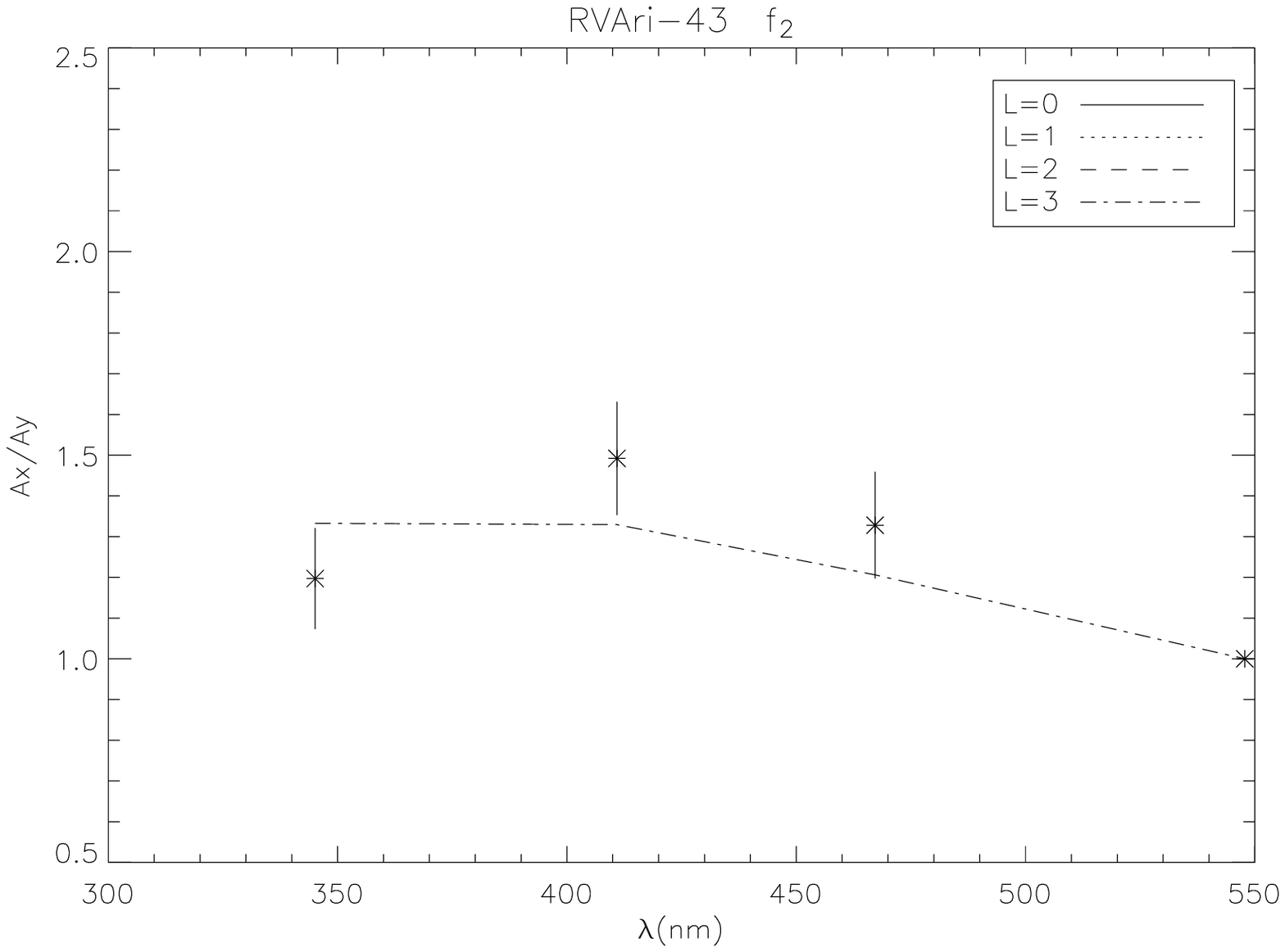}}
    \scalebox{.25}{\includegraphics{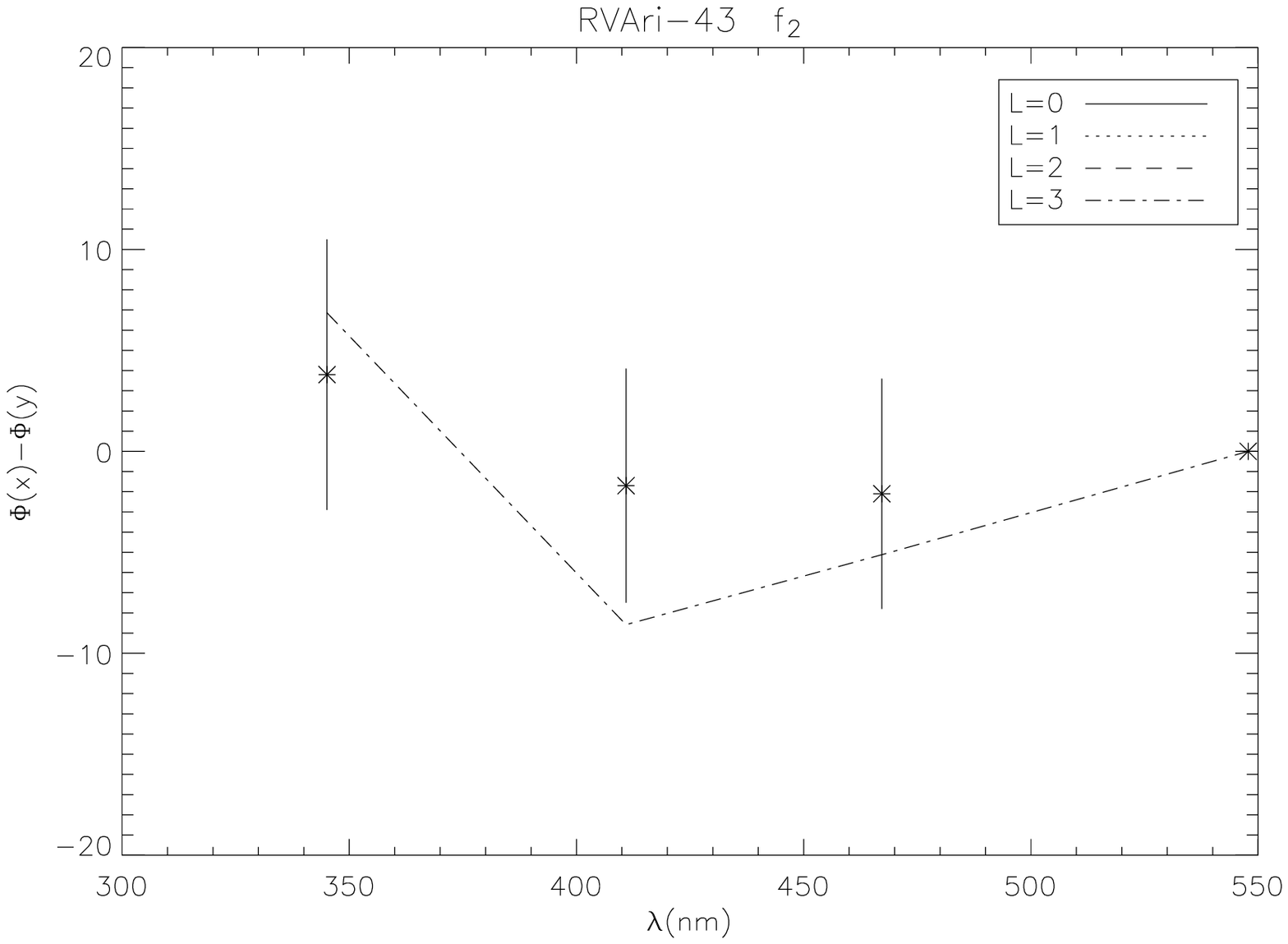}}

   \caption{Amplitude ratios and phase differences for models $i3$ ($\alpha=0.5$), $i$ ranging from 0 (top) to 4 (bottom), calculated for the observed frequencies $f_1$ (left columns) and $f_2$ (right columns), respectively. Observed values are represented by asterisks, corresponding to the four \Strom\ filters.}
   \label{fig:a-phi}
 \end{center}
\end{figure*}

In particular, insconsistencies appear when physical magnitudes are averaged over a distance $\alpha H_p$ larger than the convective zone. In other words, the mixing-length theory remains valid when $\alpha<\Delta z/H_p$ (where $\Delta z$ is the extention of the convective zone, calculated using the Scwarzschild criterion, and $H_p$ the locally-defined pressure scaleheight, calculated on the basis of the convective zone). Otherwise, the convective eddies would run a longer distance than the convective zone itself. The values of $\Delta z/H_p$, together with the fraction of the stellar radius occupied by the external convective zone, are listed in Table \ref{tab:models-instab} for all the models. In order to be physically reliable, we can establish the equality $\alpha=\Delta z/H_p$ as the very limit of validity of MLT. This is represented by a straight line in Fig. \ref{fig:a-zhp}, where $\alpha$ is displayed as a function of $\Delta z/H_p$. Therefore, it can be stated that models under the line are the only coherent ones with the MLT description, which corresponds to models computed with $\alpha=0.5$. Nevertheless, these results should be considered with care, since they may change depending on where $H_p$ is measured in the convection zone.

The ultra-precise asteroseismic observations of the forthcoming space mission COROT represent an important opportunity to study the sensitivity of these borders to $\alpha$ because of the larger number of frequencies expected to be detected at very low amplitude. This would provide a better understanding of the external convective zone when the MLT is used.

\begin{figure*}
 \begin{center}
  \scalebox{.4}{\includegraphics{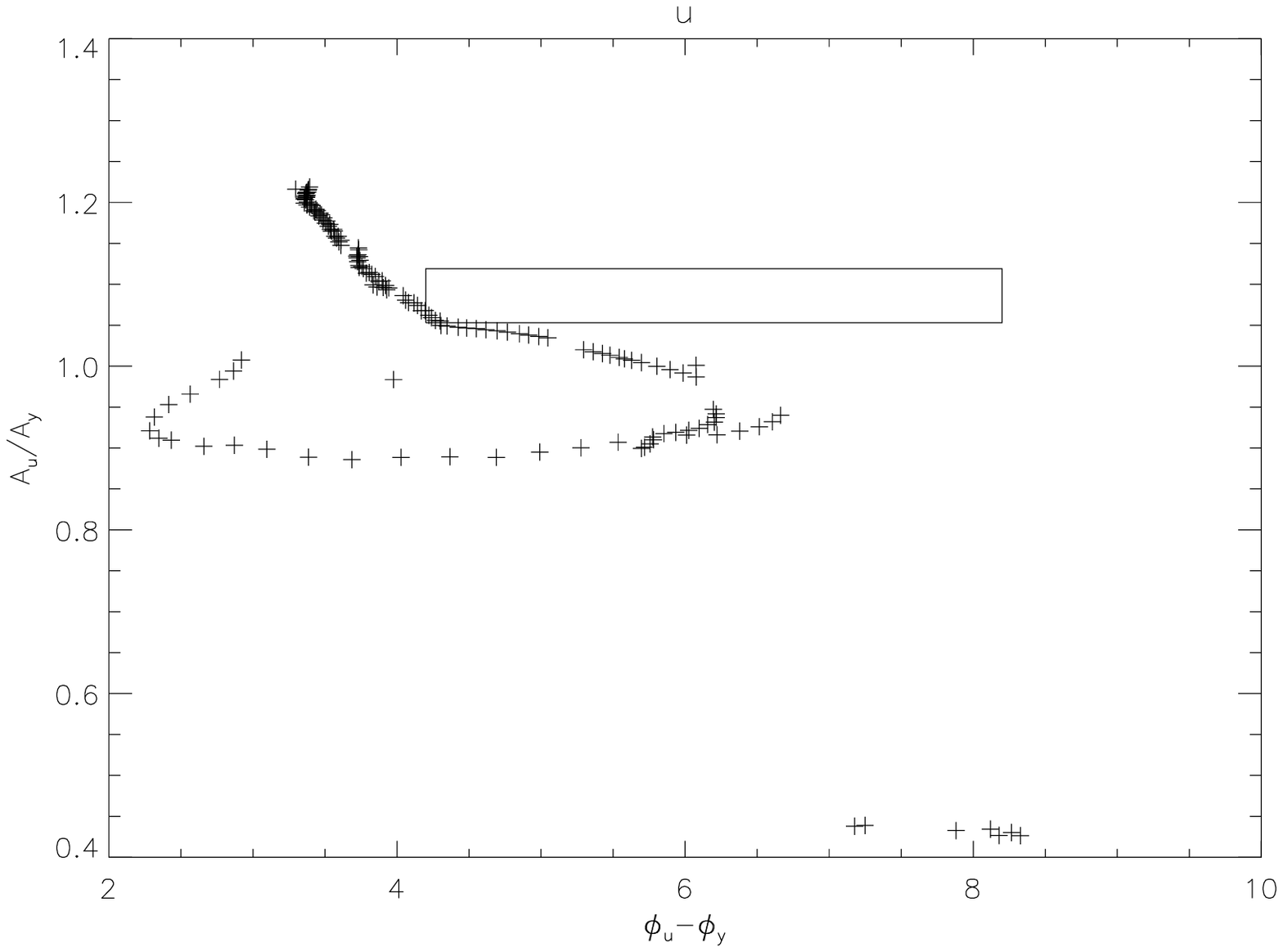}}
  \scalebox{.4}{\includegraphics{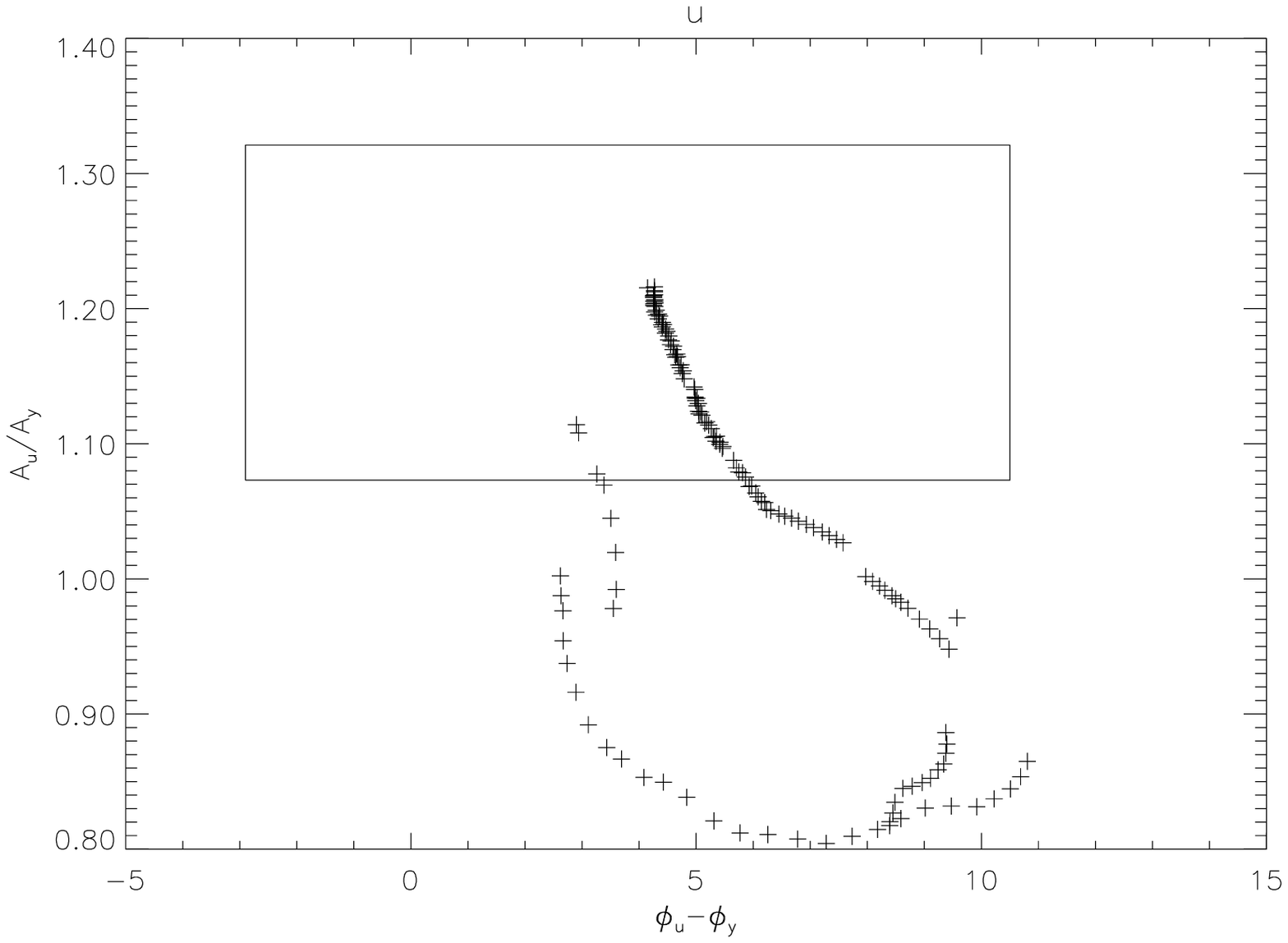}}
  \scalebox{.4}{\includegraphics{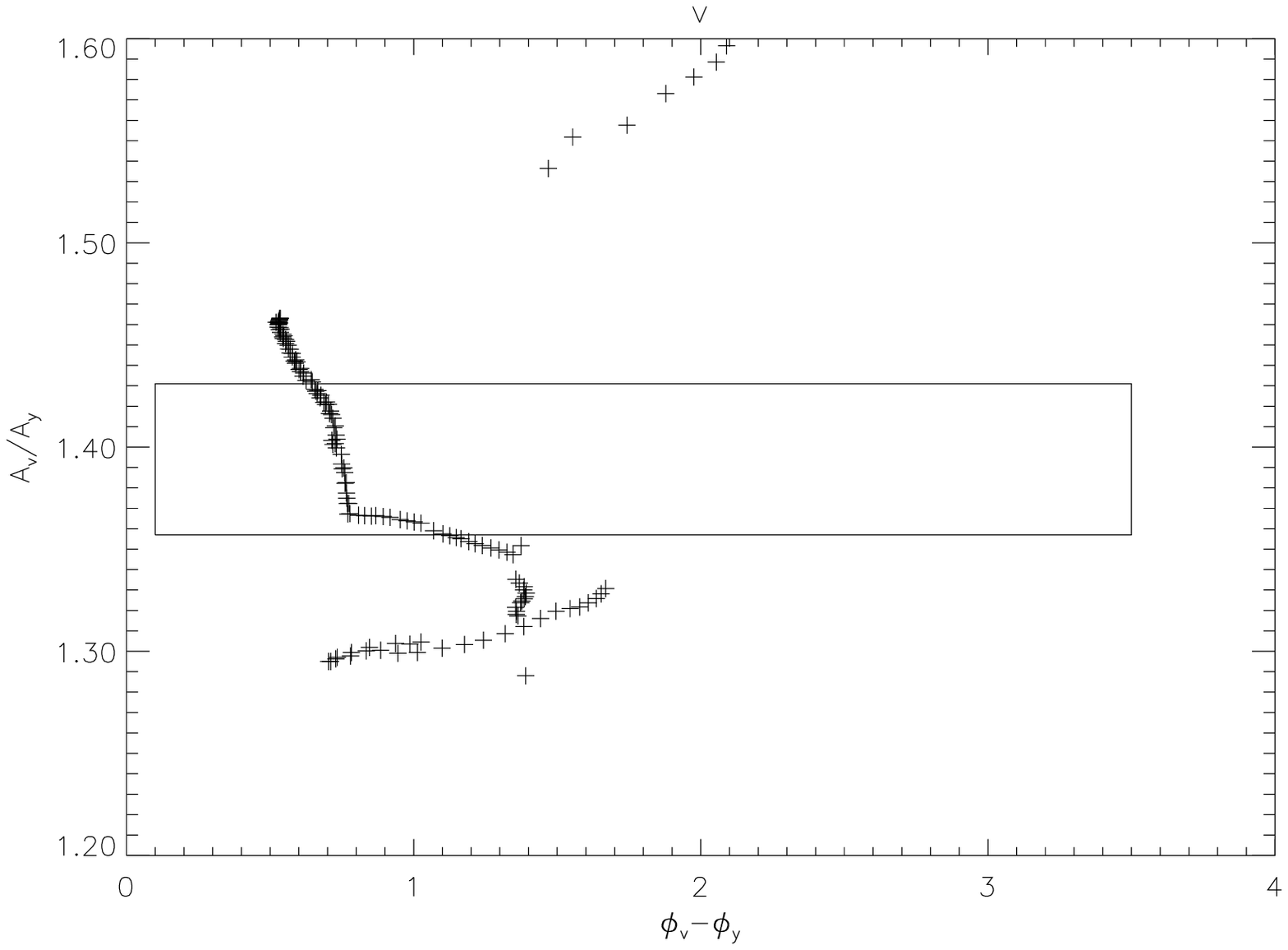}}
  \scalebox{.4}{\includegraphics{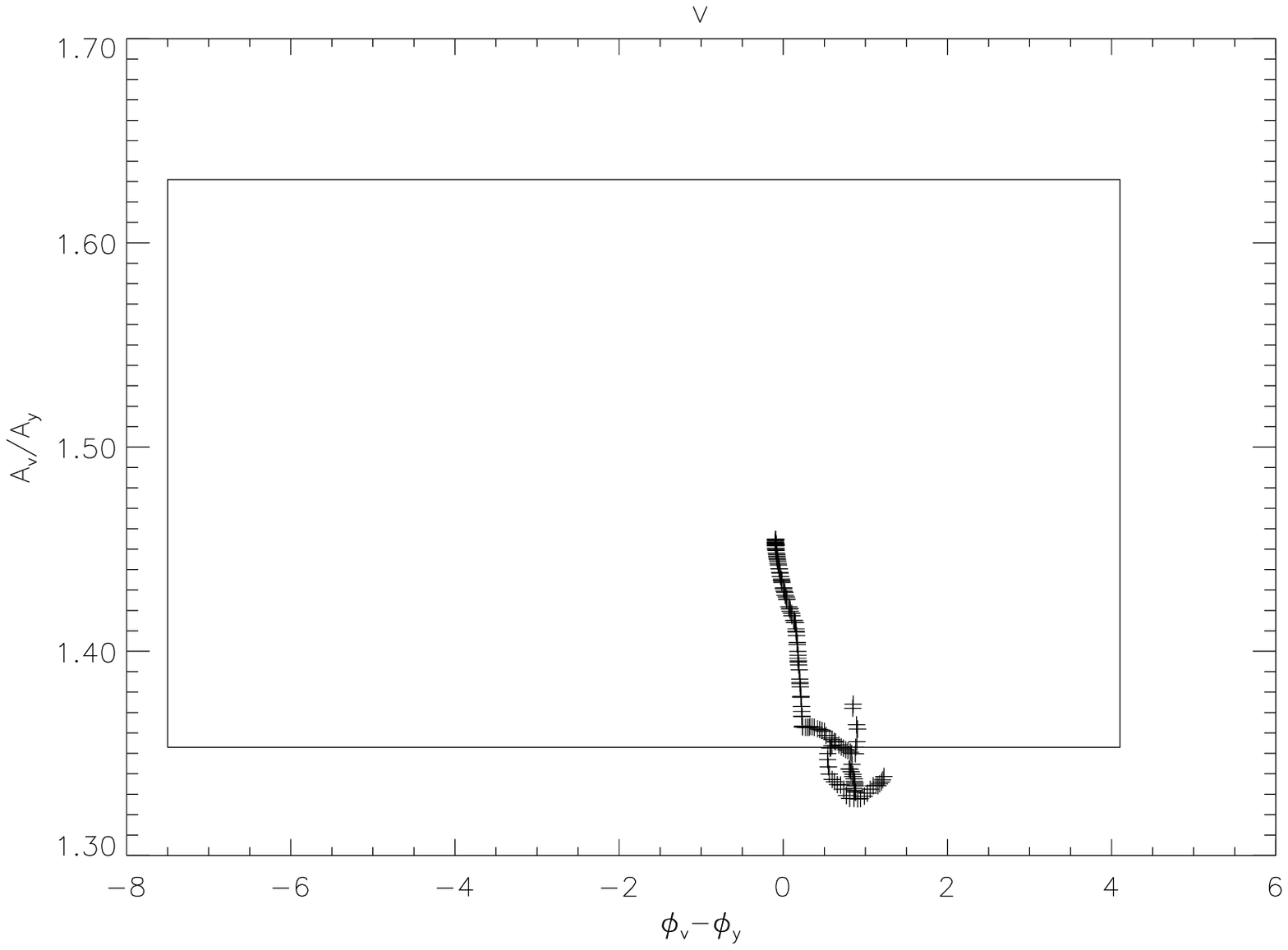}}
  \scalebox{.4}{\includegraphics{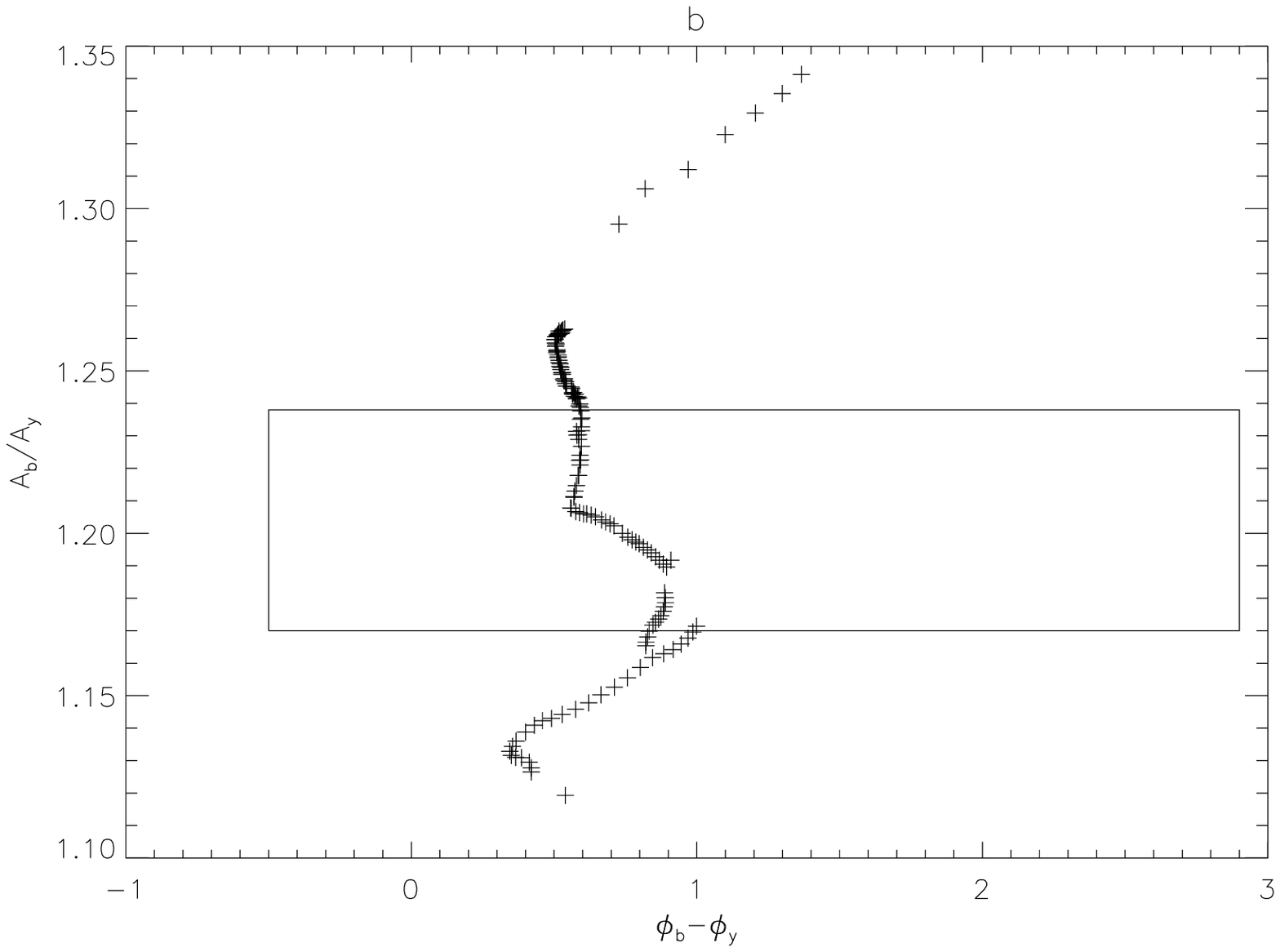}}
  \scalebox{.4}{\includegraphics{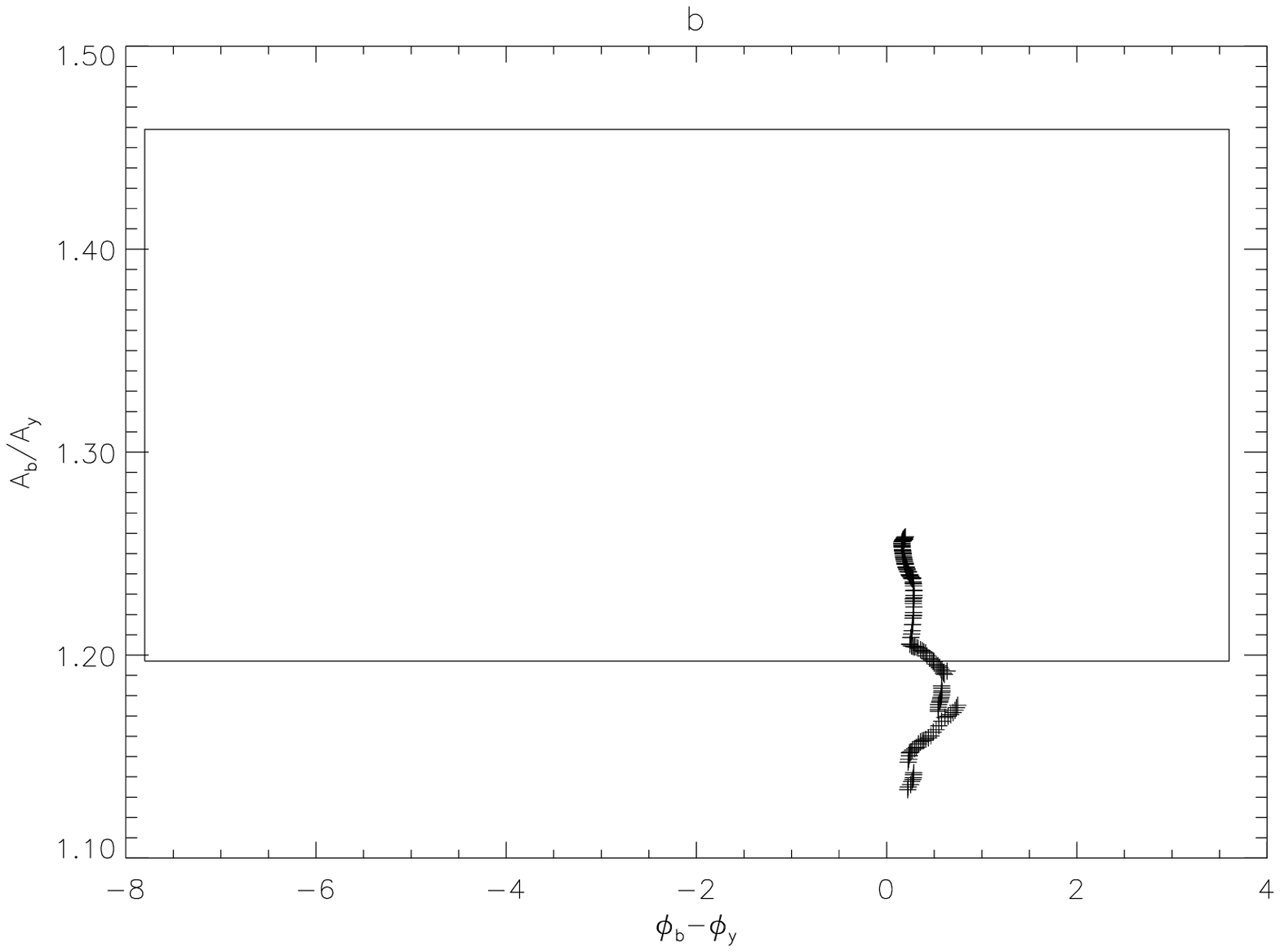}}
   \caption{Phase-amplitude diagrams showing the evolution of the fundamental radial mode (left panels) and first overtone (right panels) for the model $02$ ($\alpha=1$). From top to bottom, the different panels depict the results for the \Strom\ filters $u$, $v$, and $b$, referred to $y$.}
   \label{fig:phi-A_evolfund}
 \end{center}
\end{figure*}

\section{Towards a mode identification\label{sec:identif}}

Two of the three detected frequencies in \rvari\ are identified as the fundamental radial mode and its first overtone. The only plausible argument to suggest such an identification comes from the ratio between frequencies, 0.772, which is typically found in Pop. I HADS. This possibility is examined in detailed in the framework of the Petersen diagrams in Sect. \ref{sec:pd}. In the following we use the available multicolour photometric observations to provide additional information on the degree $\ell$ of the spherical harmonic associated to each observed pulsation frequency.

\subsection{Amplitude ratios vs. phase differences\label{ssec:a-phi_diag}}

The linear approximation to non-radial flux variations of a pulsating star was first derived by \citet{Dziem77} and later reformulated by \citet{BalonaStobie79} and by \citet{Watson88}. Then, \citet{Garrido90} show that "$v$" and "$y$" \Strom\ filters can be used for distinguishing the degree $\ell$ of the spherical harmonic associated to each observed pulsation frequency. This technique, based on non-adiabatic calculations, supplies amplitude ratios and phase differences for the different modes as a function of $\ell$ and is basically dependent on the modes and on the atmospheric physical conditions \citep[more details in][]{MoyaThesis,Moya04}. In particular, pulsation is highly non-adiabatic in stellar surface layers, in which thermal relaxation time is either of the same order or even lower than the pulsation period. Accurately determinating the eigenfunctions in these layers therefore requires using a non-adiabatic description that includes the entire atmosphere. This procedure then makes it possible to relate multicolour photometric observables with such eigenfunctions and therefore allows direct constraining of some unknown physical parameters through direct comparison with observations (see Table \ref{tab:freqobs_uvbV}).
\begin{table*}
  \begin{center}
  \caption{Main characteristics of computed models representative of \rvari\ for the two rotational velocities considered. Labels $i$ from 0 to 4 indicate the location of the models in the HR diagram given in Fig. \ref{fig:HR-rvari}. From left to right, $\Omega$ represents the rotational velocity in $\kms$, $M$ the stellar mass in solar masses $\msol$, $\teff$ the effective temperature in K (on a logarithmic scale), g the surface gravity in cgs (on a logarithmic scale), $X_c$ the central Hydrogen fraction, the age in Myr, $\nuf$ the frequency of the fundamental radial mode (in $\muHz$), and finally, $\nu_\Omega$ represents the rotational frequency (in $\muHz$) of the model.}
    \vspace{1em}
    \renewcommand{\arraystretch}{1.2}
    \begin{tabular}{ccccccccccc}
      \hline
      \hline
      & ID & $M$ & $\teff$ & $g$ & $L$ & $\xc$ & Age & $\nuf$ & $\nu_\Omega$ \\
      \hline
 $\Omega=18$ &  &  &  &  & &  & &  & &   \\

     & 02 & 1.75 & 3.859 & 3.94 & 1.13 & 0.304 & 1182 & 130.85 & 1.8 \\

     & 12 & 2.10 & 3.868 & 3.70 & 1.48 & 0.135 & 858  & 82.00  & 1.3 \\

     & 22 & 2.00 & 3.850 & 3.70 & 1.39 & 0.126 & 994  & 82.26  & 1.3 \\

     & 32 & 1.57 & 3.867 & 4.19 & 0.86 & 0.542 & 834  & 208.33 & 2.5 \\

     & 42 & 1.49 & 3.850 & 4.20 & 0.76 & 0.554 & 920  & 202.46 & 2.6 \\

 $\Omega=52.6$ & &  &  &  & &  & &  & &   \\

     & 02 & 1.76 & 3.859 & 3.94 & 1.14 & 0.305 & 1162 & 129.09 & 5.2 \\

     & 12 & 2.10 & 3.868 & 3.70 & 1.48 & 0.141 & 854  & 82.18  & 3.6 \\

     & 22 & 2.01 & 3.850 & 3.69 & 1.40 & 0.127 & 980  & 81.36  & 3.7 \\

     & 32 & 1.58 & 3.868 & 4.19 & 0.87 & 0.543 & 815  & 210.00 & 7.3 \\

     & 42 & 1.49 & 3.850 & 4.20 & 0.76 & 0.560 & 890  & 203.33 & 7.6 \\
      \hline
    \end{tabular}
    \label{tab:rotmodels}
  \end{center}
\end{table*}

The computed models obtained in the previous section provide predictions not only about the region of the unstable modes but also on non-adiabatic quantities that are necessary for calculating phase differences and amplitude ratios for the different Str\"omgren colours (Sect. \ref{sec:intro}). Theoretical predictions of both quantities are then compared with the observed ones obtained for $f_1$ and $f_2$. For the sake of clarity and brevity, diagnostic diagrams are depicted in Fig. \ref{fig:a-phi} only for the sets of models $i3$. The reference filter for amplitude ratios and phase differences is taken to be $y$.

For each $ij$ model, the theoretical frequencies used to build these diagrams are the closest to the observed ones, within an uncertainty of 5\% (in frequency) approximately, which includes effects coming from rotation, $\alpha$ and metallicity variations around the observed value $\mathrm{[M/H]}=0.01$. Nevertheless, the uncertainties coming from rotation largely dominates.

Let us first examine the results for the $0j$ models. Analysis of amplitude ratios and phase differences obtained for $\ell=0$ modes reveals that the predictions for these modes are not sensitive to $\alpha$. Accordingly, the \emph{best} identification corresponds to the fundamental mode and first overtone. However, non-radial modes cannot be discarded. For instance, the observed frequency $f_2$ is found to be compatible with a $\ell=2$ mode. Indeed, the smaller $\alpha$, the closer the theoretical predictions. In contrast, the predictions for the radial modes are almost insensitive to $\alpha$. Then, for this set of models, only non-radial modes could provide a possible discrimination of this parameter.

For $1j$ models, the predictions for $f_1$ are quite different from those for $f_2$. While none of the models reproduces the observed colour indices for the former, for the latter, four models can be considered as representative. As for $0j$, the smaller the value of $\alpha$, the better the predictions. In particular, the best results are obtained for $\alpha=0.5$, with which $f_2$ is identified as $\ell=3$. Multiple mode identification was obtained for $f_2$, for which, nevertheless, a small set of solutions was obtained.

Only one mode for the $2j$ models is found to be compatible with observations for $f_1$, which would identify it as the second overtone of the fundamental radial mode. In the case of $f_2$, when considering low values for $\alpha$, all modes are compatible with observations within the error bars, so that a multiple mode identification can be done. In the case of $3j$ and $4j$ models, we found again that the best predictions are found for models evolved with low $\alpha$. Analysis of diagrams reveals that, for $\alpha\lesssim1$, $f_1$ and $f_2$ are identified only as $\ell=1$ and $\ell=3$, respectively.

The results obtained for the coolest models should be interpreted with care since the effective temperature of \rvari\ is within the limit of validity of the \emph{frozen convection} approximation adopted in this work. At these temperatures, the interaction between convection and pulsation described by
\citet{Ahmed05cpc1} and \citet{Dupret05cpc2} may change the non-adiabatic eigenfunctions significantly.

\subsection{Colour index variations\label{ssec:chemical}}

Studying the evolution of colour indices using the amplitude/phase diagrams can also be useful for constraining the evolutionary stage and eventually the metallicity of the models. To do so, we examined the evolutionary tracks of $0j$ models for which a mode identification can be reliably performed (see Sect. \ref{ssec:oscil-insta}). In particular, the observed $f_1$ and $f_2$ for $02$ were identified
as the fundamental radial mode and its first overtone. The evolution of the theoretical \Strom\ indices $u$, $v$, and $b$ for these two modes is depicted in Fig. \ref{fig:phi-A_evolfund}.

If we keep only the models lying inside the photometric error box for each filter and identify $f_1$ and $f_2$ as the fundamental radial mode and its first overtone within a $\pm5\%$ in frequency, it is possible to constrain the range of fundamental parameters given by the observations. Left and right panels indicate that both frequencies provide similar constraints on the age and effective temperature of the models. In particular, the effective temperature and evolutionary stage are constrained to [3.849, 3.860] dex and [1190, 1270] Myr, respectively. Note that these ranges are unusually narrow compared to the typical asteroseismic modelling of \dss.

\section{Effect of rotation \label{sec:effrot}}

The effect of the low-rotational velocity on the determination of fundamental parameters of \rvari\ remains within the observational typical uncertainties for \dss. Since no constraints of its inclination angle have ever been provided, we will only consider two extreme values in the following for the rotational velocity of this star: $18\,\kms$, which represents the minimum possible value (i.e., $i=90^\circ$) and $52.6\,\kms$, corresponding to an angle of $i=20^\circ$. These values reasonably cover the range of rotational velocities observed in HADS.

As shown by \citet{DG92}, by \citet{Soufi98}, and more recently by \citet{Sua06rotcel}, the effect of rotation on adiabatic oscillation modes should not be neglected as regards, for instance, the effect of near degeneracy (see Sect. \ref{ssec:oscil-insta}) or differential rotation \citep{Sua06rotcel}. Indeed, near degeneracy not only affects the value of the oscillation frequency, but also the corresponding angular components are somehow mixed up. As explained in Sect. \ref{ssec:oscil-insta}, near-degenerate modes cannot be represented by a single spherical harmonic, but a linear combination of them is involved. One of the most important consequences of this physical effect concerns the identification of mode degrees using multicolour photometry by means of the phase/amplitude diagrams discussed in the previous section. This was studied for the first time by \citet{Pagoda02}, who claimed that, for near-degenerate
modes, such diagrams depend on the rotational velocity and on the inclination angle of the star.
\begin{figure*}
 \begin{center}
  \scalebox{.44}{\includegraphics{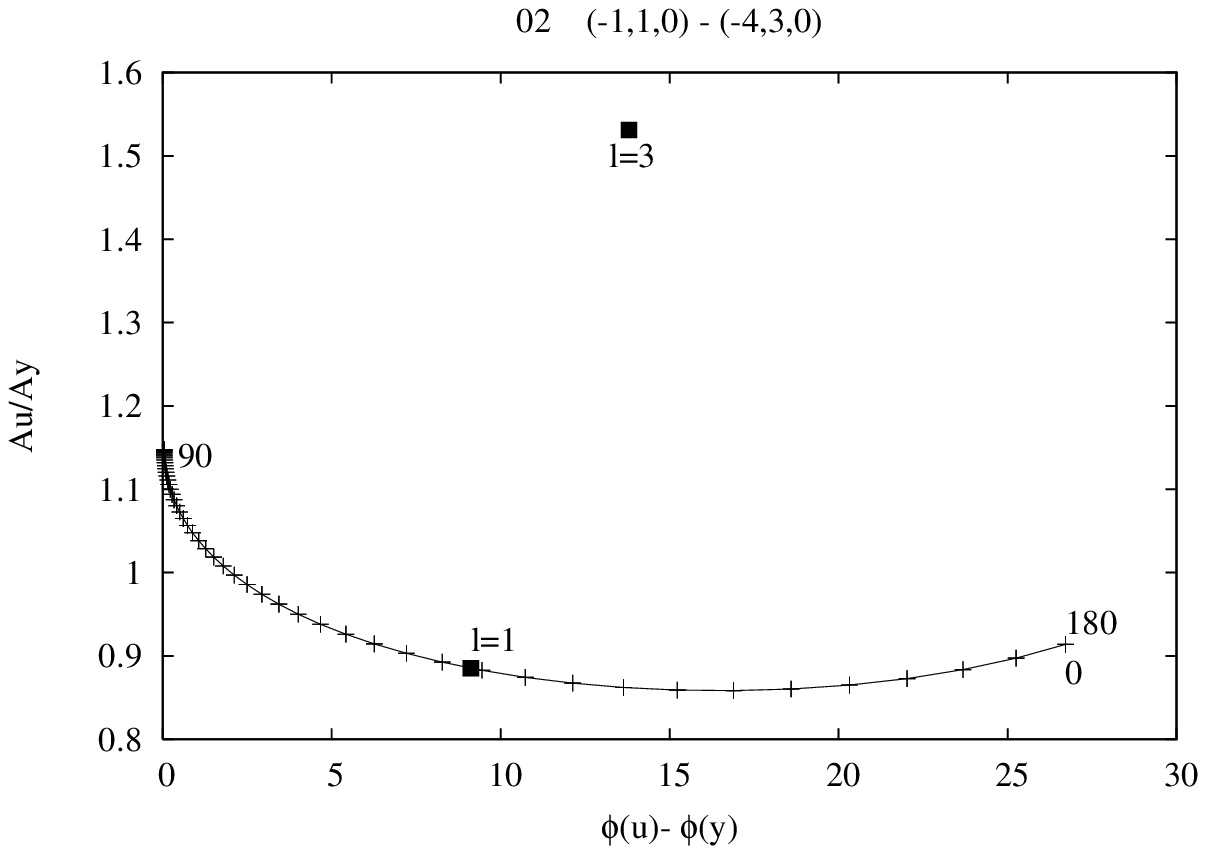}}
  \scalebox{.44}{\includegraphics{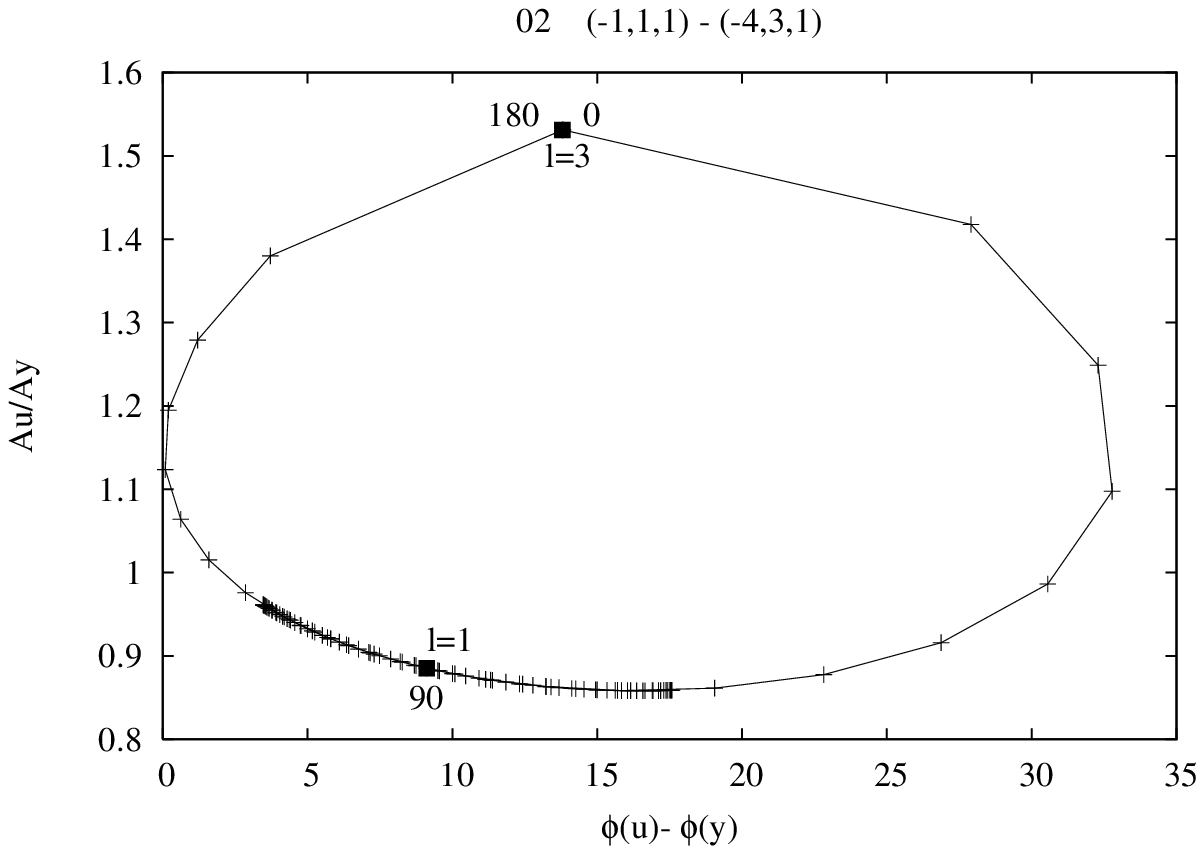}}
  \scalebox{.44}{\includegraphics{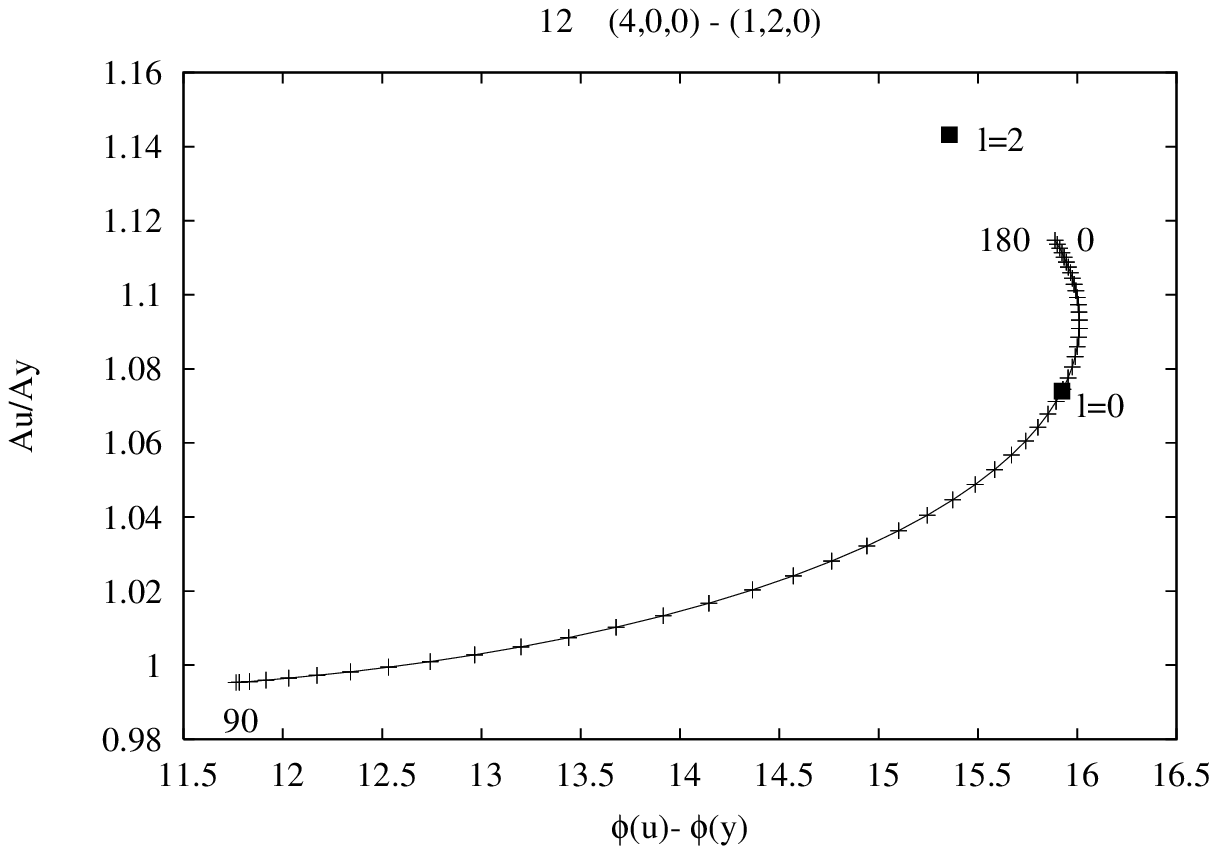}}
  \scalebox{.44}{\includegraphics{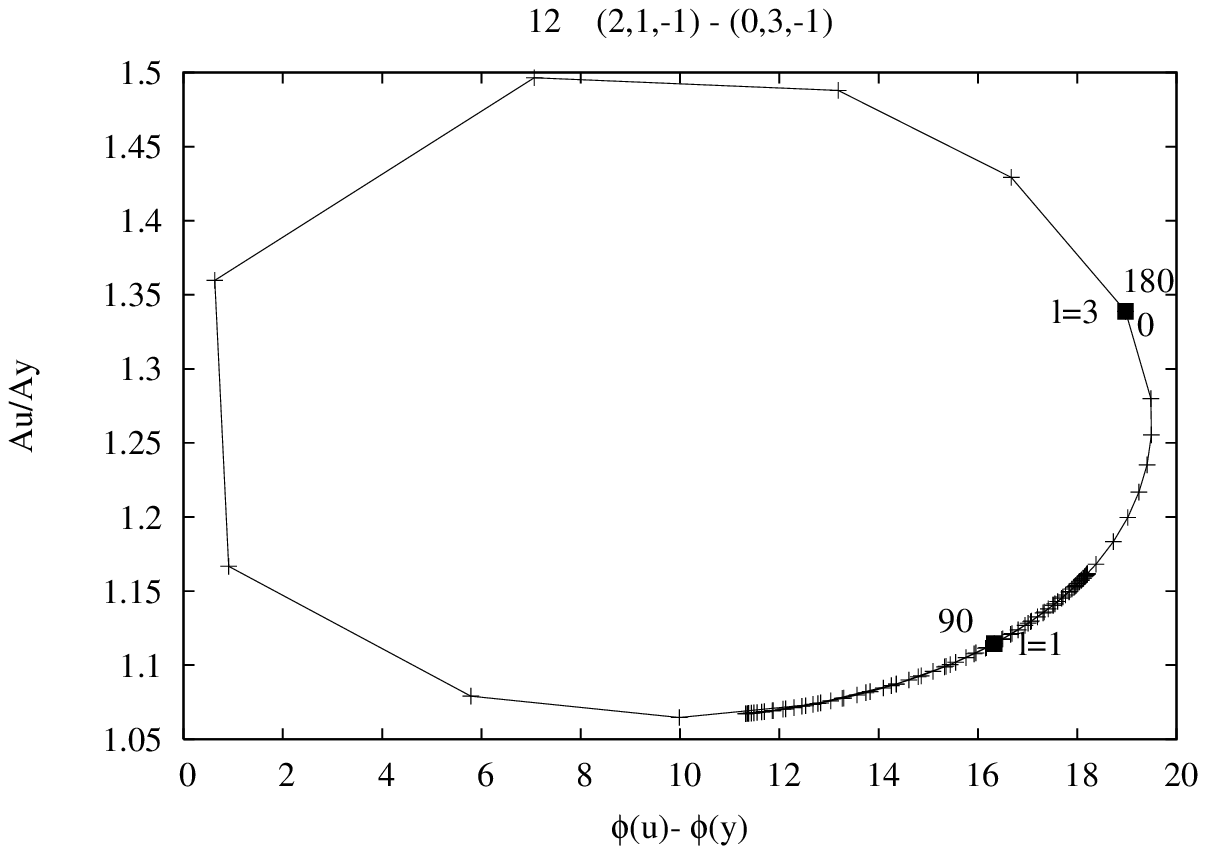}}
  \scalebox{.44}{\includegraphics{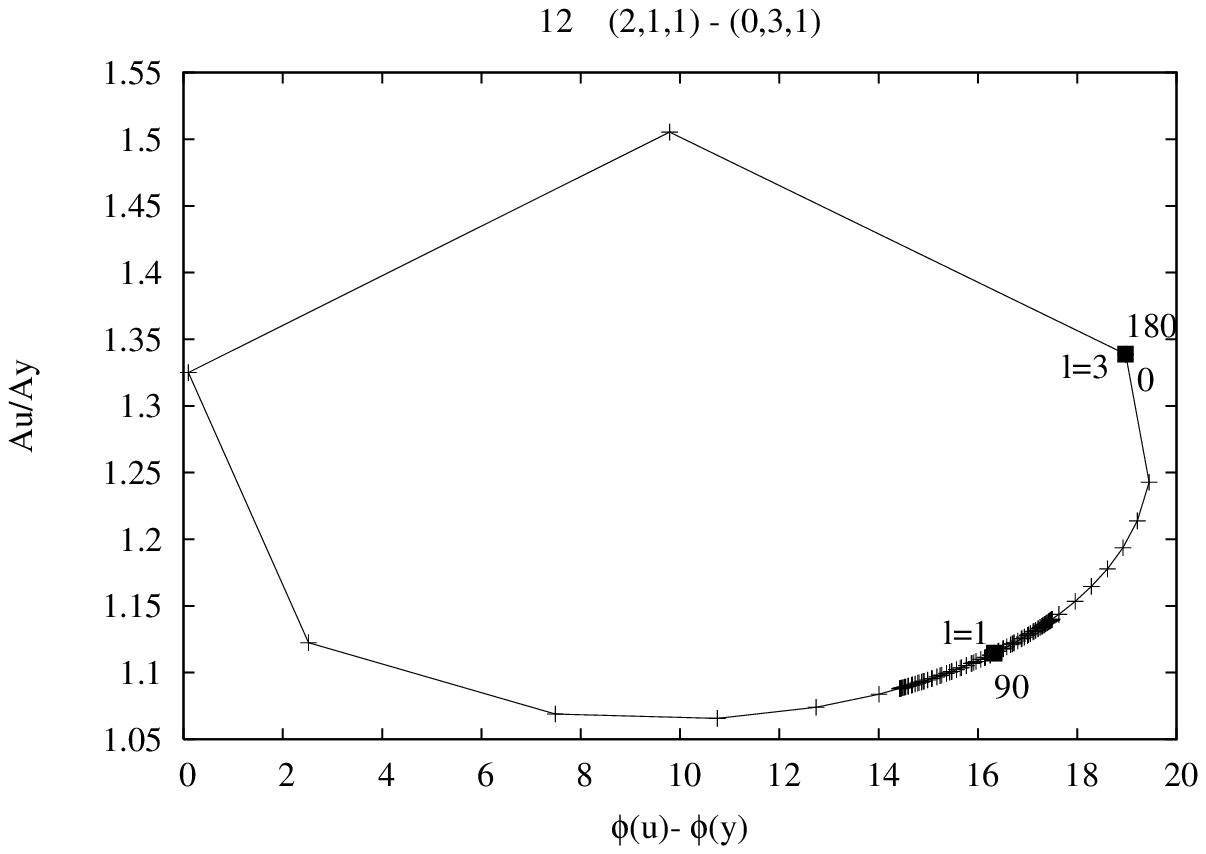}}
  \scalebox{.44}{\includegraphics{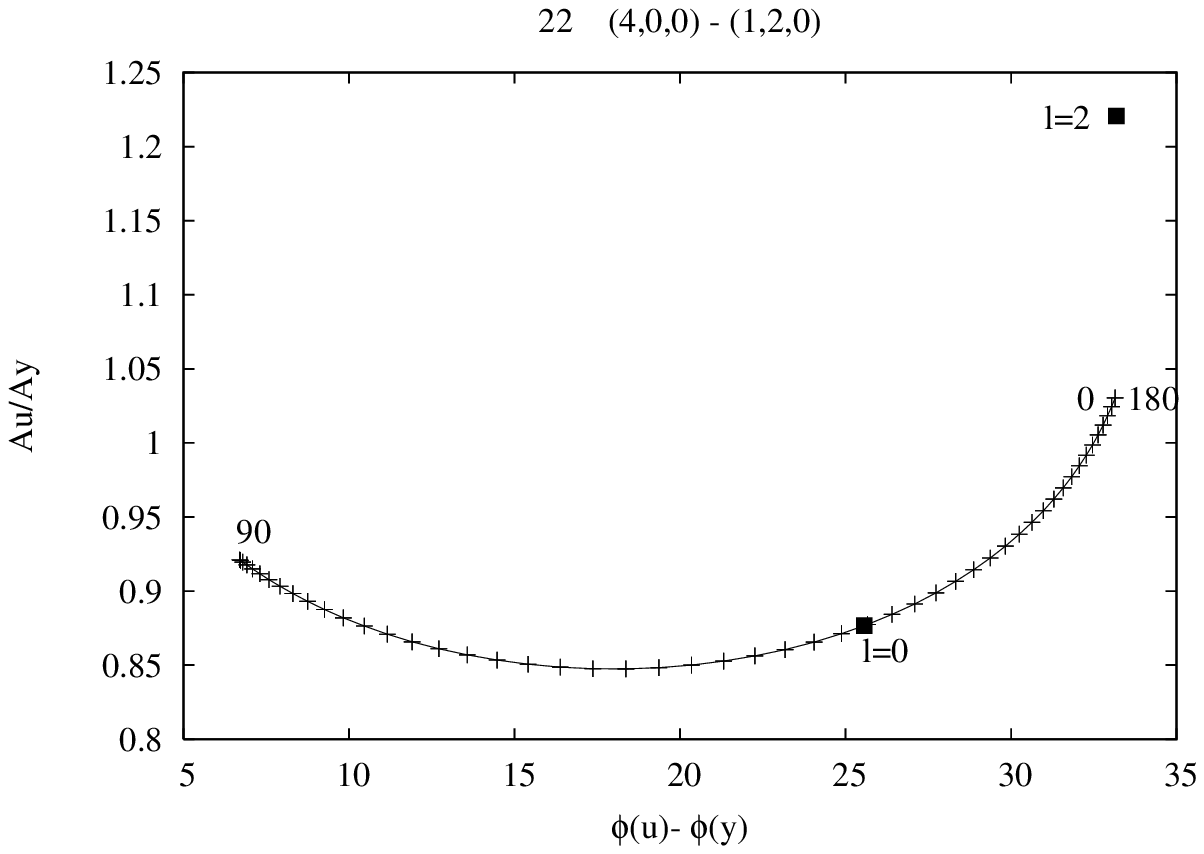}}
  \scalebox{.44}{\includegraphics{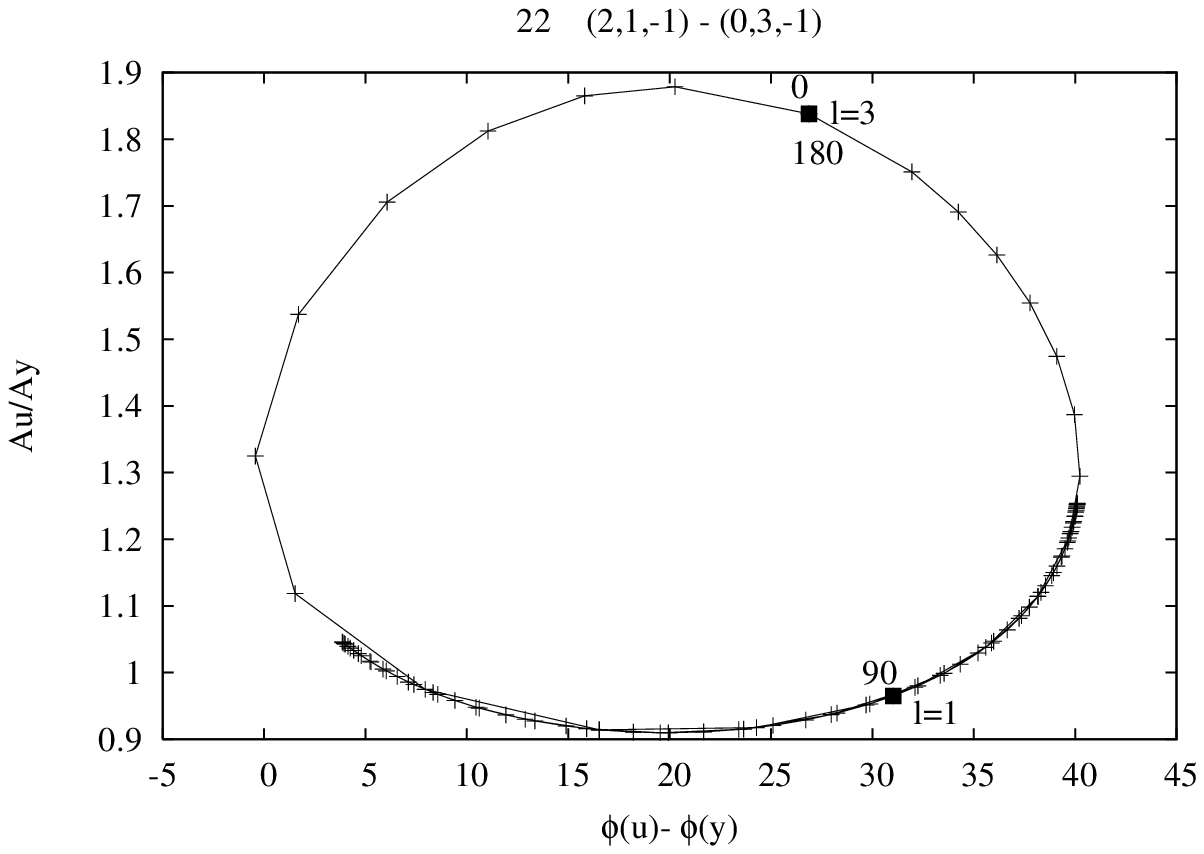}}
  \scalebox{.44}{\includegraphics{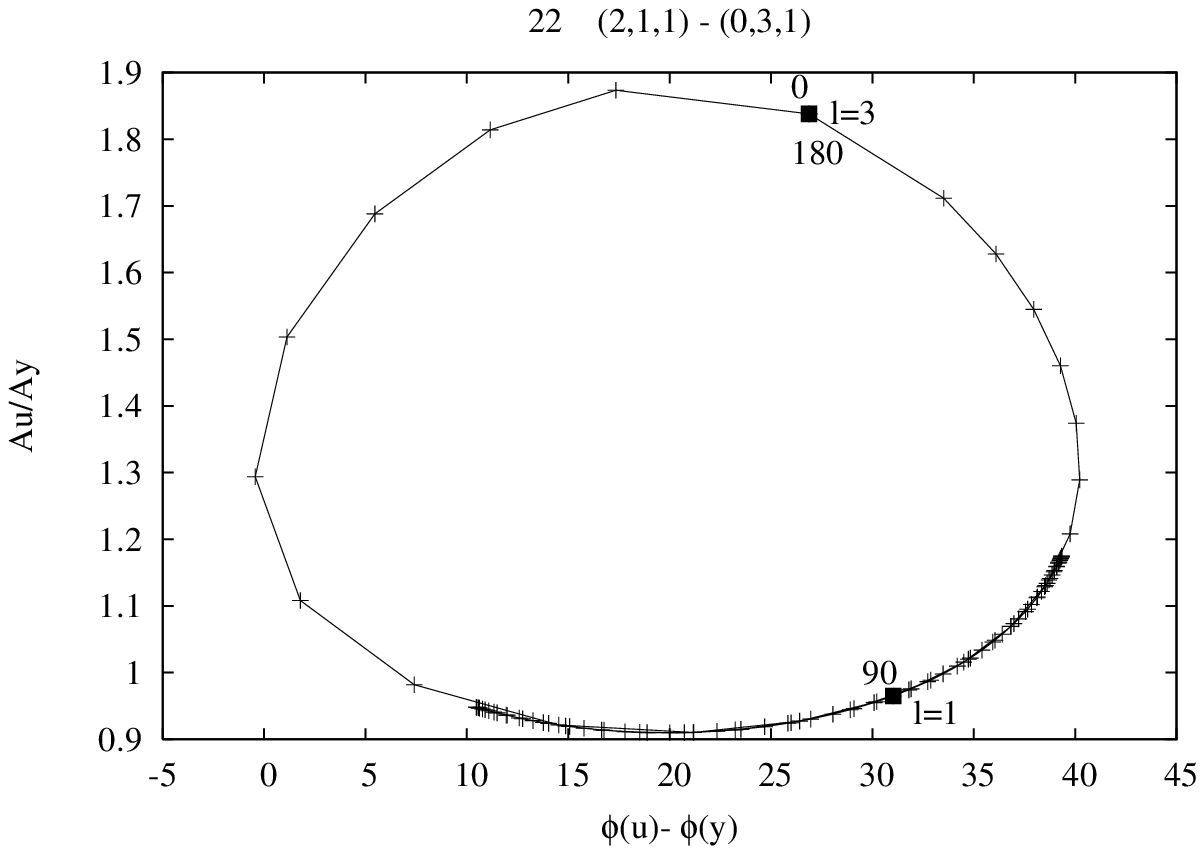}}
  \scalebox{.44}{\includegraphics{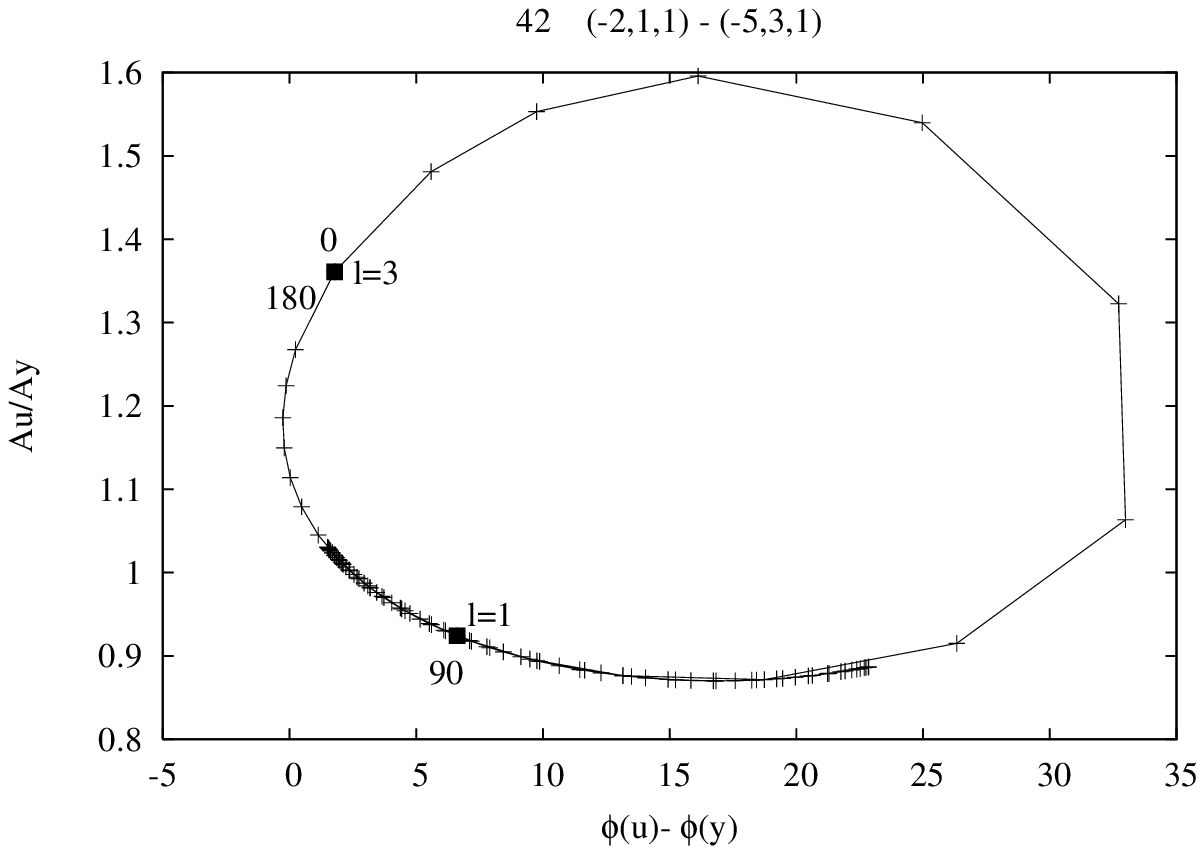}}
   \caption{Phase--amplitude diagrams for the \Strom\ filter $u$ (relative to $y$), for theoretical near-degenerate modes whose frequencies are close to the observed ones. The squares represent the location of modes obtained when rotation is not taken into account.} 
   \label{fig:f-a_yagoda}
 \end{center}
\end{figure*}

In order to take the aforementioned effects into account, we first built pseudo-rotating models representative of \rvari\ in the manner described in Sect. \ref{sec:models} for the two
rotational velocities selected.
\begin{table}
   \begin{center}
    \caption{Coupled modes found for the $i2$ models with $\Omega=52.6\,\kms$ (see Table \ref{tab:rotmodels}). The mode identification is given by its radial order $n$, spherical degree $\ell$, and azimuthal order $m$. $\sigma_a$ and $\sigma_b$ (in $\muHz$) represent the oscillation frequency. $\beta_j$ is the contamination coefficient (see text for definition).}
    \vspace{1em}
    \renewcommand{\arraystretch}{1.2}
    \begin{tabular}[ht!]{cccccc}
    \hline
    \hline
    Model & $n_a$, $\ell_a$, $m_a$ & $\sigma_a$ & $n_b$, $\ell_b$, $m_b$ & $\sigma_b$  & $\beta_j$ \\
      \hline
      02 & -1, 1, 1 & 127.63 & -4, 3, 1 & 128.24 & 0.18 \\
      02 & -1, 1, 0 & 133.52 & -4, 3, 0 & 133.97 & 0.17 \\
      02 & -4, 3, 1 & 116.27 & -1, 1, 1 & 116.88 & 0.82 \\
      02 & -4, 3, 0 & 131.00 & -1, 1, 0 & 131.45 & 0.83 \\
      \hline
      12 & 4, 0, 0 & 159.37 & 1, 2, 0 & 160.17 & 0.20 \\
      12 & 2, 1, 1 & 159.16 & 0, 3, 1 & 160.33 & 0.11 \\
      12 & 2, 1, -1 & 166.07 & 0, 3, -1 & 167.11 & 0.21 \\
      12 & 1, 2, 0 & 158.12 & 4, 0, 0 & 158.92 & 0.80 \\
      12 & 0, 3, 1 & 155.91 & 2, 1, 1 & 157.08 & 0.89 \\
      12 & 0, 3, -1 & 162.56 & 2, 1, -1 & 163.6 & 0.79 \\
      \hline
      22 & 4, 0, 0 & 159.95 & 1, 2, 0 & 160.27 & 0.20 \\
      22 & 2, 1, -1 & 164.95 & 0, 3, -1 & 166.07 & 0.15 \\
      22 & 2, 1, 1 & 157.89 & 0, 3, 1 & 159.19 & 0.11 \\
      22 & 1, 2, 0 & 159.15 & 4, 0, 0 & 159.47 & 0.80 \\
      22 & -2, 3, 1 & 129.53 & -3, 1, 1 & 130.52 & 0.87 \\
      22 & 0, 3, -1 & 162.10 & 2, 1, -1 & 163.18 & 0.85 \\
      22 & 0, 3, 1 & 155.38 & 2, 1, 1 & 156.68 & 0.89 \\
      \hline
      42 & -2, 1, 1 & 120.94 & -2, 1, 1 & 124.51 & 0.24 \\
      42 & -5, 3, 1 & 123.91 & -2, 1, 1 & 127.48 & 0.76 \\
      \hline
      \hline
    \end{tabular}
    \label{tab:degmodes}
  \end{center}
\end{table}
The main characteristics of the selected pseudo-rotating models are listed in Table \ref{tab:rotmodels}. Notice that the same labels are used to represent the different models for both the 18 and the $52.6\,\kms$ rotational velocities. This was conveniently chosen for the sake of clarity (the location of models evolved with a rotational velocity of $18\,\kms$ and the equivalent with $52.6\,\kms$ are quite close in the HR diagram).

\subsection{Near degeneracy and phase-amplitude diagrams\label{ssec:neardeg}}

Using the pseudo-rotating models described in Table \ref{tab:rotmodels} we compute the corresponding oscillation frequencies in the manner described in Sect. \ref{ssec:oscil-insta}. In principle, the presence of non-radial modes (up to $\ell=3$) is not discarded and the investigation is extended to degrees up to $\ell\leq3$. Higher $\ell$ values are thought to suffer from cancellation effects, thereby affecting its visibility. As expected, for models with a rotational velocity of $18\,\kms$, the effect of rotation on the oscillation frequencies (including the effect of near degeneracy) is found to be small as far as the position in the HR diagram, metallicity, or $\alpha$ value, is concerned. Thus for this rotational velocity, the results given in the previous sections remain valid. However, if the star is not observed from the equator, for example with $i=20^\circ$ (i.e., $\Omega=52.6\,\kms$), the effect of rotation cannot be neglected. Since we are interested in studying the impact of the rotation effects in the whole method and, more particularly, in the phase-amplitude diagrams, only models rotating at $\Omega=52.6\,\kms$ are considered hereafter.

A complete list of degenerate modes close to the observed frequencies is reported in Table \ref{tab:degmodes}. We calculate the ratio between their \emph{horizontal} components for each degenerate mode couple. Following the formalism given in \citet[][hereafter SGM]{Sua06rotcel} for two coupled modes $j\equiv(a,b)$, this relation is written in the form:
\eqn{\beta_{a,b}=
  \displaystyle\frac{m_{ab}^{(1)}+m_{ab}^{(2)}} {m_{aa}^{(1)}+m_{aa}^{(2)} + \displaystyle\frac{\delta\omega_0}{2} + \displaystyle\frac{\delta\omega_0^2}{8\,{\bar \omega}_0}},}
where $m_{ij}^{(1)}$ and $m_{ij}^{(2)}$ ($i,j=a,b$) represent the elements of the first- and second-order matrix ${\cal M}_{ij}$ given, respectively, by Eqs. (22) and (23) of SGM, obtained from the oscillation equation in the presence of near degeneracy (SGM, Eq. 20). The non-perturbed eigenfrequency of modes are included in the terms ${\bar \omega}_0$ and $\delta\omega_0$ as defined in SGM (Eqs. 16 and 17, respectively). Hereafter, $\beta_{j}$ quantities are called \emph{contamination} coefficients. They roughly provide the ratio of the original horizontal component in the coupled mode. Analysis of the contamination coefficients of modes listed in Table \ref{tab:degmodes} reveals that, systematically, the lowest value of $\beta_{j}$ is assigned to the lowest $\ell$ value and vice-versa. Inspired by genetics, such modes can be called \emph{recessives}, and those with the highest $\beta_{j}$ value, \emph{dominants}.

As claimed in \citet{Pagoda02}, the coupled-mode positions in the amplitude/phase diagrams become both aspect- and $m$-dependent. Following that work, we calculated amplitude/phase diagrams for modes of Table \ref{tab:degmodes}, varying the angle of inclination $i$ from $-180^\circ$ to $180^\circ$, in steps of $\cos\,i=0.02^\circ$. The results are depicted in Fig. \ref{fig:f-a_yagoda}, in which only the filter $u$ (as compared to $y$) is used. For the sake of clarity and brevity, only \emph{recessive} modes
are considered, because \emph{dominant} modes show similar results and do not provide additional information. For comparison, the values of the amplitude ratios and phase differences obtained for the non-rotating  case are also depicted. Note that the density of points varies along the curves. This means that the amplitude and phase predictions are found to be very close each other for a significant number of inclination angles. Points of curves corresponding to $\ell=2$ and 3 modes are systematically located close to $i=\pm180^\circ$ regions. The concerned \emph{dominant} modes present contamination coefficients close to 1; therefore, their contribution to the horizontal component of the coupled mode is almost the maximum possible. However, if the star is observed at an intermediate $i$ angle, the situation becomes less dramatic, and it is possible to identify the \emph{recessive} modes correctly. In the present case, it means that the $\ell$ identification performed using amplitude/phase diagrams without taking the rotation effects into account remains valid for $\ell=0$ and 1 modes.

\section{Petersen diagrams for \rvari \label{sec:pd}}

The Petersen diagrams (PD) show the ratio between the fundamental radial mode and the first overtone periods, $\ratio$, as a function of the fundamental mode period, $\log\po$, for a given stellar model. These well-known diagrams are particularly useful for constraining the mass and metallicity of double-mode radial pulsators like HADS. In the case of \rvari, the ratio between the observed frequencies $f_1$ and $f_2$ is $0.772$, which lies in the range of values typically observed in Pop. I double-mode pulsators (see Sect. \ref{sec:intro}). Our previous modelling points towards the possibility that $f_1$ and $f_2$ correspond to the radial fundamental mode and its first overtone (for $0j$ models). It is thus worth investigating the observed periods in the framework of the PD.

\begin{figure}
 \begin{center}
    \scalebox{.40}{\includegraphics{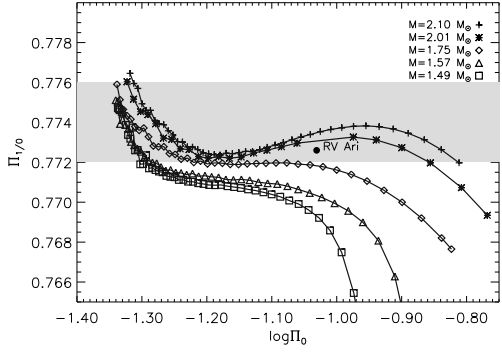}}
   \caption{Typical PD ($\po$ in $\mathrm{d}$) containing different tracks that correspond to the evolutionary sequences of $i2$ models, computed with a metallicity of [M/H]=0.01. The shaded area corresponds to typical values found for Pop. I stars. The filled symbol represents the observed $\ratio$ for the double-mode high-amplitude \ds\ star \rvari.}
   \label{fig:classic_PD}
 \end{center}
\end{figure}

In Fig. \ref{fig:classic_PD}, a typical PD for $i2$ non-rotating evolutionary tracks is depicted. The shaded region indicates the typical $\ratio$ values found for Pop. I stars in the range of $\ratio=[0.772, 0.776]$. The five tracks were computed with the observed metallicity [M/H]=0.01\,dex, for which the range of masses $[1.49, 2.10]\,\msol$ was found. The mass for \rvari\ is then predicted between 1.75 and $2.01\,\msol$, in agreement with the results obtained by \citet{Poretti05hads} when considering solar metallicity.

\begin{figure}
 \begin{center}
    \scalebox{.40}{\includegraphics{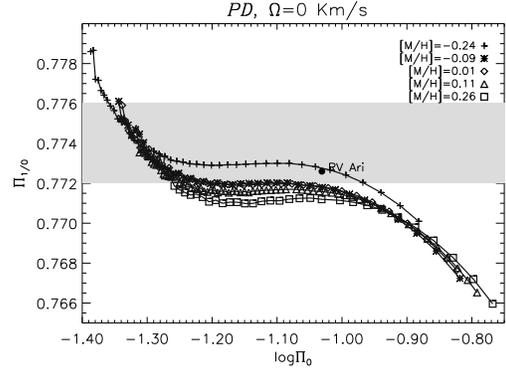}}
   \caption{PD ($\po$ in $\mathrm{d}$) showing evolutionary sequences of the \emph{central} non-rotating model $02$, computed for 5 different metallicities around the observed [M/H]=0.01. As in Fig. \ref{fig:classic_PD}, the filled symbol represents the observed $\ratio$ for \rvari, and the shaded area corresponds to typical values found for Pop. I stars.}
   \label{fig:z_PD}
 \end{center}
\end{figure}

Furthermore, PD diagrams can also be used to constrain the metallicity of representative models. As shown by \citet{Petersen73, PetDalsgaard96, PetDalsgaard96_2}, the period ratios increase when decreasing the stellar initial metal content. This property is illustrated in Fig. \ref{fig:z_PD}, where tracks of the non-rotating model $02$ are depicted for the five metallicities studied in the previous sections. Accordingly, each track corresponds to models with a different mass, from $1.62\,\msol$ for [M/H]=-0.24\,dex to $1.86\,\msol$ for [M/H]=0.26\,dex. From this figure, the best agreement between models and observations is obtained for the track of [M/H]=-0.24\,dex, for which the closest model has a mass of $1.62\,\msol$, $\log\teff=3.858$, $\log L/L_\odot=1.09$, and an age of 1478 Myr. In principle, this would correspond to $3j$ models (see Table \ref{tab:models-instab}) rather than $0j$ models.
\begin{figure}
 \begin{center}
    \scalebox{.40}{\includegraphics{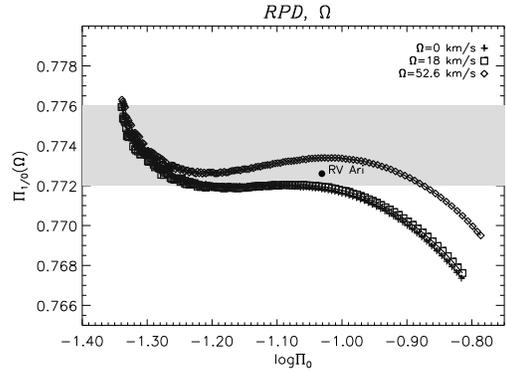}}
   \caption{PD ($\po$ in $\mathrm{d}$) showing evolutionary sequences of the \emph{central} pseudo-rotating model $02$, computed for the two rotational velocities considered in this work: 18 and $52.6\,\kms$. A third track corresponds to the non-rotating case shown in Fig. \ref{fig:classic_PD}.}
   \label{fig:rot_PD}
 \end{center}
\end{figure}
To fit the $0j$ models, it would thus be necessary to increase the value of the metallicity of the models (and therefore the mass) and/or modifying their rotational velocity. Recently, \citet{Sua06pdrot} showed that rotation can also modify the PD, and must be taken into account when performing accurate predictions for these parameters, even for slow rotators. Indeed, as shown in \citet{Sua06pdrot}, rotation increases the $\ratio$ period ratios for a given mass and metallicity. In contrast, for a given mass (and rotational velocity), such period ratios increase when decreasing the stellar metal content. This is illustrated in Fig. \ref{fig:rot_PD}, where, for the $0j$ models computed with the observed metallicity, we compare the non-rotating track shown in Fig. \ref{fig:classic_PD} with two other pseudo-rotating models with $\Omega=18$ and $52.6\,\kms$. As expected, when considering $\Omega=18\,\kms$, models are predicted to be located in the PD very close (larger $\ratio$ ratios) to those in the absence of rotation. On the other hand, if we consider $\Omega=52.6\,\kms$, models are predicted to show higher period ratios than the observed ones. Thus, we can argue that the best location would be somewhere in between, that is, in the range [$1.75, 2.01]\,\msol$ with a rotational velocity around $35\,\kms$ (which corresponds to $i=30^\circ$ approximately) and with a metallicity approximately in the range $[-0.09,0.11]$.

\section{Conclusions\label{sec:conclusions}}

We have performed a comprehensive theoretical study of the HADS star \rvari\ using some of the most modern techniques available for modelling intermediate-mass pulsating stars. Updated Kurucz models were included in order to take the atmosphere-pulsation interaction into account. Different approaches were followed to calculate the oscillation frequencies. Non-adiabatic observables were calculated in order to perform instability analysis and mode identification within the framework of multicolour photometry. Moreover, adiabatic oscillation spectra corrected for the effect of rotation up to the second order (in a perturbative theory) were also calculated. This allowed us to examine the effect of rotation on previous studies, in particular, for multicolour diagnostic diagrams and for Petersen diagrams.

Non-adiabatic calculations point toward models located at the high-luminosity regions of the observational error box. Detailed analysis reveals strong constraints on the efficiency of the convection. In particular, we showed that the mixing-length parameter, $\alpha$, is constrained to values close to $\alpha=0.5$.

Multicolour diagnostic diagrams are found to be compatible with the assumption that the observed frequencies $f_1$ and $f_2$ are radial. However, the error bars of the observed amplitudes and phases for the different \Strom\ filters do not allow modes to be identified in an unmistakable way. Nevertheless, all the amplitude ratios variations are predicted to be compatible with the calculated theoretical models that fulfill the above constraints. The study of colour index variations has allowed us to significantly constrain the effective temperature and evolutionary stage of the models to [3.849, 3.860] dex and [1190, 1270] Myr, respectively. This constitutes one of the most accurate modellings ever performed for a \ds\ star, and for \rvari\ in particular.
\begin{table*}
  \begin{center}
    \caption{Compilation of the most restrictive ranges of physical parameters obtained in the present work for RV Ari. The first line corresponds to the results obtained using non-rotating, non-adiabatic models, and the second line shows results from using rotating, adiabatic models.}
    \vspace{1em}
    \renewcommand{\arraystretch}{1.2}
    \begin{tabular}[ht!]{cccccccc}
    \hline
    \hline
      & $M$ & $\log T_{\mathrm{eff}}$ & $\log L/L_\odot$ & $\log g$ & $\Omega$ & Age & [M/H] \\
      &  $\mathrm{M}_\odot$ & dex & dex & dex & $\kms$ & Myr & dex \\
    \hline
    NRNAM & $1.70-2.00$ & $3.860- 3.849$ &  $1.13-1.14$ & $3.94-3.89$ & 0 & $1190-1270$ & $0.01\pm0.20$ \\
    RAM & $1.75-2.01$ & $3.903-3.821$ & $1.38-1.15$ & $3.93-3.77$ & $18-53$ & $795-1430$ & $-0.09-0.11$ \\
    \hline
    \end{tabular}
    \label{tab:rangesmethods}
  \end{center}
\end{table*}

As expected, rotation was found to affect the amplitude/phase diagrams depending on the inclination angle $i$ of the star for near-degenerate modes. However, analysis of the variations of amplitude ratios and phase differences as a function of $i$ (using the \emph{contamination coefficients}) has revealed that the amplitude/phase diagrams are still useful for the most probable $i$ configurations of the star.

Petersen diagrams were studied including the effect of rotation. Frequency ratios were found to be compatible with the theoretical predictions obtained with other modelling techniques discussed previously.

Interestingly, the coherence and completeness of the results obtained (see Table \ref{tab:rangesmethods}) constitute strong evidence of the capabilities and robustness of the method presented here. Indeed, the full capabilities of the method will be tested once more frequencies with lower amplitudes have been detected by the forthcoming space mission COROT\footnote{http://corot.oamp.fr} \citep{Baglin02}.

\acknowledgements{This work was partially financed by the Spanish "Plan Nacional del Espacio", under project ESP2004-03855-C03-01. JCS acknowledges the financial support of the European Marie Curie action MERG-CT-2004-513610 and the Spanish "Consejer\'{\i}a de Innovaci\'on, Ciencia y Empresa" from the "Junta de Andaluc\'{\i}a" local government.}

\bibliography{5054bib.bib}
\bibliographystyle{aa}

 \end{document}